\documentclass[runningheads]{svmult}

\usepackage{makeidx}   
\usepackage{graphicx}  
\usepackage{subeqnar}  
\usepackage{multicol}  
\usepackage{epstopdf}	

\usepackage{physprbb}  
\makeindex             

\def\ni{\noindent }
\def\etal{{\it et al.}}

\def\y3{{v}}

\usepackage{amsfonts}   
\usepackage{epsfig}  
\begin{document}
\title*{Effects of rotation on  stellar  p-mode frequencies }
%
%
%
%
\titlerunning{Effects of rotation on  stellar  p-mode frequencies }
%

\author{Mariejo Goupil}
\authorrunning{Mariejo Goupil}

\institute{Observatoire de Paris,  \\ 
5 place Jules Janssen, 92190 Meudon principal cecex, France \\
E-mail: \texttt{mariejo.goupil@obspm.fr}}


\maketitle              

\begin{abstract}
\noindent Effects of stellar rotation on adiabatic oscillation frequencies of pressure modes are discussed. 
Methods to evaluate them are briefly exposed and  
some of their main results are presented. 
\end{abstract}


\section{Introduction}

Many pulsating stars oscillate with pressure (or p-) normal modes 
for which  pressure gradient is the main restoring force. A nonrotating  star  
keeps its spherical symmetry; 
consequently,  
 frequencies of  its free vibrational modes\index{vibrational mode} are  $2\ell+1$
  degenerated when using a description  by means of spherical harmonics, $ Y_{\ell}^m$ 
 ($\ell$ the degree, $m$ the azimuthal indice) for the angular dependence
 of the variations.   However, stars are rotating  and their rotation 
 affects their oscillation frequencies. This is an advantage  since 
 rotation breaks the spherical symmetry of the star; as a consequence,   
the  $2\ell+1$  degeneracy  is lifted and  gives access to information about the internal rotation.
The information is relatively easy to extract 
if the  star rotates slowly provided observations are of high quality enough 
that they enable  to achieve the necessary 
high accuracy  of  frequency measurements. On the other hand, when the star is rotating fast,
  one must properly  take into account   the consequences of the spheroidal shape of the star 
on the oscillation frequencies before extracting seismological information on the 
internal structure or the rotation profile of
the star. 
Three methods have been used to investigate the effect of rotation on oscillation
frequencies of stellar  pressure modes: a variational principle, perturbation
techniques and direct numerical integration of a  two dimensional eigenvalue system. We  briefly describe the 3 methods and their
main results. All these methods assume the knowledge of the structure of the distorted star, 
which is obtained either by perturbation or by 2D calculations. 
This  is also  sketched when appropriate.

This lecture is  priorily adressed to  PhD students who possess some basic background in stellar seismology but wish to
learn with some details how to handle the impacts of the rotation
 on the oscillation frequencies of a star. 
The present lecture cannot be exhaustive neither in the credit to authors nor to the numerous effects of  rotation on stellar
pulsation. Indeed, due to the importance of stellar rotation in a more general astrophysical context, 
a tremendous amount of work   has been performed on stellar rotation and its interaction with stellar oscillations 
since the 1940-1950s.  For more details, 
 we refer the reader to standard text books such as Ledoux \& Walraven (1958); Cox (1980); Unno et al (1989) and to 
 litterature on the subject:   Gough,Thompson (1991), Gough (1993),  Christensen-Dalsgaard (2003, CD03) and references therein.

The organization of this contribution is as follows: 
as a start, some definitions and orders of magnitude are given in Sect.2. 
Sect.3 introduces the basic background in deriving 
the  wave equation   and eigenvalue problem modelling the  adiabatic pulsations of a rotating star.  
Sect.4 discusses the existence and consequence of a variational principle for rotating stars. 
Sect.5  and 6 respectively concern calculation of
the oscillation frequencies of respectively slowly and modetately  rotating stars 
by means of perturbation methods.
Sect.7 then turns to fast rotating stars and the computation of
their oscillation frequencies with nonperturbative techniques. 
Some comparisons of  the results of perturbative and nonpertubative
methods  and comments on the range of validity of the first ones are given  in the case of polytropic models. 
Sect.8  briefly discusses
observations of real stars and  forward inferences  on stellar internal rotation    sofar derived. 
Sect.9 turns to inverse methods  specifically applied to  obtain the internal rotation profile.  Finally 
results of
inversions for the rotation profiles of some stars are presented in Sect.\ref{inversion}.   To keep up with the allowed number of pages, 
the choice has been to show almost no figures  and rather to 
refer in the text  to plots in the litterature.  Some  illustrations of the 
consequences of  rotation on pulsations of
 stars can also be found in Goupil et al. (2006).


\section{Definitions and orders of magnitude}

 We will denote $\omega^{(0)}_{n,\ell}$ 
  the pulsation\index{pulsation} of an oscillation  mode 
  with radial order $n$  and degree $\ell$ for  a nonrotating star and  
  $\nu^{(0)}_{n,\ell} =\omega^{(0)}_{n,\ell}/2\pi$ the 
 associated  frequency in $Hz$. 
 For a symmetric star of radius $R$ and mass $M$ and for any given eigenfrequency $\omega$,
  we define the dimensionless frequency, $\sigma$:
   \begin{equation}
   \sigma= {\omega \over (GM/R^3)^{1/2} }
   \label{sigma}
   \end{equation}

Consider now  a star which rotates 
with a rotational period $P_{rot}$. We denote  $ \Omega = 2\pi /P_{rot}$, its angular rotational velocity. 
One also defines the equatorial velocity  
$v_{eq}= R_{eq} \Omega $  where  $R_{eq}$ is the equatorial stellar radius, this is the velocity of the fluid 
at the equatorial level generated by rotation.
One also  uses the projected velocity $v \sin i$ where 
 $i$ is the angle between the line of sight  and the rotation axis. 

\subsection{Why are the oscillation frequencies of a rotating  star modified compared to those a nonrotating star
?}

\ni ({\it i)} A first effect is easy to understand as it is    purely  geometric. 
  Let consider  a uniformaly rotating star which 
 oscillates  with a pulsation frequency $\omega^{(0)}_{n,\ell}$ 
 in the rotating frame. 
    In an inertial frame, an observer will see 
    oscillations with frequencies given by  $\omega_{n,\ell,m} = \omega^{(0)}_{n,\ell} + m \Omega$ with  $m \in [-\ell,\ell]$
with progade ($m>0$) et retrograde   ($m <0$) modes. Each $(n,\ell)$ mode appears in a power spectrum as 
a frequency multiplet split by rotation  composed of $2 \ell +1$ components.

\ni {\it (ii)} At the same level (i.e. linear in the rotational angular velocity $O(\Omega)$), 
one must also take into account intrinsic  effects  of rotation in the rotating frame  related to 
the Coriolis  force  which   directly acts on the oscillating fluid velocity, deflecting the wave.
  Ledoux (1951) has shown that 
  the Coriolis acceleration modifies the  pulsation   such that in the observer frame, it becomes:
  \begin{equation}
   \omega_{n,\ell,m} = \omega^{(0)}_{n,\ell}  + m \Omega (1-C_{n,\ell}) +O(\Omega^2)
   \end{equation}  
  where $C_{n,\ell}$ is the Ledoux  constant.  The rotation rate then  is  simply given by 
   \begin{equation}  \Omega= (\omega_{n,\ell,m} - \omega^{(0)}_{n,\ell})/ (m (1-C_{n,\ell}))   \end{equation}  
   where the Ledoux constant is assumed to be known from a
    nonrotating stellar model. 
 However,   the rotation of a star is not necessarily uniform: it can vary with latitude and depth; 
 this again results in
   modifications of the frequencies. Actually these modifications are  those one really wishes to identify. Indeed
   current open issues  are:  
     what are  the regions inside the star  where the rotation is uniform  and where it is not? 
   how strong  are rotational  gradients? 
   what are the mechanisms which  make the region uniform or in contrary generate  differential  rotation?
    how to model them ot test the validity of  their  modelling?
   
  The magnitude of the above   effects are proportional to 
  $\Omega$. For a slowly rotating star indeed,  effects of the centrifugal force, proportional to  $\Omega^2$, on oscillation
  frequencies  are small and one usually ignores them.  
  The {\it rotational splitting}\index{rotational splitting} of a mode with given $(n,\ell)$ then is  defined as:
\begin{equation}
\delta \omega_{n,\ell} = {\omega_{n,\ell,m} - \omega_{n,\ell,0}  \over m}  
\label{split0}
\end{equation}
where $\omega_{n,\ell,0}$  can  reasonably be identified  with $\omega^{(0)}_{n,\ell}$  for a slow rotating star.

\ni {\it (iii)} For   stars rotating faster, one must take into account 
the various effects due to the  distorsion of the star by the centrifugal force:

\ni {\it`mechanical' impact}:  the centrifugal force\index{centrifugal force} 
affects the wave propagation  by deforming the resonant  cavity where 
 they  propagate. Pioneer studies  were due  to 
 Ledoux (1945) and Cowling, Newing (1949).  Ledoux (1945) studied radial oscillations of  a uniformaly rotating star 
by means of a variational principle and assuming  uniform dilation
and compression motions. He  obtained the squared frequency of the fundamental radial mode under the
influence of the centrifugal force as
 $$\omega^2 = -(3\Gamma_1-4) {E_g \over I_0}+(5-3\Gamma_1)2T$$
 where $I_0, \Gamma_1,  E_g, T$ respectively denote the moment 
 of inertia, the adiabatic exponent (assumed constant throughout the
 star),   gravitational potential  and rotational kinetic energies.
For a homogeneous spheroid in uniform rotation (Simon, 1969),
  $E_p/I_0 \sim 4\pi G \rho$  decreases whereas 
  $T = {\cal M}/I_0 \sim \Omega^2/3$ increases (${\cal M}$ is the  total angular momentum) when $\Omega$ increases.
 Cowling, Newing (1949)  extended the study to 
  nonradial oscillations  of a rotating  configuration
    and obtained   numerical results for radial
  oscillations of a polytrope.

 In order to obtain the rotational profile, 
 one defines the {\bf generalized  rotational splitting }  of a mode with given $(n,\ell)$ as:
\begin{equation}
S_m = {\omega_m -\omega_{-m}\over 2 m}
\label{split1} 
\end{equation}
This 2nd definition  eliminates second order perturbation effects (cf Sect.\ref{third})
 that-is  eliminates the mechanical effects of the nonspherically symmetric distorsion\index{distorsion} 
 of the star  
 on the oscillation frequencies.

\ni {\it`thermal' impact}: Works by  Von Zeipel (1924ab),
 Eddington (1929), Vogt (1929), Sweet (1950)  showed that 
  the non spherical  shape of a rotating star   causes a   departure from 
thermal equilibrium  which generates   large scale meridional currents. 
 As already emphasized by Ledoux (1945), this meridional circulation
can influence the pulsations. Indeed it  generates a differential rotation which 
 causes  hydrodynamical  instabilities. The net effect results 
  in  transport of chemical  elements and angular momentum in radiative regions 
 and  modifications of 
  the thermodynamical state of the star.  All these   processes
  influence the star evolution  and  structure   
   (Tassoul (1978), Zahn (1992); see also  
   Meynet , Maeder, (2000); Mestel (2003); Zahn (2003)  for  reviews), in particular the adiabatic sound speed and the rotation profile, 
  the two main quantities {\bf accessible} with seismology. 
      Hence it is important to note that  even for a star which is assumed to be in near 
    spherical symmetry and with a given rotation profile at the time of observation,
   there exists a nonzero difference 
between  the zeroth order  eigenfrequencies of a rotating model which has been evolved 
 taking into account the effect of rotation  and that of an 'equivalent'  model  
at the same age evolved without rotation in which effect of rotation law is  imposed afterward 
or which has been evolved with rotation
included only through the effective gravity.
For a slow rotating star however, this difference can often be neglected.

\subsection{How to determine the frequency changes due to rotation?}
 
   The methods which must be used in order to determine precisely how and how much  
   the oscillation frequencies are affected by
   rotation  can be divided into 2 groups  (apart from the variational technique): 
the  {\it perturbative methods} on one hand and the 
 {\it direct numerical approaches or  non perturbative methods} on the other hand. 
 The choice will depend whether the star is a  rather slow or rather fast rotator. This classification can be loosely made
 according to values of dimensionless parameters. 
 The ratio of the time scales related to Coriolis force and pulsation is  roughly given: 
\begin{equation}
\mu =   {P \over P_{rot}}= {\Omega \over \omega}
\label{mu}
\end{equation}
Another  important parameter is the ratio of rotational kinetic energy to the 
gravitational one. As an order of magnitude, it can be evaluated as 
\begin{equation}
\epsilon_{eq} = {\Omega^2 R_{eq}^2 \over GM/R_{eq}} =  {\Omega^2 \over
(GM/R_{eq}^3)} =\mu^2 ~ \sigma^2
\end{equation}
where $M$ is the  mass of the star ; $R_{eq}$ its radius at the equator; $G$ is the 
 gravitational constant. One also uses the flatness, $f$,  as defined by 
$ f= 1-{R_{pol}/  R_{eq}}$
where $R_{pol}$ is the radius at the pole.
 
In the following sections, we will then   distinguish fast, moderate and slow rotators  with respect to the
pulsations of the stars according to these parameters.  Stars  for which these parameters 
 obey  ($\mu,\epsilon <<1$ ), ~($\mu,\epsilon  < 0.3$)
  or  ( $\mu$ or $\epsilon > 0.3$)
 are classified respectively  as slow, moderate  or fast rotators. For slow rotators, a first order perturbation is enough; for a moderate 
 rotator,  a perturbation method can still be valid provided 
   higher order contributions are included. Studying oscillations of fast rotators
   requires the use of   2D equilibrium   models as well as solving a 2D eigenvalue system.
 However one must keep in mind that the above  classification   depends on the star as it depends on the frequency range  
of its excited modes. Perturbation methods cease indeed to be valid whenever the wavelength of the mode   
reaches the order of $\sim \epsilon ~R$ where $R$ is the stellar radius that-is for p-modes with  high enough
frequencies.

In the solar case,  $P_{rot,\odot}$ varies from $\sim  25$ days at the
equator up to $P_{rot,\odot} \sim 35$ days at the poles then
 $\epsilon_{eq,\odot} = 2.1~10^{-5}$  is very small. 
 The splitting is of the order $0.456 \mu$Hz  
 with a Coriolis strength   $\mu = 1,4 ~10^{-4}$. 
With $\mu,\epsilon << 1$, the Sun is considered as a slow rotator with respect to its
  pulsations.  However, measurements of the rotational splittings\index{rotational splitting} 
  are so accurate for the Sun-  
 one is able  to  determine  the internal rotation velocity of the Sun as a function of radius and 
 latitude with unprecedented spatial resolution and accuracy-  that  
 one measures  second order effects in the rotational splittings which must be taken
into account in order to  obtain information on the solar magnetic field
    (Gough, Thompson, 1990 (GT90); Dziembowski \& Goode, 1992 (DG92))  and to determine the 
    gravitational quadrupole moment of the Sun  with much higher accuracy 
(Pijpers, 1998).

\medskip 
For a $\delta$ Scuti-type  variable star   ($~2M_\odot, 2R_\odot$), 
observations
show that   
$v \sin i = 70- 280$ km/s  that-is 
$P_{rot}\sim 1,45$ days to $\sim 8$ hours ; this yields  $ \Omega= 5$ to $  20~10^{-5}$ rad/s and
a rotational splitting of the order of 
$ \nu_{rot}= 8$ to $32 \mu$Hz. Typical values for the  dimensionless parameters  then are
$\epsilon= 0.025-0.41 $  and $\mu= 0.06-0.6 $  for an oscillation frequency range
of $200-500 \mu$Hz. Altair is an exemple of a rapidly rotating $\delta$ Scuti star  with a rotation
velocity between 190 and 250 km/s (Royer \etal~ 2002)  which has been discovered
to pulsate over at least 7 frequencies in the range 170-350  $\mu$Hz (Buzasi \etal~ 2005).
With   $M=1.72\pm 0.05 M_\odot$ and $R=1.60\pm0.12$ for  its mass and radius as found in the litterature,
one obtains $\epsilon \sim 0.305$ and $\mu \sim 0.09-0.2$ which definitely classes this star as a rapid rotator 
with respect to its pulsations.

\medskip 
A  $\beta$ Cepheid-type variable star 
 ($~9M_\odot, 5 R_\odot$) can  reach 
$v \sin i =300$ km/s that-is  
$P_{rot} < 20$ hours.This yields  $\Omega >8.6 ~10^{-5} $rad/s and $\nu_{rot} > 14\mu$Hz.
 For the dimensionless parameters, one derives
$\epsilon >  0.26  $ and for 
$\nu \sim  500  \mu Hz$ , $\mu > 2.7 10^{-2} $

A much more evolved star such as a Cepheid\index{Cepheid} is a  radially
pulsating star and a slow rotator.   However  period ratios of radial modes 
 can be quite significantly affected by
 rotation as  mentionned  by Pamyatnykh (2003, Fig.6) for a $\delta$ Scuti star\index{$\delta$ Scuti star} 
and quantified by Suarez \etal ~ (2006a, 2007) for a Cepheid.

     The above  figures  and classifications  must be seen as illustrative and only 
     a comparison between perturbative and nonperturbative
 approaches can give more precise delimitations between the different classes  (Sect.\ref{valid}).

\section{A wave equation  for a rotating star}

A derivation of the wave equation for a rotating star can be found  in  Unno et al. (1989). 
The basic equations for a non magnetic, self gravitating fluid  are written as : 
\begin{eqnarray}
{\partial  {\bf v} \over \partial t} &+& {\bf v} \cdot {\bf \nabla} {\bf v} 
= -{1\over \rho} {\bf \nabla} p - {\bf \nabla} \psi  \\
{\partial \rho \over \partial t} &+&  \nabla \cdot (\rho {\bf v}) = 0 \nonumber\\
\rho T {\partial \vec S\over \partial t} &+& \rho T ({\bf v \cdot  \nabla})S = \rho \epsilon -{\bf \nabla  \cdot {\cal F}}  \nonumber\\
\nabla^2 \psi &=& 4\pi G \rho \nonumber
\label{basic}
\end{eqnarray}
where the quantities take their usual meaning and
$$ \psi(\vec r,t)= G\int_V ~{\rho(\vec r',t)\over |\vec r-\vec r'|} ~\vec d^3r'$$
In an inertial frame, the effect  of rotation appears through
 the inertial term  ${\bf v \cdot \nabla  v}$. In the frame rotating with the star, 
 the inertial term acts as 2  fictive forces: the Coriolis  force 
$(-2 \rho {\bf \Omega} \times {\bf  v})$  which modifies the dynamics
and insures the conservation of angular momentum and the centrifugal force 
 ($-\rho  {\bf \Omega} \times  {\bf \Omega} \times {\bf  r}$) which 
 affects the structure of the equilibrium configuration of the star, i.e. 
 the star is distorted and loses its 
 spherical  symmetry.   The oscillations are described  as a
  linear perturbation about an equilibrium or steady-state
configuration which is assumed to be known.

\subsection {Equilibrium  structure for a rotating star}
We assume a steady state configuration i.e. all local time  derivatives vanish, 
the  quantities describing this equilibrium  are  given a subscript 0, for instance
 the steady state velocity field is noted  $\vec v_0$.
The assumption of stationarity leads to: 
   \begin{eqnarray}
{ 1\over \rho_0} {\bf \nabla} p_0 + {\bf \nabla} \psi_0 &=& - {\bf v_0 \cdot \nabla \vec v_0}  \\
{\bf \nabla \cdot}  (\rho_0 {\bf v}_0) &=& 0 \nonumber\\
 \rho_0 T_0 ({\bf v}_0 {\bf \cdot \nabla})S_0 &=& \rho_0 \epsilon_0 -{\bf \nabla \cdot 
\vec {\cal F}}_0  \nonumber\\
\nabla^2 \psi_0 &=& 4\pi G \rho_0 \nonumber
\label{eq}
\end{eqnarray}

We consider an axisymmetric rotation
 ${\bf \Omega} $. In spherical coordinates, one has:
$$\vec \Omega(r,\theta) =  \Omega(r,\theta) \left(\cos \theta ~{\bf e_r} -\sin \theta ~
{\bf e_\theta} \right) $$


The large scale velocity field $\vec v_0= v_\Omega+\vec U$ is
composed of the rotational velocity and the meridional circulation. We consider here that the meridional
circulation has no direct effect on the dynamics of the pulsation ($U <<c_s$), hence one  neglects it in the velocity field 
which  therefore is only due  to rotation and has the expression:
\begin{equation}
{\bf v}_0({\bf r})= {\bf \Omega} \times {\bf r}= \Omega(r,\theta) ~r \sin \theta ~{\bf e_\phi}
\end{equation}
The steady state configuration is then given by:
\begin{equation}
-{1\over \rho_0} {\bf \nabla} p_0- {\bf \nabla} \psi_0 = {\bf \Omega} \times {\bf \Omega}
\times {\bf r}  =- r \Omega^2 \sin \theta ~{\bf e_s} 
\label{equil}
\end{equation}
where ${\bf e_s}  = \sin \theta ~{\bf e_r} +\cos \theta ~{\bf e_\theta} = \partial{\bf e_\phi}/\partial\phi$. 
Note that $ - \vec \Omega \times \vec \Omega
\times \vec r = r \Omega^2  {\bf e_r}  $ for  $\theta =\pi/2$
is directed toward  the $r>0$  and we recognize the 
centrifugal acceleration.

\subsubsection{Particular cases: barotropy and uniform rotation}

When the centrifugal force can be  considered as deriving from a potential
 (conservative rotation law),
 it can be included  similarly to the gravity 
and  the equilibrium equation becomes:
$$ -{1\over \rho_0} {\bf \nabla} p_0 - {\bf \nabla} (\psi_0- |{\bf \Omega} \times {\bf r}|^2/2)=0$$
One sets  $\psi_{eff}\equiv  \psi_0- |{\bf \Omega} \times {\bf r}|^2/2$ and obtains 
$${dp_0\over dr}= - \rho_0 ~g_{eff} ~~~~~~~~~~~~~~{\bf \nabla \psi}_{eff}=g_{eff}$$
The star  keeps it spherical symmetry.  
This happens for a uniform rotation or for a cylindrical rotation $d\Omega/dz = 0$
in a cylindrical coordinate system (Tassoul, 1978). The Von Zeipel (1924) law states that for a barotrope and a conservative
rotation law, isobars and  isopycnics
coincide with the level surfaces (constant potential):
$\rho =\rho (\psi_{eff}); T=T(\psi_{eff})$ with $\psi_{eff}= \psi + r \sin  \theta \Omega$.
 Consequently the heat flux is proportional
  to local effective gravity $T_{eff}^4 \propto g_{eff}$ and poles are hotter than the equator
(gravity darkening\index{gravity darkening}; see Tassoul 1978 for details). 

\subsubsection{Shellular rotation  $\Omega(r)$ }

\ni Zahn (1992) considered that in stellar  radiative regions,  
highly anisotropic turbulence  developes: this leads to  an efficient 
homogeneization in the horizontal plane 
 but not in the vertical direction due to the stable  stratification which inhibes  vertical
motions 
 hence generating a shellular rotation. This was verified by Charbonneau (1992)  by means of 2D 
numerical simulations.    The combined effects of
rotation and turbulence induce a process of  advection/diffusion  of angular momentum   
and diffusion of chemical elements.  
The  turbulence itself can arise from a dynamical shear instability generated by the differential 
rotation $\Omega (\theta)$ 
 due to the  advection of angular momentum by the meridional  circulation 
 (see for a  review, JP. Zahn (2005), web site of the Brasil Corot meetings, CB2; see also Talon, 2006). 
For rotators with no angular momentum loss,  a  weak meridional circulation exists only to balance  
 transport by differential rotation induced turbulence.  
 For rotators   which lose  angular momentum through a wind,  the meridional circulation must adjust so as to   
 transport momentum toward the surface.
 Hence effects of a complex rotationally induced  2D process can be described in a 1D framework    
(Zahn, (1992); Maeder, Zahn, (1998);   Mathis, Zahn (2004))

\subsubsection{ 2D rotation:  $\Omega(r,\theta)$ }\label{rapideq}

Several studies have developped numerical schema in order to build  more and 
more realistic 2D rotating stellar models. 
As the full problem is quite difficult, simplifying assumptions have 
been made depending on the purpose of the study  
(Clement's series of papers from  1974 to 1998; Deupree from 1990 to  2001;
see for a detailed historical review, M. Rieutord (2007), web site of the Brasil Corot meetings, CB3).

With the initial purpose of modelling the rapidly rotating, oblate Be star, Achernar, 
Jackson et al. (2004, 2005)  neglected the evolution and dynamics of
 the star and focused their studies  
on a specified conservative rotation law.
 This allows them to solve a 2D partial differential 
equation for Poisson equation while dealing 
with only ordinary differential equations 
for the other equilibrium quantities. They were
able to build stellar rotating models for a wide mass range ($2 M_\odot$ to $9 M_\odot$) and 
for equatorial velocity up to 250 km/s for the most massive ones.  Roxburgh (2004) on the other hand  computed 
2D  uniformaly rotating, barotrope  zero age main sequence models again over a
wide mass range and rotational velocities. 
 The determination of the adiabatic nonradial 
oscillation frequencies of a rotating star only  depends on  the knowledge of $p_0(r,\theta)$,
$\rho_0(r,\theta)$  and  $\Gamma_1(r,\theta)$ and   does not require that the
thermal equilibrium be satisfied. Roxburgh (2005) takes advantage of this property to build 
2-dimensional acoustic models of rapidly rotating stars with a prescribed rotation 
profile $\Omega(r,\theta)$.  The initial density profile at a given angle $\theta_m$ can be taken as 
 the density profile of a 1D stellar model (which can include  effects  of
  transport and mixing due
rotation as described in the above section). Roxburgh (2005) then solves iteratively the 2D hydrostatic 
and Poisson equations; the adiabatic index $\Gamma_1$ is then obtained through the equation of state and the
knowledge of the hydrogen profile (again possibly derived from the 1D spherically averaged stellar model).


In all the above studies,  the $\Omega$ profile is prescribed and stellar models 
are built at
a given evolutionary stage with no inclusion of any feedback of rotation on evolution and
structure. To proceed a step further with the purpose of 
 including  effects of evolution and internal dynamics,
 Rieutord (2005, 2006), Espinosa Lara, Rieutord (2007)  worked at building 
 more and more realistic rotating  
models using spectral methods for both directions $r$ and $\theta$. 
 Espinosa Lara, Rieutord (2007)  succeeded in computing a   fully radiative, baroclinic model
in which the microphysics is treated in a simplified form  and
the star is assumed to be  enclosed in a rigid sphere.  Some  interesting conclusions could 
nervertheless be drawn.
For  a  stellar model  rotating  at $82 \%$ of the  rotational  break up velocity, 
the baroclinic model is much less centrally
 condensed  than a radiative polytrope. 
The steady state is characterized by poles hotter 
but rotating less rapidly  than the
equator.  The authors found that the temperature contrast
between poles and equators is less than that given by the Von Zeipel model although 
this might be reduced with the use of a  more realistic
surface boundary condition. This is also found by Lovekin \& Deupree (2006) using the 2D rotating stellar code of Deupree.  
 With increasing rotation,  the equator cools down enough that
a convective region  seems to be able to develope. 
It is also found that the isothermals are more spherical that the isobars. 
As a consequence and in constrast with the Boussinesq case, the meridional
currents  circulate from equator to poles: the meridional velocity $ U_\phi$ component decreases outward,
 because  $T$ decreases on a isobar from
pole to equator. Worth to be noting also, 
differential rotation in latitude seems to keep 
the same general form  for increasing $\Omega$ between 0.01 to 0.08 
 of the critical rotation rate  (Fig.10, Espinosa
Lara, Rieutord, 2007).

\subsection{Oscillations: linearization about the equilibrium}

For  the  oscillating, rotating fluid  of interest here, 
 the velocity field in each point of the space can be split into 2 components
$ \vec v  = \vec v_0 + \vec v'$
where  $\vec v'$ is the Eulerian  perturbation of the velocity due to the oscillation. 
Similarly for any scalar quantity $f$ (such as $p,\rho,\psi$), one writes $f=f_0 +f'$.
The  quantities with a  prime represent the oscillations which are modelled 
 as linear Eulerian perturbations about the equilibrium. 
One inserts   these decompositions into the basic equations Eq.9 and then linearizes  
with respect to the perturbations  $\vec v',\rho'$ etc... 
about the equilibrium quantities. Only adiabatic
oscillations are studied here, the system is then closed by using 
 the linearized adiabatic  relation:
\begin{equation}
 {\delta p\over p_0} =\Gamma_1 \left({\delta \rho\over \rho_0} \right) ~~~ {\rm where} ~~~~ 
\Gamma_1 \equiv \left({\partial p \over \partial \rho}\right)_S 
\label{adia}
\end{equation}
 since $\delta S=0$  ($S$ entropy) where $\delta$ denotes Lagrangian variations
or in the Eulerian form:
$${p'\over p_0} = \Gamma_1 \left( {\rho' \over \rho_0} + {\vec \xi \cdot A} \right) 
~~{\rm where}~~ 
 {\bf A } = {1\over \Gamma_1}{\vec \nabla p_0\over p_0} - {\vec \nabla \rho_0\over \rho_0} $$
 
The system of linearized equations  
for adiabatic nonradial oscillations of a rotating star takes
the form:
 \begin{eqnarray}
{\partial {\vec v'} \over \partial t} &+ &{\vec v' \cdot \vec \nabla \vec v}_0 + {\vec v}_0 { \cdot \vec \nabla
\vec v'}= -{1\over \rho_0} {\vec \nabla} p'+{\rho'\over \rho_0^2} {\vec \nabla} p_0 - {\vec \nabla} \psi' \nonumber\\
{\partial \rho' \over \partial t} &+ & {\vec \nabla \cdot} (\rho' {\vec v}_0 + \rho_0 {\vec v}') = 0   \\
{p'\over p_0} &=& \Gamma_1 \left({\rho'\over \rho_0} -  {\vec \xi \cdot  A} \right)  \nonumber \\
\nabla^2 \psi' &=& 4\pi G \rho' \nonumber
\label{pert1}
\end{eqnarray}

 The displacement is related to the Lagrangian velocity by 
 $${\vec \delta v} = {D{\vec \xi }\over Dt}={\partial {\vec \xi }\over \partial t} + {\vec v_0 \cdot \vec \nabla \vec \xi}  $$
where  ${D/Dt}$ is the Lagrangian derivative   (Lebovitz, 1970).
Using the  linearized relation  between  the      Lagrangian and Eulerian  velocities is:
${\vec v'} = \delta {\vec  v} -({\vec \xi \cdot \vec \nabla)  \vec v_0 }$
the Eulerian velocity  perturbation  $ {\bf v'}$ is related to the displacement  ${\vec \xi}$
by:
\begin{equation}
{\vec v'} = {\partial {\vec \xi} \over \partial t} + ({\vec v_0 \cdot \vec \nabla) \vec \xi  - (\vec \xi \cdot \vec \nabla)
\vec v_0 }
\label{vp}
\end{equation}

\subsection{ Time and azimutal dependances  of the oscillation}
 One adopts a spherical coordinate system with the  polar axis coinciding
with the star rotation axis. For an axisymmetric  star, the system Eq.13
 is separable in $\phi$   and one can then seek  quite generally for a solution of the form:
  $${\vec \xi}({\vec r},t)\propto {\vec \xi}(r,\theta,t) ~e^{i m \phi}$$
In a description based on spherical harmonics, an eigenfunction will be decomposed on the 
 spherical  harmonics  with the  \underline{same} $m$. Because of the assumed existence of a 
steady state configuration, one can represent
all scalar perturbations by :
$f'({\bf r}, t) = f'(r,\theta) ~ e^{i(\omega t + m \phi)}$
and for the fluid displacement  vector 
$\vec{\xi} ({\bf r}, t) = \vec {\tilde \xi} (r,\theta)    ~ e^{i(\omega t + m \phi)}$
The relation between ${\bf v'}$ and the displacement  (Eq.\ref{vp}) becomes:
 \begin{equation}
 \vec v' =i (\omega + m \Omega) \vec {\tilde \xi} -(\vec {\tilde \xi} \cdot \vec  \nabla \Omega) r \sin \theta \vec e_\phi
\label{vp2}
\end{equation}
From now on, we drop the tilde for the displacement vector field.

\subsection{The eigenvalue system}
 One  assumes that the configuration is axisymmetric and that the motion is purely 
rotational i.e. one then has:
$B= {\bf v_0 \cdot \nabla} =\Omega ~{\partial / \partial \phi}$ and
$B f\equiv {\vec v_0 \cdot \nabla} f = \Omega{\partial f\over \partial \phi}$ for a scalar $f$.   
Using Eq.13, Eq.\ref{vp}, Eq.\ref{vp2},  and  the  equality:   
 \begin{eqnarray}
B {\vec \xi } \equiv {\bf \vec v_0 \cdot \vec \nabla} {\vec \xi}  &=& i m \Omega {\vec \xi} 
+ {\vec \Omega \times \vec \xi} 
\label{BB}
\end{eqnarray}
for a vector field  ${\vec \xi}$, one obtains the wave equation:  
 \begin{equation}
{1\over \rho_0}   {\cal L} {\vec \xi} -\hat \omega^2  {\vec \xi} +2 \hat \omega  
   ~i~ {\vec \Omega} \times {\vec \xi} - ({\vec \xi}  {\bf \cdot} {\vec \nabla} \Omega^2)~r \sin \theta ~{\vec e_s} =0
\label{wave0}
\end{equation}
where we have defined
 \begin{equation}
   {\cal L} \vec \xi = \vec \nabla p'- {\rho'\over \rho_0} \vec \nabla p_0 + \rho_0~ \vec \nabla \psi'
\label{wave1}
\end{equation}
and $\hat \omega = \omega+m\Omega(r,\theta)$. 
The continuity equation in Eq.13 becomes : 
\begin{equation}
\rho'+ \vec \nabla \cdot (\rho_0 {\vec \xi}) = 0
\label{ro2}
\end{equation}
The system is completed with the linearized adiabatic relation Eq.\ref{adia}.
The first two terms in Eq.\ref{wave0} susbist in absence of rotation and provide 
the eigenvalue problem for a nonrotating star:
 $ {\cal L} {\vec  \xi} - \omega^2  \rho_0 {\vec \xi}  =0$.
The first additional term in Eq.\ref{wave0} is due to the Coriolis force, 
 the second additional term  only   exists in presence of a nonuniform rotation.
The centrifugal  force comes into play  through its effects on the structure 
i.e. through the quantities $p_0,\rho_0,..$. The usual boundary conditions  are the requirements that the solutions must keep a regular behavior 
in the center and at the surface. More specifically one usually  asks that  $\delta p=0$ 
 at the surface and the 
 gravity potential goes to zero at infinity.
At the center the regularity conditions are expressed   as 
$f' = O(r^\ell)$  for a scalar quantity for instance  (see Unno et al (1989)).

Eq.\ref{wave0}, \ref{wave1}, \ref{ro2} together with  the  perturbed adiabatic and  Poisson equations plus the boundary conditions 
give rise to an eigenvalue problem for the oscillation of a rotating star. 
One looks for  $\omega$ et ${\bf \xi}$, given $\Omega$ and the equilibrium structure. 
Although it is not necessary, the angular dependence of the 
eigenfunctions  are most commonly seeked under
the form of a decomposition over the basis formed with the spherical harmonics (with a given $m$):
\begin{eqnarray}
 \vec \xi_m ({\bf r}) &=& \sum^{+\infty}_{\ell \geq |m|} \left( \right.
\xi_{r,\ell}(r) ~Y_\ell^m(\theta,\phi) ~{\bf e_r} + \xi_{h,\ell}(r) ~{\vec \nabla} ~ Y_\ell^m(\theta,\phi) \nonumber \\
& & ~~~~~~~~~~~~~+ \tau_\ell(r)  ~{\bf e_r \times}  ~{\vec \nabla}  Y_\ell^m(\theta,\phi)
\left. \right) \label{eigenxi}\\
f'_m({\bf r})&=&  \sum^{+\infty}_{\ell \geq |m|}  f^{'m}_\ell(r)~ Y_\ell^m(\theta,\phi)
\label{eigenf}
\end{eqnarray} 

The first two terms in $ \vec \xi_m ({\bf r})$  represent the  spheroidal  part of the oscillation whereas 
the 3rd   represents the toroidal part. If the star possesses 
the equatorial symmetry  $\theta  \Longrightarrow \pi-\theta$
 (with $\Omega(r,\pi-\theta)= \Omega(r,\theta)$), the eigenmodes can be classified into 2 groups:
symmetric    (or even) modes   and  anti symmetric (or odd)   modes with respect to the equator.
One then obtains  two  different systems of   differential equations, one   for each
parity (see Unno \etal~ 1989; Hansen \etal~  2006). 
Symmetry with respect to the equatorial plane induces the property
\begin{equation}
\omega_{n,\ell,-m} (\Omega)= -\omega_{n,\ell,m} (\Omega) 
\label{symequat}
\end{equation}

\subsection{Symmetry properties of the operators and orthogonality relations}
Symmetries of the operators in Eq.\ref{wave0}, \ref{wave1}, \ref{ro2} and their consequences were
 studied   by Lynden-Bell and Ostriker (1967); 
Dyson, Schutz (1979),  Schutz (1980), Clement  (1989), see also Unno et al (1989) and 
Reese (2006); Reese \etal~ (2006) for a polytrope in uniform rotation.
One first  introduces the inner product between two arbitrary vector fields  $\vec \eta$ and ${\vec \xi}$:
\begin{equation}
<{\vec  \eta}|{\vec \xi}> \equiv \int_V   \left({\vec \eta}^*({\vec r}) {\bf \cdot \vec \xi}({\bf r})
\right)  {\bf d^3r}
\label{inner}
\end{equation}
where $\vec \eta^*$ is the complex conjugate of  $\vec \eta$.  We also recall 
that the adjoint operator ${\cal Q}^+$   of an
operator ${\cal Q}$ is defined such that 
$< {\cal Q}^+  ({\vec \eta}) |{\vec \xi}>\equiv <{\vec  \eta}|{\cal Q} ({\vec \xi})> $
for any ${\vec \xi} \in $ domain of  ${\cal Q}$. An operator ${\cal Q}$ 
is  symmetric with respect to an inner product
if for any nonsingular vector fields ${\vec \eta}$ and ${\vec \xi}$ defined in the unperturbed volume and having 
continuous first derivatives
everywhere,  one has 
\begin{equation}
  <{\cal Q} ({\vec \eta})| {\vec \xi}> =<{\vec \eta}|{\cal Q} ({\vec \xi})>
\label{eqQ}
\end{equation}
This is equivalent to 
$$ \int \vec \eta^* \cdot {\cal Q}(\vec \xi) {\bf d^3r} =  \left(\int {\cal Q}(\vec \eta) \cdot \vec \xi^* {\bf d^3r} \right)^*    $$ 

  Let start  from the fluid motion  equation 
 $${D {\bf v} \over Dt} =({\partial \over \partial t}+ B) {\bf v}= -{1\over \rho}\vec \nabla p +\vec \nabla \psi$$
with $B$ is  defined in Eq.\ref{BB} above.
If the configuration is assumed to be in steady state, $B$ commutes  with ${\partial/\partial t}$.
Using this property, Lynden-Bell and Ostriker (1967) have shown that  the wave equation Eq.\ref{wave0}
 can be cast under the form: 
$${D^2 {\vec \xi} \over Dt^2}= \delta \left(-{1\over \rho} {\vec \nabla} p +{\vec \nabla} \psi\right)
$$
where ${\vec \xi}$ is the fluid Lagrangian displacement, $\delta$ represents a Lagrangian variation.
As \begin{eqnarray}
{D^2 {\vec \xi} \over Dt^2}  &=& {\partial^2 {\vec \xi} \over\partial t^2}+ 
2 ~B  {\partial {\vec \xi} \over \partial t} + B^2 {\vec \xi}
\end{eqnarray}
one must then  solve:
\begin{equation}
{\partial^2 {\vec \xi} \over \partial t^2}+ 
2~ B  {\partial  {\vec \xi}\over \partial t}  =  ~C({\vec \xi})
\label{eqa}
\end{equation}
where the Lagrangian expression for $C({\vec \xi})$  can be found in Lynden-Bell and Ostriker (1967)  or 
Dyson and Schutz (1979) for instance.  It is also given by 
$${\cal C} ({\vec \xi}_m)= -B^2({\vec \xi}_m) -{1\over \rho_0} {\cal L} ({\vec \xi}_m) + K({\vec \xi}_m)  $$
where
${\cal L}({\vec \xi}) $ is defined by  Eq.\ref{wave1}  
and 
$$K(\vec \xi)= \vec \xi\cdot \vec \nabla  \left(-{1\over \rho_0} \vec \nabla p_0 + \vec \nabla \psi_0  \right) =\vec \xi\cdot \vec \nabla 
\left(-\Omega^2  r \sin \theta \vec e_s\right) $$
where we have used  Eq.\ref{equil}.  Eq.\ref{eqa} is valid for any 
axisymmetric rotation law for a steady state rotating configuration. 
 Assuming again a time dependence $e^{i \omega t}$ for ${\vec \xi}$ 
 and the  scalar variables, the linear adiabatic perturbations of 
 a differentially rotating, axisymmetric  stellar model  
Eq.\ref{wave0},\ref{wave1} are then  cast into the form (Dyson, Schutz 1979):
\begin{equation}
{\cal Q}({\vec \xi}_m)\equiv  -\omega^2  \rho_0 {\vec \xi}_m+ 
 i ~\omega  {\cal B} ({\vec \xi}_m) + C({\vec \xi}_m) = 0
  \label{Dyson}
  \end{equation}
where we have defined
\begin{equation}
{\cal B}({\vec \xi}_m) = 2  \rho_0   \left( {\vec  \Omega}  \times {\vec \xi}_m  +  i m \Omega {\vec \xi}_m
 \right) 
 \label{BLO}
 \end{equation}

Lynden-Bell \& Ostriker (1967) have shown that $ C$ is symmetric and $B$ is antisymmetric  i.e,
for any ${\vec \eta}$ and ${\vec \xi}$, 
$$<{\vec \eta}|C({\vec \xi})>= <C({\vec \eta} )|{\vec \xi}>  ~~;~~<{\vec \eta}|{\cal B}({\vec \xi})>= - <{\cal B}({\vec \eta}
)|{\vec \xi}>$$
and actually  $C$ is  real, ${\cal B}$ is purely imaginary.
 Dyson \& Schutz (1979) and Schutz (1980)  studied the general properties of eigenfunctions of a rotating star. 
We focus here on a much simpler issue, for later use. 
Let $\vec \xi_1$ and $\vec \xi_2$ two eigenfunctions associated with 2 eigenvalues 
$\lambda_1$ and $\lambda_2$, they verify the
equalities:
\begin{eqnarray}
 {\cal Q} ({\vec \xi}_1) &=& \lambda_1^2 \rho_0  {\vec \xi}_1 + \lambda_1  {\cal B} ({\vec \xi}_1)+
{\cal C} ({\vec \xi}_1)=0 \label{Q1} \\
{\cal Q} ({\vec \xi_2}) &=& \lambda_2^{2} \rho_0  {\vec \xi}_2 + \lambda_2  {\cal B} ({\vec
\xi}_2) +{\cal C} ({\vec \xi}_2)=0
\label{Q2} 
\end{eqnarray}
Taking the inner product of Eq.\ref{Q1} with $< \vec \xi_2|$, one gets:
\begin{equation}
 \lambda_1^2 <\vec \xi_2 | \rho_0   {\vec \xi}_1> 
+ \lambda_1 <\vec \xi_2  |{\cal B} ({\vec \xi}_1)> +
<\vec \xi_2 |{\cal C} ({\vec \xi}_1)>=0
\label{lamb1}
\end{equation}
Similarly taking the inner product of Eq.\ref{Q2} with $<\vec \xi_1|$, one obtains:
\begin{equation}
\lambda_2^{2}< \vec \xi_1  \rho_0  |{\vec \xi}_2> + \lambda_2 <\vec \xi_1  |{\cal B} ({\vec
\xi}_2)> +
<\vec \xi_1|{\cal C} ({\vec  \xi}_2)>=0
\label{lamb2}
\end{equation}
We now make use of the symmetry properties of $C$ and $B$, ie
$<\vec \xi_1|{\cal C} ({\vec \xi}_2)>= <{\cal C} ({\vec \xi}_1)| \vec \xi_2>$
and  $<\vec \xi_1|{\cal B} ({\vec \xi}_2)>= -<{\cal B} ({\vec \xi}_1)| \vec \xi_2>$
which give for the last equality Eq.\ref{lamb2}
$$\lambda_2^{2}<\vec \xi_1  \rho_0  |{\vec \xi}_2> - \lambda_2 <{\cal B} ({\vec \xi}_1)| \vec \xi_2>  + 
<{\cal C} ({\vec \xi}_1)|\vec \xi_2>
=0$$
Taking the complexe conjugate of this equation yields:
$$\lambda_2^{*2} <\vec \xi_2 |  \rho_0  \vec \xi_1> - \lambda_2^{*} <\vec \xi_2 |{\cal B} ({\vec \xi}_1)>  
+< \vec \xi_2| {\cal C} ({\vec \xi}_1)>
=0$$
which can be compared to Eq.\ref{lamb1}. The difference between these 2 equations yields the relation:
\begin{equation}
 (\lambda_1^2 - \lambda_2^{*2}) <\vec \xi_2 |  \rho_0  {\vec \xi}_1>
+ (\lambda_1  + \lambda_2^{*}) <\vec \xi_2 |{\cal B} ({\vec \xi}_1)>  =0
\label{orth}
\end{equation}
For a pure  imaginary eigenfrequency $\lambda=i \omega$, this reduces to
$ (\omega_1^2 - \omega_2^{2}) <\vec \xi_2 |  \rho_0  {\vec \xi}_1>
+ i(\omega_1  - \omega_2) <\vec \xi_2 |{\cal B} ({\vec \xi}_1)>  =0$.
For $\omega_1  \not = \omega_2$, one obtains the orthogonality relation\index{orthogonality relation}:
 \begin{equation}
 <\vec \xi_2 |  (\omega_1 +\omega_2)  \rho_0 + i{\cal B} ({\vec \xi}_1)>  =0
\label{orth}
\end{equation}
which is valid for any rotating, axisymmetric star. A similar relation although in the context of 
a perturbative third order approach  was derived in Soufi et al (1998).

\section{A variational principle}\label{variat} 

Chandrashekhar (1964) has  shown that nonradial oscillations of a nonrotating star 
 can be treated as a variational  problem\index{variational principle}. This was extented to uniformaly rotating bodies 
 by  Chandrasekhar and Lebovitz  (1968)  and  nonuniform axisymmetric rotation for a steady state configuration 
 by  Lebovitz  (1970).  
The symmetry properties of the operators in Eq.\ref{wave0},\ref{wave1}
indeed leads to the existence of a {\it variational principle}\index{variational principle} which states:
let a real number  $\zeta$  and a function 
 $\eta(r)$  related by the  relation 
 $$\zeta^2 <\eta|\rho_0 \eta> +\zeta<\eta|B \eta>+ <\eta|C \eta> =0$$
where ${\cal B, C}$ are antisymmetric  and  symmetric operators. 
One assumes that $\eta(r)$  is a trial function 
close enough but  not equal to the unknown true eigenfunction
  $\xi(r)$  and one sets $\eta = \xi+ \delta \xi$.
 The value of  $\zeta$  will be close but different of the true unknown eigenvalue 
 $\lambda$ with $\zeta = \lambda + \delta \lambda$.  
 Inserting the  expressions pour $\zeta$ et $\eta$  in the above equation and  linearizing, one gets: 
   \begin{eqnarray}
  & & 2  (\delta \lambda) \lambda <\xi|\rho \xi>+\lambda^2  \delta (<\xi|\rho \xi>)= \nonumber \\
   & & (\delta \lambda) <\xi|B\xi> +\lambda (\delta <\xi|B\xi>) + \delta (<\xi|C\xi>)
    \end{eqnarray}
 
which is rewritten as: 
     \begin{eqnarray}
 \delta \lambda  &=&{2 ~{\cal R}(<\delta  \xi|(-\lambda^2  \rho \xi  +\lambda B \xi + C\xi)>)
   \over  (2 \lambda<\xi|\rho \xi> - <\xi|B\xi>)}
   \label{var}  
      \end{eqnarray}
where ${\cal R}$ means real part.
 The numerator in the  right hand side of Eq.\ref{var} vanishes  since  $\xi$ is the eigenfunction  
associated to  $\lambda$; one then has  $\delta \lambda =0$ (i.e.
 $\zeta = \lambda$)  for any arbitrary  $ \delta \xi \not=0$. 
The reciprocal is also true.
 Hence the eigenfrequencies are invariant to first order to a change in the function $\xi$. 
 The existence  of such a  variational  principle has the 
 consequence that  any trial function  used instead of the true 
 eigenfunction  gives an estimation  of the eigenvalue 
  accurate  enough: the error on  $\lambda$  is quadratic in the error on the 
eigenfunction.


 \subsubsection{Variational expressions for the eigenfrequencies}
 
It is possible to use Eq.\ref{Dyson} to derive an integral expression for the eigenfrequency.
Taking the inner product with ${\bf \xi}^*$  and integrating over the entire volume of the star, 
one gets $-\omega^2 + \omega  b  + c  = 0$
where 
$$b= {i \over I} ~ \int_0^R ~\vec \xi_m^* \cdot {\cal B} (\vec \xi_m) {\bf d^3r}  ~~~
~~~~~~~;~~~~~~~~~c = {1\over I} \int_0^R ~ \vec \xi_m^* \cdot  C(\vec \xi_m)  {\bf d^3r}$$
and
\begin{equation}
I=  \int_0^R ~(\vec \xi_m^* \cdot \vec \xi_m) ~ \rho_0 {\bf d^3r}
\label{inertia}
\end{equation}
 $b$ and $c$ are real since the operators are Hermitian. Then one obtains 
$ \omega = \left(b \pm \sqrt{b^2+4c} \right)/2$
For $\Omega=0$, one recovers $ \omega^2 = c= 
 {1\over I} \int_0^R ~ \vec \xi_m^* \cdot {\cal L} ({\bf \xi}_m)   {\bf d^3r}  \equiv \omega^2_0 $.
 Since $c$ does not depend on m $(-\ell<m<\ell)$, the
eigenfrequencies of nonradial pulsations in a nonrotating star are $(2\ell+1)$-fold degenerate. 

  Clement (1984)  applied the variational  principle to compute eigenfrequencies of polytropes.
   Clement (1989) again used a variational approach to compute the nonaxisymmetric modes for a 
   $15 M_\odot$  uniformly rotating 2D model. For the p-mode eigenfunctions 
   entering the integral expression,  as they  have a dominant
contribution from the poloidal part,  Clement chose  trial functions in the form of  the gradiant of a longitudinal potential 
${\bf \xi_p} \sim {\bf \nabla} \phi_1$, the potential $\phi_1$ is then expanded in powers of Legendre polynomials and the expansion
coefficients are determined by the stationary condition on the eigenfrequency. Saio (2002)   neglected  the perturbation  of  gravitational  potential $\psi'$
in  Eq.\ref{Dyson} (which is justified  in the case of high frequency p-modes) and derived simplified expression for the
coefficients b and c. As the error on the frequencies is much smaller than that of the eigenfunctions, it is possible to use the 
numerical eigenfunctions (obtained by solving numerically the eigensystem Eq.\ref{wave0}) in the integrals for the
coefficients $b$ and $c$ defined above. The
resulting eigenfrequencies $\omega_{var}$  can be compared to the numerical ones.
 If  an error $\delta \xi$ exists for  $\xi$ in the numerical calculation,  the resulting error on 
  $\omega$ will be : $\omega_{var}=\omega+O(||\delta \xi||^2)$. This property has been 
 used for instance by   Christensen-Dalsgaard and  Mullan (1994)  to check the accuracy of their numerical 
   computations of eigenfrequencies of nonrotating polytropes. Reese \etal ~ (2006) used it to check the numerical
    accuracy of the results of their  2D  eigenvalue system for a
rotating polytrope.


\section{Slow rotators}

At the lowest order, $O(\Omega)$, only the Coriolis force\index{Coriolis force} plays a role and one can 
 ignore the direct effect of the centrifugal   oblatness on the oscillation frequencies for slow rotators like the Sun for  exemple. 
 From here on , we  use the DG92 notations.
 The eigensystem Eq.\ref{wave0}, Eq.\ref{wave1} then reduces to:
 $$ {\cal L} ({\bf \xi}) -\hat \omega^2  \rho_0 ~\xi +2  \hat \omega \rho_0  ~i {\bf \Omega} \times  {\bf \xi}   =0$$
The scalar  perturbations $f=(p',\rho',\psi')$ are expanded up to first order  $f'=f'_0+ f'_1$
and similarly  for the fluid displacement: ${\bf \xi} = {\bf \xi_0}+ {\bf \xi_1}$, 
this yields ${\cal L}={\cal L}_0+{\cal L}_1$ where the operators ${\cal L}_0,{\cal L}_1$ take the same
form than ${\cal L}$ with  the perturbed quantities subscribed with 0 and 1 respectively.
Note that the equilibrium quantities entering ${\cal L}$ can come from a rotating stellar model,
although strictly speaking, this would be somewhat inconsistent. 
The eigenfrequency  is written as $\omega = \omega_0 +\omega_1$ and
$\hat \omega = \omega + m\Omega = \omega_0 + \omega_1 + m\Omega  $.
 The subscript 0 refers to solutions of the problem at the zeroth order which is specified  by 
 \begin{equation}
  {\cal L}_0 ~{\vec \xi}_0 - ~\omega^2_0 \rho_0~{\vec \xi}_0  = 0
  \label{zerothorder}
  \end{equation}
  this yields  the zeroth order eigenvalues $\omega_0$ and eigenfunctions ${\vec \xi_0}$ for a given
$(\ell,n)$ set. 
We recall that the zeroth order operator  $  {\cal L}_0$  is 
selfadjoint
\begin{equation}
 <{\vec \xi}_0|{\cal L}_0 ~ {\vec \xi}_0>=  <{\cal L}_0  {\vec \xi}_0| ~ {\vec \xi}_0>
\label{zerothorder1}\end{equation}
 consequently  the right- and left-eigenfunctions are identical and orthogonal to one  another. 
$$< {\vec \xi}_{0i}| {\vec \xi}_{0j}> = \int \rho_{0}  ({\vec \xi}_{0i}^* {\vec \cdot  \xi}_{0j})  {\bf d^3r}= 
I ~\delta_{ij}$$
with $I$ the oscillatory  moment of inertia  Eq.\ref{inertia}
and $\delta_{ij}$ is the  Kronecker symbol, $i,j$ label  eigenfunctions and are shortcuts for the subscripts 
$(n_i,\ell_i)$ and $(n_j,\ell_j)$.
Using the integral form, the zeroth order solution is then given by
\begin{equation}
  \omega^2_0= {1\over I}  <\vec \xi_0|{\cal L}_0 ~\vec \xi_0>  
  \label{omeg0}
  \end{equation}
In practice, $\omega^2_0$ is obtained as the  numerical solution of the zeroth order 
 eigensystem. At the first order in $\Omega$,  one must solve 
\begin{equation}
 {\cal L}_0 ~ {\vec \xi}_1 - ~\omega^2_0 ~\rho_0 ~{\vec \xi}_1  = 2 \omega_0 ~\rho_0 ~(\omega_1 + K) ~{\vec \xi}_0
 \label{om1}
  \end{equation}
with $K = m \Omega - i {\vec \Omega}  \times$. One seeks a solution for the correction $\omega_1$. 
The integral solution is  again  obtained 
by projecting  Eq.\ref{om1}  onto the eigenfunction ${\vec \xi_0}$ 
 using the inner product Eq.\ref{inner}: 
$  <{\vec \xi}_0|{\cal L}_0 {\vec \xi}_1> -  \omega^2_0 <{\vec \xi}_0^*|\rho_0 {\vec \xi}_1>  = 2 \omega_0 
 <{\vec \xi}_0|\rho_0 (\omega_1 + K) {\vec \xi}_0>$
which admits a solution provided that $< {\vec \xi}_0 |(\omega_1 + K) {\vec \xi}_0>=0$
is satisfied, which yields
  \begin{equation}
  \omega_1 =  - {1\over I} <{\vec \xi}_0 | K ~{\vec \xi}_0> 
\label{om1sol}
\end{equation} 
Due to the existence of the variational  principle discussed in Sect.4,
 if one is interested only in the eigenfrequency, 
  one needs not to know the correction to the eigenfunction ${\vec \xi_1}$.
Note  however that one  obtains the correction $\omega_1$ numerically  by integrating  the
O($\Omega$) eigensystem   as shown by Hansen et al. (1978). Comparison of the variational expression and the
numerical one  can be used to check the results.

\medskip
 \ni{\it Rotational splitting:} 

\ni The integral expression of 
Eq.\ref{om1sol} is:
  \begin{eqnarray}
  \omega_1 &=&   {1\over I} ~ \int \rho_{0}  \left( m\Omega |{\vec \xi}_0 |^2- i ~{\vec \xi}^*_0 {\bf \cdot} ({\vec \Omega \times 
  \xi}_0)  \right)   {\bf d^3r} 
  \label{omeg1}
\end{eqnarray} 
which requires the knowledge of the zeroth order displacement eigenfunction ${\vec \xi}_0$.
Each  mode  is usually determined as  a sum of spherical  harmonics Eq.\ref{eigenxi}.
 Solving the zeroth order eigensystem Eq.\ref{zerothorder} shows 
that actually the zeroth order solution can be written with a single spherical harmonics: 
\begin{equation}
 {\vec \xi}_0 (\vec r) =    \xi_r(r) Y_{\ell}^m~ {\bf e_r} + \xi_h(r) {\bf \nabla}_h
 Y_{\ell}^m 
 \label{xi0}
 \end{equation}
where the gradient $\vec \nabla_h$ is defined as:
$$ \vec \nabla_h= {\bf e_\theta} {\partial \over \partial \theta} + {\bf e_\phi} {1\over \partial \sin \theta} {\partial \over \partial \phi}$$
Using this expression, it is straightforward  to show that 
\begin{eqnarray}
 m\Omega  |{\vec \xi}_0 |^2 & &- i ~{\vec \xi}^*_0 {\bf \cdot} ({\vec \Omega \times \vec \xi}_0)  =  \nonumber \\ 
  & &  m \Omega \left[ \right. (|\xi_r|^2 -(\xi^*_r \xi_h+cc)) |Y_{\ell}^m|^2 
 + \nonumber \\
 & & |\xi_h |^2 ( {\vec  \nabla}_h Y^*_{\ell,m} {\bf  \cdot \vec \nabla}_h Y_{\ell}^m-{\cos \theta\over \sin \theta}
 {\partial |Y_{\ell}^m|^2\over \partial \theta})
  \left.\right] \nonumber 
\end{eqnarray}
Inserting this expression into Eq.\ref{omeg1} yields the rotation splitting, Eq.\ref{split0}, 
 in the  compact form: 
\begin{equation}
\delta \omega_m=  {\omega_{1,j}\over m}= \int_0^R \int_0^\pi  K_{n,\ell,m}(r,\theta) ~\Omega(r,\theta) ~r  d\theta dr
\label{split}
\end{equation}
where  $K_{n\ell}$ is called {\it rotational kernel}\index{rotational kernel}; its  expression can be found
 in  Schou et al (1994a,b), Pijpers (1997)  or 
CD03.
For instance, 
the sectoral modes ($|m|=\ell$) for  increasing  $\ell$,  get increasingly confined toward the equator. 
Hence the  measure of the  splittings of sectoral  modes  provide a measure of the equatorial velocity. 

\subsubsection {Shellular rotation $\Omega(r)$}

When the rotation  can be assumed   independent of $\theta$, the expression of the rotational
splitting reduces to 
\begin{eqnarray} 
\delta \omega_{n,\ell}   =  \int_0^R  ~K_{n,\ell}(r) ~\Omega(r) ~dr 
\label{split4}
\end{eqnarray}
with  $$K_{n\ell} (r) ={1\over I}  \left( \xi_r^2 +\Lambda \xi_h^2 -2 \xi_r \xi_h + \xi_h^2\right) ~\rho_{0}~ r^2~$$ 
and
$$I=\int_0^R  \rho_{0} r^2   ~(\xi_r^2 +\Lambda \xi_h^2) dr $$
and  $\Lambda = \ell(\ell+1)$ and $R$  stellar radius.
One also finds in the litterature:
\begin{eqnarray} 
\delta \omega_{n,\ell}   = m \bar \Omega (1-C_{n,\ell} - J_{n,\ell})  
\end{eqnarray}
in the observer frame with $\bar \Omega$ an averaged or 
 the surface rotation rate.  The Ledoux   constant is:
$$ C_{n,\ell}  = {1\over I} \int_0^R ~ \rho_{0} r^2 ~  (2 \xi_r \xi_h + \xi_h^2)  dr$$
and
$$J_{n,\ell} = {1\over I} 
\int_0^R ~ \rho_{0} r^2 ~  (\Omega(r)-\bar \Omega)~ (\xi_r^2 +\Lambda \xi_h^2- 2 \xi_r \xi_h - \xi_h^2)  dr$$
A particular case is that 
of {\it solid-body rotation}  for which one derives 
$  \delta \omega_{n,\ell}  = m  \Omega (1-C_{n\ell})$.
It is worth noting that   for  $|\xi_h| <<|\xi_r|$, $C_{n,\ell} \rightarrow 0$. As 
$$ {\xi_h \over |\xi_r|} \propto {1\over \sigma^2}$$ 
 where  $\sigma$ is  the normalized frequency,
 $C_{n\ell} \sim 0$  for  high  frequency  p-modes and   the measure of the rotational splitting  
$\delta \omega_{n,\ell,m}   = \Omega $ then is a quasi direct measure 
 of the rotational angular velocity.


Forward techniques\index{forward techniques} compute the splittings  with Eq.\ref{split}, \ref{split4} 
 by assuming a rotation profile.
The rotational kernels $K_{n\ell} $  are assumed to be known and
  calculated for the appropriate  stellar model. 
  The results are compared with the observed splittings.
The integral relations Eq.\ref{split}, Eq.\ref{split4} can be inverted 
to provide $\Omega$. 
The observed splittings  for several differents modes $(n,\ell)$  
constitute the data set (see Sect.\ref{inversion}).

\subsubsection {Latitudinal dependence }

It is convenient to assume  a rotation of the type:
\begin{equation}
\Omega(r,\theta)=  \sum_{s=0}^{s_{max}} ~ \Omega_{2s}(r) ~(\cos\theta)^{2s}
\label{latit}
\end{equation}
The expression for the rotational splitting  then becomes 
(see for instance Hansen \etal ~ 1977; DG92; CD03):
$$ \delta \omega_m=  \sum_{s=0}^{s_max} ~  \int_0^R ~ \Omega_{2s}(r) ~K_{n,\ell,m,s}(r) ~\rho_0 r^2 dr$$
with
$$K_{n,\ell,m,s}(r) =  |\xi_r|^2-(\xi^*_r \xi_h+cc) {\cal S}_s
+|\xi_h|^2 ((2s-1){\cal S}_{s-1}-\Lambda {\cal S}_s $$
where
$${\cal S}_s = \int_0^\pi d\theta \sin \theta  (\cos\theta)^{2s} ~|Y_{\ell}^m(\theta,\phi)|^2$$
and ${\cal S}_{-1}=0; {\cal S}_{0}=1$. For later purpose, we also give  
\begin{eqnarray}
 {\cal S}_{1} &=& {1\over 4\Lambda-3} (-2m^2+2\Lambda-1)  \\
{\cal S}_2 &=& {1\over 4\Lambda-15} \left[{\cal S}_{1}  (-2m^2+2\Lambda-9)+1 \right]  \nonumber
\label{SS}
\end{eqnarray}
 More generally 
$ {\cal S}_s$ for any $s$ is given by a recurrent relation (Eq.31 in DG92). 
Results of 2D inversion in the solar case are discussed in Sect.\ref{suninvers}.

\section{Moderately rotating pulsating stars}

For moderate rotators, one needs to include second order 
corrections to the frequency $\omega_2$
(not to  say higher order contributions, see Sect.\ref{third}).   
At the second order level, several contributions must be included.

\ni 1-  The eigenfunction is modified by the Coriolis  acceleration   and becomes 
${\bf \xi} = {\bf \xi}_0+ {\bf \xi}_1$ 
with   the eigenfunction  correction including a spheroidal 
(${\bf \xi}_{1p}$) and a toroidal (${\bf \xi}_{1t}$) components respectively:
${\bf \xi}_1= {\bf \xi}_{1p} + {\bf \xi}_{1t}$
which result in frequency corrections labelled $\omega_{2P}$ et $\omega_{2T}$.

\ni 2- Oscillatory inertia  is also modified  $I= I_0+I_2$
where $I_2= <{\bf \xi}_1|{\bf \xi}_1>$.
We recall that $<{\bf \xi}_0|{\bf \xi}_1>=0=<{\bf \xi}_1|{\bf \xi}_0>$  
and  $<{\bf \xi}_0|{\bf \xi}_2>=0=<{\bf \xi}_2|{\bf \xi}_0>$.
This causes a  second order frequency  correction $\omega^I_2$.

\ni 3- The centrifugal force generates  changes in the structure of the star:

\ni  i) geometrical distorsion  which can be split into 2 components :

                   -  {\it spherically symmetric}  distorsion due to 
		   the latitudinally averaged centrifugal force  which generates 
		   an effective 
		   gravity 
\begin{equation}
g_{eff}(r) =   g(r) - (2/3) r \Omega^2
\label{geff}
\end{equation}  
Its principal effect is to  decrease slightly  the central density and to modify 
		   the radius of the star. The corresponding frequency change  is proportional to a quantity which we denote
		       $Z_1$, 

\medskip 
 - {\it non spherically symmetric} distorsion. 
		    This is the dominant second order effect  
                   and is denoted  $\omega_{2D}$   as in DG92. \\

\ni ii) rotationally induced transport and mixing  
 modify  the internal  stratification and influence   the   evolution  of the steady configuration. This 
 effect is responsible for a frequency change which is proportional to a quantity denoted 
  $Z_2$.
The frequency correction  induced by modifications of the the steady configuration  is then  
proportional to $Z_1+Z_2$. Finally the (second order)  frequency correction   is given by :
\begin{equation}
\omega_{2,n,\ell,m}=   {\Omega^2\over \omega^{(0)}} ~(Z_1+Z_2)
+\omega_{2I}+\omega_{2P}+\omega_{2T}+\omega_{2D}+ {\omega_1^2 \over \omega_0}
\label{om2a}
\end{equation}
In order to obtain explicit expressions and quantitative estimates for these corrections, one 
starts again with Eq.\ref{wave0}. The expansions are carried out up to second order 
$f'= f'_0+f'_1+f'_2$ for any scalar perturbed quantity and  for the 
displacement vector  field ${\bf \xi=\xi_0+\xi_1+\xi_2}$.   Assuming that the zeroth order and first order systems have
  been solved, it remains to solve for the second order system of 
  equations. The knowledge of the first order correction to the {\it eigenfunction} is  indeed required to
   compute $\omega_2$. 
   One also needs to include the  
 effects of the centrifugal force on the equilibrium structure.
 For moderate rotation, 
 it  is enough to use a perturbative technique
 to compute the centrifugal distorsion.
 Two approaches have been used: 
  a mapping technique and  a direct perturbative  method.

\subsection{ Mapping technique}\label{mapping}

Because the shape of the star is distorted,  
one must in principle work in a {\it spheroidal} coordinate system  and take into account the oblateness of the
surface in the boundary conditions. It is possible however to  work in a {\it spherical} coordinate system  with simple
(classical) boundary conditions by defining a mapping between 
the  coordinate system   $(r,\theta,\phi)$  in the spheroidal volume of the star 
and  the corresponding one  $(\zeta,\theta,\phi)$  in  a spherical volume. 
This mapping then consists in defining a transformation $(r,\theta,\phi) \Longrightarrow (\zeta,\theta,\phi)$
where $r$ and $\zeta$ are related through the transformation:
\begin{equation}
r(\zeta,\theta) = \zeta ~(1 -h_2(\zeta, \Omega) P_2(\cos\theta)) 
\label{rzeta}
\end{equation}
where $P_2(\theta)$ is the second order  Legendre polynomial.
The function $h_2$  is  chosen so that surfaces of constant $\zeta$ are surfaces of constant pressure
in particular the surface of the star is given by $\zeta =R$. This gives for the oblatness of the star 
$r_{eq}-r_{pole}=  (3/2) h_2(\zeta,\Omega) ~\zeta $
between the equatorial and polar radii.

This approach was carried out  by Simon (1969) and  Lebovitz (1970). 
Gough, Thompson (1990)  developed the formalism under  the Cowling approximation 
 -  which consists in  neglecting the Eulerian
perturbation to the gravitational potential- for a general stellar model and a prescribed  rotation law $\Omega(r,\theta)$. 
The authors  considered the wave equation Eq.\ref{wave0} from Lynden-Bell and Ostriker (1967)
 which in our notation is (see also Sect.\ref{variat}):
$${\cal L} (\vec \xi) - \rho_0 \omega^2 (\vec \xi) = i \omega {\cal B} (\vec \xi) + {\cal N}(\vec \xi)$$
where
\begin{eqnarray}
 {\cal L}(\vec \xi) &=& \vec \nabla \left( \right. (p_0-\rho_0 c_{s0}^2) \nabla \cdot \vec \xi  \label{Lmap} \\ 
 & -& \vec \xi \cdot \nabla p_0  \left. \right) 
- p_0 \vec \nabla(\vec \nabla \cdot \vec \xi) 
+\vec \xi \cdot \vec \nabla(ln \rho_0 ) \nabla p_0  \nonumber 
  \end{eqnarray} 
with  
$ {\cal N}\xi =  \rho_0 \left[\vec \xi_0 \cdot  \vec \nabla  (\vec v_0 \cdot \vec \nabla \vec v_0) 
- (\vec v_0 \cdot \nabla)^2 \vec \xi \right]$ and  ${\cal B} (\vec \xi) $ is defined in Eq.\ref{BLO}.
GT90 actually were interested in the effects of a magnetic field on adiabatic oscillation frequencies and included the Lorenz force
contribution to the above equation which we disregard here. 
The equilibrium quantities $p_0,\rho_0$ and the adiabatic sound speed $c_{s0}^2$ can be expanded 
about their values in the spherical volume, 
so for exemple:
\begin{equation}
p_0({\bf r})=p_{00}(\zeta)+\epsilon^2 p_2(\zeta) ~P_2(\cos \theta)
\label{p0p2}
\end{equation}
This procedure has the advantage of retaining simplicity in the boundary conditions which  are  $\delta P =0 $ at surface 
and $\psi'=0$  matching the solution of Laplace equation away outward 
from the surface.  GT90 found that for slow rotation, these boundary conditions are far enough. However, they also 
 discussed in detail the boundary conditions obtained by matching the interior with an isothermal atmosphere in
 hydrostatic equilibrium.   \\
The price to pay   for the change of coordinate system Eq.\ref{rzeta} is   additional terms coming from the
decomposition of the gradients such that 
$\nabla = \nabla_0+ \nabla_\Omega$ where
\begin{eqnarray}
\nabla_0 &=& {\bf e_r} {\partial \over \partial \zeta} +
 {\bf e_\theta} {1\over \zeta }{\partial \over \partial \theta } + {\bf e_\phi} {1\over \zeta \sin \theta }{\partial \over \partial \phi } \\  
\nabla_\Omega &=& {\bf e_r}~{d (\zeta h_2)\over  d\zeta} ~P_2(\theta)  {\partial \over \partial \zeta} + {\bf e_\theta} \left( h_2
{d P_2(\theta) \over d\theta }{\partial \over \partial \zeta}+{1\over \zeta} h_2 P_2(\theta)  {\partial \over \partial \theta}\right)  
\nonumber \\ 
&+& {\bf e_\phi} h_2 P_2(\theta) {1\over \zeta \sin \theta} {\partial \over \partial \phi}
\end{eqnarray}
so that ${\cal L}={\cal L}_0+{\cal L}_\Omega$. An inhomogeneous differential equation and 
boundary conditions for $h_2(\zeta)$ are obtained 
by matching the gravitational potential onto the vacuum potential in $\zeta>R$ 
 and requiring the transformation  be regular at $\zeta=0$.

\medskip 
 Simon (1969)
  carried the expansion  up to 2nd order.
He derived an integral expression for the second order  eigenfrequency correction for  nonradial
p-modes for  a prescribed rotation law $\Omega(r,\theta)$. 
He then studied the particular case of a uniformly rotating  homogeneous spheroid  
and computed the frequency for 
the radial fundamental mode which was found to be of the form:
 $\sigma = \sigma_0 + \beta ~\hat \Omega^2$, 
$\bar \Omega$ is the dimensionless rotation rate $\Omega/(GM/R^3)^{1/2}$,
 retrieving results  similar to those of  Ledoux (1945)  and  Chandrashekhar and Lebovitz (1962). 
Chlebowski (1978) derived the expressions for the correcting coefficients for non  radial modes
and wrote  the frequencies under the form
 \begin{eqnarray}\sigma = \sigma_0 -m (1-C_{n,\ell})~\hat \Omega+ {1\over 2\sigma_0} (P_{n,\ell} - m^2  ~Q_{n,\ell}) ~\hat \Omega^2
\label{chleb}
\end{eqnarray}
 Numerical results were obtained for g-modes  of white dwarfs. 
 Later  Saio (1981) developed an equivalent procedure and 
 applied it to the study  of a polytrope in uniform rotation.
   He put  the scaled eigenfrequency  under the convenient form
\begin{eqnarray}
 \sigma_{n,\ell,m} &=& \sigma_{0,n,\ell}+ m ~\hat \Omega ~(1-C_{nl}) +~  {\hat \Omega^2\over \sigma_0} 
\left((Z+X_1+X_2)  \nonumber \right. \\ 
&+& m^2  (Y_1+Y_2) \left. \right)
\label{saioc}
\end{eqnarray}
in the observed frame.  $\sigma_0$ is the dimensionless frequency of a nonrotating polytrope 
having the same central pressure and density than
the rotating one.  Saio  found a   significant departure from equidistance  due to both spherical
 Z and non spherical X2,Y2   distorsion.
  GT90  applied this formalism to a shellular rotation  for a solar model. 
Burke, Thompson (2006)  used the same approach and  
correct some errors in GT90. They did not include the poloidal
 correction of the first order correction to eigenfunction ${\bf \xi_{1p}}$. 
 They  computed  p-mode  eigenfrequencies for 
 stellar models with   a variety of masses and ages in the Cowling approximation. 
 
\subsection{Perturbative approach }
This  procedure has been first   developped by Hansen et al. (1978) and  DG92.

\subsubsection{Perturbative approach for the structure}
 Eq.\ref{equil} is decomposed as 
\begin{eqnarray}
-{1\over \rho_0}{ \partial p_{0}\over \partial r} &=& {\partial \psi_0 \over \partial r} +  {2\over 3} r
\Omega^2(r,\theta)  ~(1-P_2(\cos \theta))  \label{st11}\\
-{1\over \rho_0}{ \partial p_{0}\over \partial \theta} &=& {\partial \psi_0 \over \partial \theta} + {1\over 3} r^2
\Omega^2(r,\theta)  ~{dP_2(\cos \theta)\over d\theta}  
\label{st12}
\end{eqnarray}
where the acceleration term $ -r \Omega^2 \sin \theta \vec e_s $ has been written in terms of Legendre polynomials. 
Pressure, density and gravitational  potential
 are then  expanded in terms of  even  Legendre polynomials
due to the equatorial symmetry. One keeps only  the first two terms i.e. for the pressure, for instance  one sets
\begin{equation}p_0(r,\theta) = p_{00} (r)+ p_2(r) ~ P_2 (\cos \theta) 
\label{p0}
\end{equation}

Inserting this expansion into    Eq.\ref{st11},\ref{st12}, one gets
on one hand one equation for  the spherically symmetric perturbed part of the stellar model 
 $p_{00}, \rho_{00}, \psi_{00}$ and on the other hand equations for   the  nonspherical part of the stellar model
  $p_{2}, \rho_2,\psi_2$.

\medskip 
\ni{\it Spherically symmetric distorsion}
For a shellular rotation $\Omega(r)$, Eq.\ref{p0} and Eq.\ref{st11} yield:
\begin{equation}
-{1\over \rho_{00}} {d p_{00}\over dr} =   {d\psi_{00} \over dr} + {2\over 3} r \Omega^2
\end{equation} 
This  equation replaces the hydrostatic equation for a non-rotating star. 
The horizontally averaged centrifugal force  modifies the effect of
gravity  and  the equilibrium pressure must adjust in order to
balance an `effective' gravity $g_{\rm eff}$ (Eq.\ref{geff}
.
This effect can easily be implemented  in an evolutionary code
 by solving numerically   for a classical spherically symmetric (non-rotating) stellar model  
but using the effective  gravity  (Kippenhahn et Weigert, 1994).


\medskip
\ni{\it Nonspherically symmetric distorsion of the equilibrium structure}
For a shellular rotation $\Omega(r)$, Eq.\ref{p0},\ref{st11},\ref{st12}  yield two equations:
\begin{eqnarray} 
 {d p_{2}\over dr} &=& - \rho_{00} {d\psi_{2} \over dr}- \rho_{2} {d\psi_{00} \over dr} - {2\over 3} r \rho_{00}
 ~\Omega^2 \\
p_{2}(r)  &=& -\rho_{00}(\psi_{2} +{1\over 3}  r^2 \Omega^2)   
\end{eqnarray}
The perturbed part of the Poisson equation  writes:
\begin{equation}
{1\over r^2} {d \over dr} \left( r^2 {d \psi_{2}\over dr}\right) -{6\over r^2}  \psi_{2}= 4\pi G \rho_{2}
\end{equation} 
Details for the integration of the system can be found in DG92, Soufi et al. (1998). 
Generalisation to the case of $\Omega(r,\theta)$   is treated in DG92   and  GT90.

\subsubsection{Second order effects from uniform rotation: centrifugal force and uniform rotation}\label{s2}

For pedagogical reasons,  the effects of  the Coriolis force are ignored in this section. Including only 
the frequency corrections due to the 
centrifugal force  remains a good approximation for high frequency p-modes. Indeed the Coriolis force becomes
negligible for large $\mu$ (Eq.\ref{mu})). We also restrict the study to uniform rotation.
In this simplified case,  the eigenfrequency  and eigenfunction are expanded as 
$\omega_0+\omega_2$ and  $\xi_0+\xi_2$ respectively. One starts  again with Eq.\ref{wave0}, \ref{wave1}:
\begin{eqnarray} 
  {\cal L}_0 ({\bf \xi}_0+ {\bf \xi}_2) +{\cal L}_2({\bf \xi}_0)
  & -&  \hat \omega^2  (\rho_{00}+\rho_2) ~ ({\bf \xi_0}+{\bf \xi_2})  =0     
\label{eql02}
\end{eqnarray}  
 where
 \begin{eqnarray}
 {\cal L}_0 \vec \xi &=&  \vec \nabla p'- {\rho'\over \rho_{00}} \vec \nabla p_{00} + \rho_{00} \vec \nabla \psi'
 \label{eql00} \\
   {\cal L}_2 \vec \xi &=&{\rho_2}~ ({\rho'\over \rho_{00}} \vec \nabla p_{00} + \vec \nabla \psi') -
   {\rho'\over \rho_{00}} \vec \nabla p_2
 \label{eql22}
 \end{eqnarray}

\medskip 
\ni{\it Effect of uniform rotation on $\omega_0$}

\ni From Eq.\ref{eql02}, one first gets the zeroth order system Eq.\ref{zerothorder}
with the zeroth order  solution given by Eq.\ref{omeg0}, \ref{xi0}. However, now, second order effects  are indirectly 
inscribed in $\omega_0$ as  the equilibrium quantities $p_{00},\rho_{00}, \psi_{00}$  
are modified by the effective gravity Eq.\ref{geff}.
This effect is taken into account 
for instance in   Soufi \etal~ (1998), Goupil et al. (2000), 
Daszynska-Daszkiewicz \etal~ (2002, 2003),  Suarez \etal (2006a,b).  
It is quite small in general: it slightly  changes the  radius of the star for given
mass and age  compared to a nonrotating model. Therefore the tracks of a nonrotating equilibrium 
model  and a model with same mass but 
including spherical distorsion do not  coincide.  
Fig.1 of Goupil et al. (2000)   shows evolutionary tracks of main sequence  $1.80 M_\odot$ 
stellar models  where gravity is modified by rotation for initial rotational velocities
 50 km/s, 100 km/s and 150 km/s. Local conservation of angular momentum is assumed for the evolution of the rotation with age.
The  evolutionary track for a spherically distorted model 
is shifted to lower luminosities than  
that of  non-rotating model with same mass.
It therefore  corresponds 
 to a track of nonrotating models with
smaller masses, the larger the initial rotation velocity, the smaller the mass 
of the nonrotating model.  The rotation modification of  gravity 
 corresponds to a change of mass  smaller than  $0.02 M_\odot$ for initial velocities and masses 
  in the above ranges.

 This  has the consequence that a comparison of the  oscillation frequencies of a non-rotating model and a model including rotation, 
depends on the choice of the non-rotating model.  
 Several choices of non-rotating models are 
possible.  Indeed, for  given physical input, chemical composition and rotation rate, 
 two additional parameters must be  specified to define a model. 
 The simplest
 and meaningful choice is that these parameters be fixed at the same value for
  both models.
  Saio (1981)   compared frequencies of polytropes with and without rotation keeping the
central pressure and density 
constant (by keeping the polytropic index constant). 
For realistic stellar models, other possible choices are the mass and the radius;
the mass and the  central hydrogen content for main sequence stars; 
 the effective temperature and the luminosity. 
The frequency difference for a given mode between a nonrotating model and a 
rotating  model depends on the choice of these two constants. 
Christensen-Dalsgaard and Thompson (1999) 
kept the luminosity and the effective temperature constant  
and showed that the results  for radial modes for a polytrope 
  are quite similar to those obtained for realistic models. 
  They also found that the
frequency difference obtained between nonrotating and rotating polytropes keeping constant mass and radius 
as they chose 
can  simply be modelled by adding a constant value to the difference obtained by  keeping constant 
 central density and pressure (as in Saio(1981)) at least for asymptotic ie high $n$  radial modes :
${\delta \sigma / \sigma} = \hat \Omega^2 ~(X_1+Z) /\sigma^2 +0.33$
where Z is interpolated  from Saio's table.

\medskip
\ni {\it Behavior of $\omega_2$} The second order correction to the eigenfrequency, $\omega_2$, 
is obtained from Eq.\ref{eql02} written under the form:
\begin{eqnarray} 
  {\cal L}_0 {\vec \xi}_2 -\rho_{00}  ~\omega_0^2  {\vec \xi}_2 =   2 \rho_{00} ~ \omega_0 ~\omega_2 ~{\vec \xi}_0 
   - \left( {\cal L}_2  - \omega_0^2  \rho_{2}\right)  {\vec \xi}_0 
  \label{L2}
\end{eqnarray}  
together with the second order perturbed continuity and Poisson equations. 
Projection of ${\bf \xi_0}$ onto this equation
(recalling that 
 $<{\bf \xi}_0|{\cal L}_0 {\bf \xi}_2>= <{\cal L}_0{\bf \xi}_0| {\bf \xi}_2>
 =<\rho_0 \hat \omega_0^2 {\bf \xi}_0 |{\bf \xi}_2>$) yields:
\begin{eqnarray} 
\omega_2 = {1\over   2  \omega_0 I } ~ \int_{V0} ~{\bf \xi}_0^* {\bf \cdot}
 \left({\cal L}_2  - \omega_0^2 \rho_2 \right) {\bf \xi}_0 ~d^3r     
\label{omeg2}\end{eqnarray}  
Some algebraic manipulations leads to :
$$\omega_{2,n\ell,m} = {\Omega^2 \over \omega_{0,n,\ell}}~(D_{1,n,\ell}+m^2  D_{2,n,\ell})
$$
for each mode $(n,\ell,m)$ (DG92).  The $m, \Omega$ dependence of Eq.\ref{omeg2} 
can be understood as a consequence of symmetry properties.

\medskip
\ni {\it Consequences of symmetry properties}

\ni When only the centrifugal force is considered, the eigenfrequencies obtained by perturbation are of the form:
$\omega_{n,\ell,m}=  w^{(0)}_{n,\ell,m}+  \Omega^2 ~w^{(2)}_{n,\ell,m}+...$
More generally,  following  Ligni\`ere \etal ~(2006) and Reese \etal~ (2006), we write the frequency as an expansion in power of $\Omega$
 $$\omega_{n,\ell,m}(\Omega)=  \sum_{j\geq 0} w^{(j)}_{n,\ell,m} ~\Omega^{j}$$
 where $j $ is an integer.  
The symmetry property $\omega_{n,\ell,m}(\Omega)=\omega_{n,\ell,m}(-\Omega)$   which exists in absence of the Coriolis force
 imposes   $\omega^{(2j-1)}_{n,\ell,m}=0$. Hence the  perturbative expansion of the frequency in powers of $\Omega$  
 only involves even powers in $\Omega$.
Further the  reflexion  symmetry also imposes $ \omega_{n,\ell,-m}(-\Omega)= \omega_{n,\ell,m}(\Omega)$ 
that-is $ \omega^{(2j)}_{n,\ell,m}= \omega^{(2j)}_{n,\ell,-m}$. Hence  the even order  coefficients are even functions of $m$.
As a consequence, the generalized rotational splitting vanishes $ S_m= (\omega_{n,\ell,m} -\omega_{n,\ell,-m})/m=0$
in absence of the Coriolis force.

\medskip

\ni {\it Near-degeneracy}\index{Near-degeneracy}

\ni Chandrasekhar \& Lebovitz  (1962) 
discussed    the  effect    of  near-degeneracy in the stellar oscillation context that-is  the existence  
of small denominators which is a classical problem of the perturbation methods in quantum mechanics 
for instance.  Indeed the n-th order correction of an eigenfunction 
usually involves  small  denominators of the form  $1/(\omega_{0j}-\omega_{0k})$   which can make 
this correction  as large as the (n-1)-th order eigenfunction correction, thereby invalidating the perturbation expansion.
Hence when  2 modes, labelled say $j$ and $k$, are 
such that  $(\omega_{0j}-\omega_{0k})  \sim 0$, one must consider them as degenerate or 'coupled'.
This leads to an additional correction to the eigenfrequencies. Several works in the context of stellar pulsation 
have  included this correction: 
Simon (1969), DG92 , Suarez \etal~ (2006, 2007) , 
Soufi \etal~ (1998),   Karami \etal ~ (2005).  
%
  The eigenfunction correction, $\vec \xi_{2k}$   for the  mode labelled $k$, is obtained  by assuming that it can be
written in the  zeroth order normal mode basis as:
\begin{equation}
\vec \xi_{2k} = \sum_{j\not= k} \alpha_j \vec \xi_{0j} 
\label{xi2}
\end{equation}
 where the unknowns now are the $\alpha_{j}$  coefficients. Inserting Eq.\ref{xi2} into Eq.\ref{L2}, one obtains 
 \begin{eqnarray} 
 \sum_{j\not= k}  \alpha_j ~({\cal L}_0 &-& \rho_{00} ~\hat \omega_{0k}^2 )~ {\vec \xi}_{0j}   \\
 &=&   2  \rho_{00} ~\omega_{0k} ~\omega_{2k}  ~{\vec \xi}_{0k} 
  - \left( {\cal L}_2 {\vec \xi}_{0k} - \omega_{0k}^2 
 \rho_{2} {\vec \xi}_{0k} \right) \nonumber   
\end{eqnarray}  
Taking the inner product Eq.\ref{inner}  with   $ {\vec \xi}_{0k}$ yields  the solution for $\omega_{2k}$. 
  \begin{eqnarray}
\alpha_j= - {1\over  \left(\omega^2_{0j} - \omega^2_{0k} \right)  I_j } 
 \int_0^R  \vec \xi^*_{0j} \cdot  \left( {\cal L}_2  - \omega_{0k}^2  \rho_{2} \right) \vec \xi_{0k}   ~{\bf d^3r} 
\end{eqnarray}  
Hence the correction to the eigenfunction of  mode $k$ is given by 
 \begin{equation}
\vec \xi_{2k} = - \sum_{j\not= k} { D_{jk}  \over  \left(\omega^2_{0k} -\omega^2_{0j} \right)  } ~
\label{xi2k}
\end{equation}
where 
\begin{equation}
 D_{jk}= {1\over  I_j }  ~  \int_V {\vec \xi}^*_{0j}
 {\bf \cdot}  \left( {\cal L}_2  -\omega_{0k}^2  \rho_{2} \right) {\vec \xi}_{0k} ~{\bf d^3r}   
 \label{Djk}
 \end{equation}
 Note that $D_{jj}=2 \omega_{0j} \omega_{2j}$, the second order (nondegenerate) frequency correction for mode $j$.
Let 2 degenerate modes  such that  $\omega_a \sim \omega_b$ with
$\omega_a=\omega_{0a}+\omega_{2a}$ and  $\omega_b=\omega_{0b}+\omega_{2b}$
where  $\omega_{0a}$, et $\omega_{2a}$ 
are given by Eq.\ref{omeg0} and Eq.\ref{omeg2}. One seeks for an eigenfunction of the form: 
 ${\vec \xi}_{0ab} = c_a ~{\vec \xi}_{0a} + c_b ~{\vec \xi}_{0b}$
associated with an eigenvalue  $\omega$. Let define
 $$\bar \omega = {\omega_{0a}+ \omega_{0b} \over 2}   ~~~~~~~~~~;~~~~~~~~\delta \omega = \omega_{0a}- \omega_{0b}  $$
 Inserting these expressions into Eq.\ref{L2} and  taking the inner product with  ${\bf \xi}_{0a}$ and 
 $ {\bf \xi}_{0b}$
 respectively yield the following system of linear equations:
\begin{eqnarray}
c_a \left(\omega^2_{0a}+2\omega_{0a} ~\omega_{2a}-\omega^2\right) + c_b ~D_{ab} &=& 0  \nonumber \\ 
c_a ~ D_{ba} + c_b \left(2 \omega_{0b} ~\omega_{2b}+\omega_{0b}^2 -\omega^2 \right) &=& 0 
\end{eqnarray}
where $D_{jk}$ is given by Eq.\ref{Djk} above. Up to second order in $\Omega$, one can write $\omega^2_{0a}+2\omega_{0a} ~\omega_{2a}  \sim \omega^2_{a}$ so that 
\begin{eqnarray}
c_a   \left(\omega^2_a-\omega^2 \right)+ c_b D_{ab} &=& 0  ~;~ c_a D_{ba} + c_b \left( \omega^2_b-\omega^2 \right) = 0
\nonumber
\end{eqnarray}
The solutions must then verify:
$$\omega^4-\omega^2 (\omega^2_{a}+\omega^2_{b})+\omega^2_{a}\omega_{b}^2 - 
 D_{ab}~D_{ba} =0
$$
Accordingly, the frequencies of modified  degenerate  modes are 
\begin{equation}
\omega^2_{\pm} = {1\over 2}(\omega^2_{a}+\omega^2_{b})\pm 
 {1\over 2} \sqrt{(\omega^2_{a}-\omega^2_{b})^2+ 4 D_{ab} D_{ba} }
 \label{ompm}
\end{equation}

The corresponding eigenfunctions $ {\bf \xi}_{\pm}$ 
are composed  of  contributions from both zeroth order modes $a$ and $b$ 
in respective parts determined by the
ratio $c_b/c_a$.   For vanishing coupling coefficients or negligible in front of the
frequency difference i.e. $4\sqrt{D_{ab}D_{ba}}<<|\omega^2_{a}-\omega^2_{b}|$, 
one recovers the uncoupled second order eigenfrequencies 
$\omega_\pm = \omega_{a,b}= \omega_{0a,b}+\omega_{2a,b}$.


When the effect of the Coriolis  force is taken into account up to 3rd order  and for 
 non uniform  rotation, the calculation is slightly more complicated 
 but the procedure remains  in the same spirit (Soufi et al. 1998).

\subsubsection{Full second order and beyond and shellular rotation}

In this section, we consider both the Coriolis and the centrifugal forces and assume a shellular
rotation $\Omega(r)$, unless otherwise stated. 
At the end of this section, we also briefly discuss the 
extention of  the calculation up to third order included.
Let then  return to  the eigenvalue system Eq.\ref{wave0}, \ref{wave1} which includes both 
forces and a differential rotation. When compared to Eq.\ref{eql02}, 
additional terms are present which are   due to Coriolis; another additional  term is proportional to the gradient of $\Omega$ i.e.
\begin{eqnarray} 
{\bf f} &\equiv &  {\cal L}_0 {\bf \xi} -\hat \omega^2  \rho_{00}~ {\bf \xi}  + 2  \hat \omega \rho_{00}~   i {\bf \Omega}   \times {\bf  \xi} 
 -  \rho_{00}~ ({\bf \xi  \cdot \nabla}  \Omega^2)  ~r \sin \theta {\bf e_s}  \nonumber\\ 
 & &  
  +\left( {\cal L}_2 {\bf \xi} -\hat \omega^2  \rho_{2}  {\bf  \xi} \right)  + 2 i \hat \omega  \rho_{2} 
 \Omega \times \xi   =0 
   \label{f}
\end{eqnarray}  
where the operators ${\cal L}_0, {\cal L}_2$ have been defined in Eq.\ref{eql00},\ref{eql22}.

\medskip 
\ni {\it Effects of a shellular rotation  on $\omega_0$}

\ni When solving the zeroth order eigensystem (Eq.\ref{zerothorder}), the quantities $p_{00},\rho_{00}$  
can  be  provided by solving a 1D spherically symmetric stellar model which includes
 rotationally induced  mixing of  chemical  elements and transport of angular momentum. 
 which can  significantly change 
the evolution of a rotating  star compared to one which is not rotating.
Exemples of comparison of evolutionary tracks of models  with and without rotationally 
induced transport\index{rotationally  induced transport}
   can be found in the litterature: 
 evolutionary tracks for a  $9M_\odot$ stellar model  evolved with an initial rotational velocity 
of   $v=$ 100 and 300 km/s 
   (Talon \etal ~ 1997); 
   for  massive stars  (Meynet \& Maeder, (2000)) ;  for 
     a $~1.85 M_\odot$ stellar model evolved with an initial rotational 
   velocity of  $v= 70$km/s or 100 km/s (Goupil \& Talon, 2002); for low mass stars (Palacios \etal ~ 2003).
  The  main sequence lasts longer for the rotating model   as
mixing  fuels  fresh H  to the  burning core. This also causes for  the rotational  
 model  a larger  increase of  its luminosity  with time.  
The evolution of the rotating star can then be quite different from that of a nonrotating one. For an
intermediate mass main sequence star with a convective core, the evolution  
 is closer to that of a nonrotating model with overshoot. The inner structure in the vicinity 
of the core is therefore quite different.
 Hence, at zeroth order, the eigenfrequency 
$\omega_0$  differs from the eigenfrequency $\omega^{(0)}$ of  a nonrotating star 
or that of a uniformaly rotating star 
because one must take into account the rotationally induced  transport  which are likely to occur 
in presence of a differential rotation. 
At the same location in the HR diagram,  the Brunt-V\"aiss\"al\"a frequencies in the central regions 
 are similar for the rotating and nonrotating stars  (both with no overshoot) 
but the  Brunt-V\"aiss\"al\"a frequency for the overshoot model is quite different
  (see Fig.4. Goupil \& Talon, 2002).
Hence,  one expects large frequency differences for  low frequency  and mixed modes  which are 
 are sensitive to these layers between
 a rotating (no overshoot) model and
a nonrotating model with overshoot one.

Consequences of mixing on solar-like oscillations have been studied for several specific stars.
An exemple is a $1.5 M_\odot$ stellar model  with an initial 
velocity of 150 km/s (Eggenberger, PhD; Mathis  et al., 2006). At the same location in the HR diagram, the evolutionary
 stages  respectively  are $X_c=0.33$ and $X_c = 0.443$  for the central hydrogen relative mass content.
 The effect remains small on the averaged large separation  
 as  mass and radius are similar 
 ($<\omega_{n,\ell}-\omega_{n-1,\ell}>_n/(2\pi) \sim 70.40 \mu$Hz  
 without rotation  against $69.94 \mu$Hz with rotation).  The  difference for the 
averaged  small separation between $\ell=0,2$  ($<\omega_{n,\ell}-\omega_{n-1,\ell+2}>_n/(2\pi) \sim 5.07\mu$Hz without rotation against $5.76\mu$Hz with rotation)
  is large enough  to be detected  with the space seismic experiment CoRoT (Baglin et al. 2006).  

For lower mass stars,
undergoing angular momentum losses, the effects might be more subtil.
 {\it $\beta$ Vir} is a solar like star with a mass $1.21-1.28 M_\odot$ 
and an effective 
temperature 
$\sim 6130 K$. It was selected as one best candidate  target for the unfortunate 
seismic space missiom EVRIS (Michel \etal~ 1995, Goupil \etal ~ 1995). 
Solar like oscillations  for this star  have later  been detected from ground with Harps: 31 frequencies  between 0.7
and 2.4 $\mu$Hz.  This star is a slow rotator with a $v \sin i$  between 3 and 7 km/s. 
Eggenberger \& Carrier (2006) have modelled this star with the 
evolutionary Geneva code with the assumption of 
rotational mixing of type I (Zahn, 1992; Maeder, Zahn 1998; Mathis et al. 2004).
 The computed  frequencies show that one cannot reproduce simultaneously
the large and small separations. The authors stress that 
  a large dispersion  of the large 
  separation for the nonradial modes exists which could be attributed to 
  non resolved splittings.
Three models with initial velocity $v = 4; 6.8; 8.2$ km/s (ie with magnetic breaking) have been studied
(Fig.8 of Eggenberger \& Carrier (2006)). As the rotation is so slow, its effect on the structure, hence on $\omega_0$ and the
small separation, is very small. This small separation  however is found larger than for a corresponding nonrotating
model.  The rotationally induced transport  is not efficient enough to impose a
uniform rotation in the radiative zone. The models give $\Omega_c/\Omega_s \sim 3.12$.
Such a $\Omega$ gradient  ought to be detectable with the rotational splittings, 
provided  they could be
detected. The mean value of the splitting is found smaller 
when  the rotation is uniform (0.6 instead of 0.8-0.9 n$\mu$Hz). 
Hence if futur observations indicate that this 1.3 $M_\odot$ star, 
like the Sun,  is   uniformaly rotating, one will have to call for an additional  
mechanism to transport angular momentum as in the solar case.  

A precise knowledge of $\omega_0$  can also serve as a test for the transport efficiency of the horizontal turbulence
as reported by Mathis et al. 2006.
The investigated case is that of   a calibrated  solar model.
The initial rotational velocity is taken to be 0, 10, 30 or 100 km/s. The seismic properties are compared for 3
different prescriptions for the horizontal turbulence transport  coefficient, $D_h$ given respectively by Zahn
(1992), Maeder (2003) and Mathis et al. (2004). The  more recent prescriptions   lead 
to  increased transport and
mixing and therefore to  a larger effect on the eigenfrequencies compared to nonrotating model .
Increasing the rotation leads to an increase of the   value of the small separation but its variation  with 
frequency remains similar. 
Going from Zahn's transport coefficient to Maeder's coefficient  results in the increase 
with $\Omega$ to be larger when $\omega$ is
increased.

\medskip
\ni {\it Solving for the first order  eigenfunction correction  $\vec \xi_1$}
 
\ni The resolution for  the first order   frequency correction has been discussed in Sect.{5} above. 
 However  solving for the second order frequency correction, $\omega_2$, requires
the knowledge of the first order eigenfunction, 
 \begin{equation}
 {\bf \xi_1}={\bf \xi^P_1}+{\bf \xi^T_1}
\label{xi}
 \end{equation}
which   is composed of a poloidal part ${\bf \xi^P_1}$
and a toroidal part ${\bf \xi^T_1}$. DG92 provide a  detailed procedure  for calculating $\xi_1$ and $\omega_2$ and
 the nonspherical distortion of the star for a differential rotation $\Omega(r,\theta)$ for a prescribed rotation
 law Eq.\ref{latit}. The toroidal part is obtained by taking the radial curl of Eq.\ref{om1}. 
One can obtain  the poloidal
part by  two possible  methods:
the first one consists in  expanding the poloidal  part in terms of unperturbed eigenvectors, 
one then gets the  standard expression as Eq.\ref{xi2k}.
 However as stressed by DG92, this method  involves an 
 infinite sum which in practice must be truncated at some level. 
 An alternative approach is to solve Eq.\ref{om1} directly following 
 Hansen, Cox and Carroll (1978) and Saio (1981) in deriving equations for the
radial eigenfunctions corresponding to $\xi_1^P$. DG92  generalized it so as  to include latitudinally differential rotation.
For a shellular rotation, the poloidal  and toroidal parts are seeked under the respective form 
\begin{eqnarray}
\vec \xi^P_{1,\ell,m} &=&  \xi_{1r}(r)~Y_\ell^m~ {\bf e_r}  + \xi_{1h}(r) {\vec \nabla_h} Y_\ell^m  \nonumber \\
\vec \xi^T_{1,\ell,m} &=& \sum_{\ell,m} ~\tau_1(r) ~{\bf e_r} \times \vec \nabla_h Y_\ell^m  \nonumber 
\end{eqnarray}
For the vector field ${\vec f}$ defined in Eq.\ref{f} to vanish, one must impose: 
\begin{eqnarray}
\int  ~ Y_\ell^{m*}~ (\vec e_r\cdot \vec f) ~d{\underline \Omega} = 0 \rightarrow (\xi^P_{1})_{r,\ell} \nonumber \\
\int ~ Y_\ell^{m*}~  (\vec \nabla_h \cdot \vec f) ~d{\underline \Omega} = 0 \rightarrow (\vec \xi^P_{1})_{h,\ell} \\
\int ~ Y_\ell^{m*}~ (\vec e_r \cdot \vec \nabla \times \vec f) ~d{\underline \Omega} = 0 \rightarrow \tau_{1,\ell+1}(r), \tau_{1,\ell-1}(r) \nonumber 
\end{eqnarray}
with $d{\underline \Omega} = r^2 \sin\theta d\theta d\phi$. These conditions provide differential equations for 
the  two  components $\xi_{1r},\xi_{1h}(r)$  and analytical expressions for $\tau_1(r)$ . 
The components $\xi_{1r},\xi_{1h}(r)$ of the poloidal part are obtained numerically; the numerical 
resolution of this system also  provides  the first
order correction $\omega_1$ which can be compared to the integral value  Eq.\ref{omeg1} (Hansen et al.1978, DG92).
  For  near  degenerate   modes $a$ and $b$, the   solution is 
$$  \omega_{1\pm}= \bar \omega_1 \pm \sqrt{(\Delta \omega_1)^2+4 \omega^2_{1,ab}}$$
 where
 $$ \bar \omega_1 = {\omega_{1a}+\omega_{1b}\over 2} ~~~;~~~ \Delta \omega_1 =  \omega_{1a}-\omega_{1b}$$
 and   for the coupling coefficient:
 $$\omega_{1,ab}= -<{\bf \xi}_{0a}|(m \Omega - i {\bf \Omega} \times) {\bf \xi}_{0b}>$$ 
i.e. for a shellular rotation (see also Suarez et al. (2006)), 
\begin{eqnarray}
\omega_{1,ab}& &=  m{1\over I} \delta_{\ell_a,\ell_b} \delta_{m_a,m_b} \int_0^R dr \rho_0 r^2  ~\Omega(r) \nonumber \\ 
& &~(\xi_{ra}\xi_{rb}
+\Lambda  \xi_{ha}\xi_{hb} 
     - \xi_{ra}\xi_{hb}-\xi_{ha}\xi_{rb}-\xi_{ha}\xi_{hb}) 
  \end{eqnarray}
where $\delta_{\ell_a,\ell_b}$ and $ \delta_{m_a,m_b}$ are Kroenecker symbols and $\Lambda=\ell_a(\ell_a+1)$

\medskip
\ni {\it Solving for the second order frequency correction  $\omega_2$:}
   Once the first order system is fully solved, the second order system can be solved with the same procedure
  described in Sect.\ref{s2} above:  
    \begin{eqnarray} 
  {\cal L}_0 {\vec \xi_2} &-& \omega_0^2  \rho_{00} {\vec \xi_2}
   =(2 \omega_0\omega_2 +2m\Omega \omega_1+m^2 \Omega^2+\omega_1^2) \rho_{00} ~{\vec  \xi_0} \nonumber \\  
     &+& 2\omega_0(m\Omega+\omega_1) \rho_{00} ~ {\vec \xi_1}  
   - 2  \omega_0 \rho_{00}   i {\bf \Omega}   \times {\vec  \xi_1}  \nonumber  \\  
  & +&  \rho_{00} ({\vec \xi_0  \cdot \nabla}  \Omega^2)  ~r \sin \theta {\bf e_s}  
  - \left( {\cal L}_2 {\vec \xi_0} - \omega_0^2  \rho_{2}  {\vec  \xi_0} \right)  
   \label{f2}
\end{eqnarray}  
where the operators ${\cal L}_0, {\cal L}_2$ have been defined in Eq.\ref{eql00},\ref{eql22}. 
 DG92 finally obtained :
$\omega_2 = \omega^T_2+\omega^P_2+\omega^I_2+\omega^D_2+ {\omega^2_1 \over 2 \omega_0}$.
DG92 established the  integral expressions for each of the above contributions to $\omega_2$ in the general case of a rotation
law ~Eq.\ref{latit} and showed that 
for a given $(n,\ell,m) $ mode,  $\omega_{2,n,\ell,m}$ is a polynome in $m^2$. Following the notations of Eq.\ref{chleb}, 
Eq.\ref{saioc}, 
 the frequency computed up  to second order  for a shellular rotation can be cast under of the form:
 \begin{equation}
 \omega_{n,\ell,m} =  \omega^{(0)}_{n,\ell}+ m \bar \Omega (1-C_{n\ell}-J_{n,\ell})+  \Omega^2 ~(D_{1,n,\ell,m}+ m^2 D_{2,n,\ell,m}) 
 \end{equation}
 Suarez et al. (2006b) provided  
 the equivalence between Saio's and DG92 notations for a shellular rotation included effect of degeneracy   up to second order .

\medskip 
\ni {\it Including third order  effects  }\label{third}

\ni In order to compute the third order frequency correction, a classical perturbation procedure  
requires  the knowledge of  the  second order eigenfunction  ${\bf \xi_2}$.  However it is possible to
build a pseudo zeroth order system which avoids the lengthy computations of eigenfunctions  at 
two successive orders including degeneracy. The procedure is developped in Soufi et al (1998), see also Karami et
al.(2005).  The wave equation Eq.\ref{wave0} is 
written as ${\cal F}_0(\xi,\omega)+{\cal F}_c (\xi,\omega)=0$.
 Part of the Coriolis force is included in the pseudo zeroth order ${\cal F}_0(\xi,\omega)$.
As a consequence, the first order  frequency and eigenfunction corrections are implicitely  included 
in the pseudo zeroth order solution; first order degeneracy is implicitely included as well. one seeks for a eigenfrequency of the form: $\tilde
\omega_0+\omega_c$ where $\tilde \omega_0$ is solution of the pseudo zeroth order eigensystem and $\omega_c$ is a frequency correction. 
The solution is then expanded
as $ {\bf \xi}= {\bf \tilde \xi}_0 + {\bf  \xi}_c+O(\Omega^4)$
 where 
 ${\bf \tilde \xi}_0$ takes the form
 \begin{eqnarray}{\bf \tilde  \xi}_0 (\vec r) &=&    \xi_{r,m} (r) Y_{\ell}^m {\bf e_r} + \xi_{h,m} (r) 
 {\bf \nabla}_h Y_{\ell}^m+  \nonumber \\ 
 &+&\tau_{\ell+1,m}(r) \vec e_r \times \vec \nabla_h Y_{\ell+1}^m + \tau_{\ell-1,m}(r) \vec e_r 
 \times \vec \nabla_h Y_{\ell-1}^m 
  \label{xim}
 \end{eqnarray}
 The pseudo zeroth order system  is  ${\cal F}_0(\tilde \xi_0, \tilde \omega_0)=0$  and 
 ${\bf \tilde \xi}_0$ is then  solution of a differential  
 equation system which must be  numerically  solved and provides the pseudo
 zeroth order eigenfrequency  $\tilde \omega_0$. 
 ${\bf  \xi}_c$ is a correction to
 the eigenvector field  ${\bf \tilde \xi}_0$ and is solution of the system 
 ${\cal F}_0(\xi,\omega)+{\cal F}_c (\tilde \xi_0,\tilde \omega_0)-{\cal F}_0(\tilde \xi_0, \tilde \omega_0)=0$
  arising from  $O(\Omega^2)$ contributions. The solvability condition for this equation yields the
   frequency correction $\omega_c$.  
Part of the third order  contribution due to  Coriolis force is implicitely included in pseudo zeroth order.
The price to pay is that 1)- the pseudo zeroth order numerical eigensystem is now m-dependent and 2)- the
eigenfunctions  are no longer orthogonal with respect to the inner product Eq.\ref{inner}. 
However they are orthogonal with respect to Eq.\ref{orth}.\\
For a given nondegenerate $(n,\ell,m) $ mode, the frequency is then obtained as:
\begin{eqnarray}
  \omega_{n,\ell,m} &=& \tilde  \omega_{0,n,\ell,m} + \bar \Omega^2 ~(D_{1,n,\ell,m} + m^2 ~D_{2,n,\ell,m}) \nonumber \\
 & +&  \bar \Omega^3 ~m ~(T_{1,n,\ell,m}+m^2 ~T_{2,n,\ell,m})
\label{linpert}
 \end{eqnarray}
where $\bar \Omega$ is a constant rotation (for instance a depth  average or the surface value). 
For convenience we define a frequency  $\omega_{0,n,\ell,m}$ such that 
\begin{equation}\omega_{0,n,\ell,m}= \tilde  \omega_{0,n,\ell,m} - m \bar \Omega  (1-C_{n,\ell,m}-J_{n,\ell,m})
 \label{om00}
  \end{equation}
so that one writes the eigenfrequency in a more familiar form:
\begin{eqnarray}
  \omega_{n,\ell,m} &=& \omega_{0,n,\ell,m} + m \bar \Omega  (1-C_{n,\ell,m}-J_{n,\ell,m}) \nonumber \\
  &+& \bar \Omega^2 ~(D_{1,n,\ell,m} + m^2 ~D_{2,n,\ell,m}) \nonumber \\
 & +& \bar \Omega^3 ~m ~(T_{1,n,\ell,m}+m^2 ~T_{2,n,\ell,m})
\label{solcub}
 \end{eqnarray}
 To a good approximation, when cubic order effects are not too large, 
one has $\omega_{0,n,\ell,m} \sim \omega_{0,n,\ell} $ where 
$\omega_{0,n,\ell}$ includes only the $O(\Omega^2)$ effects of spherically symmetric distorsion.
Because of the symmetry property $$ \omega_{n,\ell,m}(\Omega)= \omega_{n,\ell,-m}(-\Omega)$$  the
coefficients in Eq.\ref{solcub} verify:
\begin{equation}
\tilde \omega_{0,n,\ell,m}=  \tilde \omega_{0, n,\ell,-m} ~;~D_{j,n,\ell,m} = D_{j,n,\ell,-m} ~; ~T_{j,n,\ell, m} = T_{j,n,\ell,-m}
 \label{symproper}
  \end{equation}
for $j=1,2$.  It is also convenient to   cast the result under the following form 
 \begin{eqnarray}
  \omega_{n,\ell,m} &=&  \omega_{0,n,\ell,m} + m \bar \Omega  (1-C_{n,\ell,m}-J_{n,\ell,m})  \nonumber \\
  & + & {\bar \Omega^2 \over \tilde \omega_0} ~\left((X_1+m^2 Y_1) +(X_2+m^2 Y_2) \right) \nonumber \\
 & +& {\bar \Omega^3 \over \tilde \omega^2_0} ~m ~({\cal S}_1+m^2 ~{\cal S}_2)
\label{solcub2}
 \end{eqnarray}
where notations similar to Saio81's are used but generalised to shellular rotation (Suarez et al 2006b).
When modes are degenerate, one uses the same procedure as described in the above paragraph  and
 for a two mode degenerate coupling,  the frequencies are then given by (Soufi et al, 1998, Daszynska-Daszkiewicz et al.
 2002)
\begin{eqnarray}
\omega_\pm &= & \bar \omega_{n,\ell,m} \pm h_{n,\ell,m}  
\label{coupl}
\end{eqnarray}
where 
\begin{eqnarray}
\bar \omega_{n,\ell,m} &=& {1\over 2} (\omega_{n,\ell,m}+\omega_{n-1,\ell+2,m}) \nonumber \\
h_{n,\ell,m} &=& {1\over 2} \sqrt{d^2_{n,\ell,m}+4 H^2_{n,\ell,m}}  ~~;~~d_{n,\ell,m} = \omega_{n,\ell,m}-\omega_{n-1,\ell+2,m}   \nonumber 
\label{coupl2}
\end{eqnarray}
This generalizes to 3 mode coupling which becomes quite common when rotation is large.

\medskip
\ni {\it Rotation splitting}

\ni Using Eq.\ref{solcub} and symmetry properties,  
one derives for the rotational splittings  Eq.\ref{split1} the following expression:
 \begin{equation}
 S_m = \bar  \Omega ~(1-C_{n,\ell,m}-J_{n,\ell,m}) + \bar \Omega^3~  (T_{1,n,\ell,m}+m^2 T_{2,n,\ell,m})
\label{Sm1}
 \end{equation}
which is free of second order nonspherically symmetric distorsion effect. 
For degenerate modes,  the expression for the splittings $S_m$ is more complicated and can be derived from
Eq.\ref{coupl}
Asymmetry\index{splitting asymmetry} of the split multiplets, or departure from equal splittings for nondegenerate modes,   is  measured by: 
 \begin{equation}\Delta \omega_{n,\ell,m} = \omega_{n,\ell,m=0}-{1\over 2} (\omega_{n,\ell,m}+\omega_{n,\ell,-m})
\label{asym}
 \end{equation}
Using Eq.\ref{solcub}, its expression becomes:
\begin{eqnarray}
\Delta  \omega_{n,\ell,m} &=& (\omega_{0,n,\ell,0}-\omega_{0,n,\ell,m}) \nonumber \\ 
&+& \bar \Omega^2 ~((D_{1,n,\ell,0} - D_{1,n,\ell,m}) 
   - m^2 ~D_{2,n,\ell,m}) \nonumber 
 \end{eqnarray}
where we have used the symmetry properties Eq.\ref{symproper}. 
When cubic order effects on $\omega_{0,n,\ell,m}$ are small, $D_{2,n,\ell,m}$  dominates over the third order  differences 
$(\omega_{0,n,\ell,m} - \omega_{0,n,\ell,0}) $  and 
$(D_{1,n,\ell,0}-D_{1,n,\ell,m}) $
so that one can most often considers:
\begin{equation}\Delta  \omega_{n,\ell,m} \sim  - m^2 ~\bar \Omega^2 ~D_{2,n,\ell,m}
\label{asym2}
 \end{equation}

\subsection{Some theoretical results :  case of a polytrope}



As mentionned in Sect.\ref{mapping}, Simon (1969) built  a mapping between a spheroidal coordinate system 
 and a spherical one  and computed the second order effects for  radial modes 
of a polytrope of index $n_{polyt}=3$ and specific heat coefficient $\Gamma_1=5/3$.
He found for the dimensionless squared frequency 
(in units of $4\pi G \rho_c$, $\rho_c$ being the central density):
$ \sigma^2 = \sigma^2_0 + \beta \Omega^2$
with $\sigma^2_0 =0.057, \beta =-3.858$ for the fundamental radial mode
which includes the effect  of the first order toroidal contribution ${8\over 3} \Omega^2$ ($m=0$, $X_1= 8/3$, $X_2=0$ in Eq.\ref{solcub2}) to 
$\sigma$ and the approximate effect of distorsion. This results agreed 
 with  the earlier work by  Cowling \& Newing (1949), 
but in a somewhat desagreement  with results of  other previous works such as
Chandrasekhar and Lebovitz (1962).  
Clement (1965) using a variational principle computed the frequency for an axisymmetric $\ell=2$ mode, 
he found  $\sigma^2=  8.014-0.255   \hat \Omega^2 $.\\
Saio (1981)  studied the effect of uniform rotation  upon nonradial oscillation  frequencies 
up to second order $O(\Omega^2)$ and  computed the frequency corrections for a
 polytrope of index $n_{polyt}=3$ and $\gamma=5/3$ and for  $\ell=0,1,2,3$ modes with 
radial order $n$ up to $n=6$ for p-modes.
  His numerical results were  in agreement with those of Simon (1969), Clement (1965), Chlebowski
 (1978).  
 For the radial fundamental mode, he found $\sigma^2 = 9.252-3.79 ~ \hat \Omega^2 $  to be
 compared to Simon's result in the same units $\sigma = 9.249-3.86  \hat \Omega^2 $.  
He wrote the corrected frequency under the convenient form Eq.\ref{saioc}
The quantity $Z+ X_2+m^2 Y_2$ arises from the distorsion of the equilibrium model.
  The first order eigenfunctions were computed by solving directly the appropriate system of equations which he gave in appendix, 
   generalising  the approach derived by Hansen et al. (1978)  in the Cowling approximation.  Perturbations of the structure were 
   obtained
 from the tabulated results of Chandrashekhar and Lebovitz (1962). 
 Eq.\ref{saioc}  shows that second order effects break the symmetry of the rotational splitting which exists at
 first order.  For non radial p-modes, he found that the effect 
 of distorsion of the equilibrium model dominates the frequency correction:  $Z$ is large and negative and dominates over $X_2$ which is
 also large and positive; $Y_2$ is negative and much larger than $Y_1$   which is positive. 
 $|Z|, X_2, |Y_2|$ increase with  radial
 order of the p-modes.  

Table \ref{tab1}  gives values of the coefficients in Eq.\ref{solcub2}  
 for a polytrope with $n=3$ and $\gamma =5/3$ computed using Soufi et al.(1998)'s approach. 
  Columns  $S_1$ and $S_2$ are the cubic order coefficients appearing explicitely in Eq.\ref{solcub2}, 
 they  are found to increase steadily with the radial order as for the second order distorsion coefficients but they are smaller
 and increase more slowly with the frequency. The last column lists the asymptotic coefficient 
  $<{\cal W}_2>$ defined in Eq.\ref{sig2m}.

\begin{table}[]
\caption{ Coefficients   of Eq.\ref{solcub}  assuming a uniform rotation
 for a polytrope with polytropic index $3$ and  adiabatic index $\gamma=5/3$. The
squared frequency $\sigma^2_0$ is the dimensionless squared frequency $\omega^2/(GM/R^3)$.  
Spherical distorsion of the polytrope
has not been included.}
\label{tab1}
\begin{center}
\begin{tabular}{llllllllll}  
\hline
      &  &             &    &  $\ell=1$   &  &  &  &   & \cr
\hline
   n &      $\sigma^2_0$   &   $C_{n,\ell}$  &  $X_1$  & $Y_1$  & $X_2$  & $Y_2$ & $S_1$  & $S_2$   & $<{\cal W}_2>$ \cr
\hline
   1 &    11.400 &     0.028 &     0.776 &     0.980 &     2.898 &    -4.347 &     0.694 &    -0.243 &0.127 \cr
   2 &    21.540 &     0.034 &     0.773 &     0.919 &     5.829 &    -8.743 &     0.345 &     0.231 &0.135 \cr
   3 &    34.896 &     0.033 &     0.773 &     0.872 &     9.677 &   -14.515 &    -0.038 &     0.767 &0.139\cr
   4 &    51.467 &     0.031 &     0.776 &     0.839 &    14.427 &   -21.641 &    -0.436 &     1.334 &0.140\cr
   5 &    71.234 &     0.027 &     0.778 &     0.815 &    20.069 &   -30.103 &    -0.835 &     1.906 &0.141\cr
   6 &    94.177 &     0.024 &     0.781 &     0.798 &    26.593 &   -39.890 &    -1.233 &     2.483 &0.141\cr
   7 &   120.280 &     0.021 &     0.783 &     0.785 &    33.995 &   -50.992 &    -1.634 &     3.067 &0.141\cr
   8 &   149.529 &     0.019 &     0.785 &     0.775 &    42.269 &   -63.404 &    -2.031 &     3.647 &0.141\cr
   9 &   181.911 &     0.017 &     0.787 &     0.768 &    51.413 &   -77.119 &    -2.426 &     4.226 &0.141\cr
  10 &   217.417 &     0.015 &     0.788 &     0.761 &    61.422 &   -92.134 &    -2.818 &     4.803 &0.141\cr
  11 &   256.040 &     0.013 &     0.789 &     0.756 &    72.296 &  -108.445 &    -3.207 &     5.377 &0.141\cr
  12 &   297.770 &     0.012 &     0.790 &     0.752 &    84.033 &  -126.050 &    -3.593 &     5.948 &0.141\cr
  13 &   342.604 &     0.011 &     0.791 &     0.748 &    96.631 &  -144.947 &    -3.976 &     6.515 &0.141\cr
  14 &   390.535 &     0.010 &     0.792 &     0.744 &   110.090 &  -165.136 &    -4.355 &     7.077 &0.141\cr
  15 &   441.560 &     0.009 &     0.793 &     0.741 &   124.410 &  -186.615 &    -4.730 &     7.633 &0.141\cr
  16 &   495.676 &     0.008 &     0.793 &     0.739 &   139.590 &  -209.386 &    -5.101 &     8.184 &0.141\cr
  17 &   552.879 &     0.008 &     0.794 &     0.736 &   155.631 &  -233.447 &    -5.466 &     8.727 &0.141\cr
  18 &   613.167 &     0.007 &     0.794 &     0.734 &   172.533 &  -258.800 &    -5.825 &     9.262 &0.141\cr
  19 &   676.539 &     0.006 &     0.795 &     0.731 &   190.296 &  -285.445 &    -6.178 &     9.787 &0.141\cr
  20 &   742.992 &     0.006 &     0.795 &     0.729 &   208.922 &  -313.383 &    -6.524 &    10.302 &0.141\cr
  21 &   812.525 &     0.006 &     0.796 &     0.727 &   228.411 &  -342.616 &    -6.861 &    10.805 &0.141\cr
  22 &   885.139 &     0.005 &     0.796 &     0.725 &   248.764 &  -373.145 &    -7.190 &    11.295 &0.141\cr
  23 &   960.830 &     0.005 &     0.796 &     0.723 &   269.982 &  -404.972 &    -7.509 &    11.772 &0.140\cr
\hline
\end{tabular}
\end{center}
\end{table}

\subsection{Some theoretical  results: realistic stellar models}

\subsubsection{Frequency comparisons  between polytropic and realistic stellar models }

Tab.2 of DG92 compares results from a polytropic and a realistic stellar models 
 for the first order splitting coefficient $C_{n,\ell}$  and  the second order coefficient 
 $D_L=Y_1+Y_2$ appearing  in Eq.\ref{solcub2}.
The authors computed $D_L$  assuming rigid rotation for a 2 $M_\odot$ at two evolutionary stages 
one with a central hydrogen content  $X_c=0.699$ (ZAMS) 
and a more evolved model with $X_c=0.313$. They compared with the results for a $n_{polyt}=3$ polytrope. 
Apart for  mixed modes, the polytropic and
realistic values for $D_L$  are comparable $D_L<0, |D_L|$ increases with $n$. 
Differences were found larger  for    $C_L$ coefficients than for $D_L$ ones. The  largest differences arise 
  for modes in  avoided crossing on the $C_l$ coefficients.

\subsubsection{The solar case and solarlike pulsators}
 GT90 computed the high p-mode frequencies up to second order 
  for a solar model assuming uniform rotation  as well as several shellular rotation laws. 
  They found that centrifugal distorsion is the dominant second order effect,  of the order of a few dozen nanoHz 
  for a surface rotation frequency $\nu_s$ 
  of about 0.5 $\mu$Hz and either a
  uniform rotation or a rotation profile with a core rotation of $\sim 2 \nu_s$ and a first order splitting of $\sim
  440$ nanoHz.  DG92   investigated these effects for a realistic solar model and a differential rotation
 $\Omega(r,\theta)$. The authors studied the 2nd order effects  on the splittings  $\delta_m$ (Eq.\ref{split}) 
 for the Sun  and   found that of the nonspherical distorsion
 $Y_2$ in Eq.\ref{solcub2} dominates (hence Eq.\ref{asym2}) with values
 as $\Omega^2 /\omega_0 \sim 0.1$ nanoHz which must be mulitplied by $\sigma^2 ~\sim 100-1000$ for
 solar p-mode,  in agreement with GT90.  
Because  the nonspherically symmetric distorsion dominates for high frequency nondegenerate modes, one can write for these modes: 
$$\omega_{2,n,\ell,m}  \sim  \omega_2^D  ~(DG92)={\bar \Omega^2 \over \omega_0}  (X_2+m^2 Y_2) ~(Saio81) $$ 
It is also convenient to write the quantity $X_2+m^2 Y_2 $ as:
\begin{eqnarray} 
X_2+m^2 Y_2 & \sim & {\cal Q}_{2\ell,m} ~ {\cal D}_{asymp}  ~~{\rm with}~~ {\cal Q}_{2,\ell,m} = {\Lambda-3 m^2 \over 4\Lambda-3} \nonumber 
 \end{eqnarray}
 For high radial order, an asymptotic analysis indeed shows that  ${\cal D}_{asymp} \sim \sigma^2_0 ~<{\cal W}_2>$  
where   $<{\cal W}_2>$ is an integral over the distorted structure quantities which 
 depends on the non spherically rotational perturbation of the gravitational potential 
 (DG92, Fig.8 of Goupil \etal ~ 2001). This explains the linear increase with the frequency, $\omega_0$,  of the
second order correction  (Saio81, GT90,  DG92, Goupil \etal~ 2004) which in dimensionless form behaves as 
\begin{equation}
\sigma_{2m} = \hat \Omega^2 ~  {\cal Q}_{2,\ell,m}  ~ \sigma_0 ~<{\cal W}_2> 
\label{sig2m}
\end{equation}
The asymptotic quantity $<{\cal W}_2> $ is listed in Tab.1 for a polytrope with $n_{polyt}=3$ and $\Gamma_1=5/3$.
Burke and Thompson (2007)  computed  the second order effects 
 for a 1 $M_\odot$ evolving along the main sequence and  for a 1.5 ZAMS model 
 They  also find    that the second order    dimensionless 
  coefficients vary little with  age for a 1 $M_\odot$ and 
  vary in a homologous way for different masses along the ZAMS.

A linear increase  with radial order  is also the case for the degenerate  second order  coupling coefficient 
(Suarez et al, 2006b): for two modes $a$ and $b$ ($\omega_{0a}\sim \omega_{0b}$):
$$H_{ab} \sim \omega_{2D,ab} \sim   {\Omega^2 \over \omega_0}~   {\cal Q}_{2,\ell,m}  ~{\cal D}_{ab, asymp}$$
with ${\cal D}_{ab, asymp} \sim  \sigma_0 ~<{\cal W}_{2}>$
 as it is  also dominated by nonspherically symmetric distorsion.

\medskip 
\ni {\it Near-degeneracy and small separation }

\ni P-mode frequency small separations, defined as $\omega_{n-1,\ell,0}-\omega_{n,\ell+2,0}$ are  of the order of a few dozen $\mu$Hz 
 for solar like main sequence low  and intermediate
 mass stars. When the star is rotating fast enough (F,G,K main sequence stars  have 
  surface projected  rotational velocities between 10 and 40 km/s), the frequencies 
  are  modified  by an amount which can be
 significant, particularly
 when they are degenerate. Close frequencies as those involved in small separations  
 favor the occurence  of 
 degeneracy induced modifications. As a  consequence, the small separation can be quite affected by rotation. This was 
  stressed by Soufi et al (1998). Quantitative estimates have been obtained by  Dziembowski \& Goupil (1998)  
  for a $1 M_\odot$ and $v=10$ and $20$ km/s
  and Goupil \etal~ (2004) for a 1.4 $M_\odot$ and a $1.54 M_\odot$ rotating stellar
models   with a  surface 
 rotational velocity of 20, 30, 35  km/s. 
  Changes in the small separations are of the order of $0.1-0.2 \mu$Hz 
  (corresponding to a change of 0.1-0.2 Gyr  for the  age of the star)
   and increase with the frequency. Provided enough components of the
rotational splittings are available, 
 it is possible to remove most of these contamination effects in order to recover a 'classical' 
 small separation   (Dziembowski \& Goupil (1998) and Fig.4 in Goupil \etal~ (2004)).
As rotation induced distorsion of the equilibrium   significantly 
affects the small separation   of high order p-modes, it    also affects the 
 shape of the ridges in an echelle diagram.
The effect is larger  in the upper part of the diagram  ie at high frequency
 (Goupil \& Dupret (2007), Fig.6)  but as one can decontaminate the small separation, it is also possible to recover  an echelle
  diagram  free of rotationally induced pollution effects (Goupil \etal ~ 2006; Lochard \etal~ 2008 in prep.)

\subsubsection{Delta Scuti stars}

DG92 computed the second order frequencies for a 2 $M_\odot$ with a uniform rotation velocity 
 of 100 km/s and discussed
 the departure from equal rotational splittings induced by  
 distorsion $\omega_{2D}$ for low frequency modes in the range of the fundamental radial mode.  
 The distorsion, hence the departure from equal splitting,
  is larger for the trapped mode (or mixed mode) in this low frequency regime 
  where all the other modes have a predominantly g-mode
  character. 
Goupil et al. (2000) investigated the effects of moderate  rotation (initial radial velocity of 100 km/s)
 on  rotational splittings of $\delta$ Scuti stars   
using the 3rd order perturbative  approach of  Soufi et al. (1998).  A $1.8 M_\odot$ stellar model 
half way on the main sequence was studied.  Effects of  successive perturbation order contributions 
 on the oscillation  frequencies   are shown 
for modes $\ell=0,\ell=2$ modes   (Fig.3). Changes in the frequency pattern appearence in a power spectrum   
are  mainly due to centrifugal distorsion  and are shown
 in the particular case of the $\delta$ Scuti star FG
Vir for 3 values of the initial  rotation velocity $10, 46, 92km/s$.  
The first order equidistant pattern  of the rotational  splittings is totally lost at 92 km/s. 
The 3rd order effects effects in the generalised  rotational splittings (Eq.\ref{Sm1}) 
computed for a uniform rotation  are found relatively small.
 Although the rotation is uniform, $S_m$ 
  show  strong variations  of the order of $\mu$Hz with the frequency  due to the presence of mixed modes
   (Fig.6).  The true (uniform)
rotation rate can however  be recovered    when combining well chosen components  of the multiplets. 
Departure from equal splittings for nondegenerate modes   as measured by Eq.\ref{asym}
is again found to be  dominated 
 by the nonspherically symmetric centrifugal distorsion contribution. The splitting asymmetry then becomes
$$\Delta \omega_m
\sim  m^2   ~\omega_0   ~{3 \over 4\Lambda-3} ~{\Omega^2 \over (GM/R^3)}  ~<{\cal W}_2>$$

Pamyatnykh (2003) give quantitative estimates of the  effect of  mode near- degenerate coupling on nonradial p-mode frequencies 
 (Fig.5)  and on period ratios of radial modes (Fig.6)    for a 1.8 $M_\odot$
  main sequence model with a surface rotational velocity of 92 km/s. The induced modifications can be quite significant.
  Suarez et al (2006a,b) studied second order effects for a 1.5 $M_\odot$ mass star, 
   representative of a  delta Scuti star and for a prescribed
   shellular rotation law with assuming a surface velocity of 100 km/s. Comparing frequencies for models assuming 
     a uniform rotation on one hand and a shellular rotation law on the other hand, they found 
     differences of about 1-3 $\mu$Hz for high frequency p-modes and larger for  lower frequencies.
 The authors ivestigated consequences of degeneracy  due to rotation  which are 
 also   illustrated  in Goupil et al. (2006).  Burke and Thompson (2007)  computed the second order frequencies 
for  a 1.98 $M_\odot$  mid main sequence star representative of a $\delta$ Scuti star. Their Fig.3 shows  that the 
 coefficients   vary little with radial order $n$ 
 except for mixed modes.

\section{Fast rotators: nonperturbative approaches}\label{fast}

 

Several types of fast rotating pulsating stars are known to exist.  
One good exemple is the nearby A-type star,  Altair. This star is rotating  fast with a 
$v \sim 227$ km/s  and is flattened with a ratio $R_{eq}/R_{pole}\sim 1.23-1.28$; $\mu =0.08-0.2$.
Interferometric observations has  revealed a gravity darkening effect in accordance with Von Zeipel theory  for this
 star   with the equatorial layers cooler ($\sim$ 6800 K instead of $\sim $ 8700K)  and $60-70\%$ darker than the poles
 (Domiciano de Souza et al 2005, Monnier et al 2007).  Altair is a delta Scuti stars and
its  power spectrum  shows 7   frequencies from WIRE observations (Buzasi \etal ~ 2005). Modelling of these
pulsations has been attempted by Suarez et al (2005).  For such a star, it is likely that a perturbative approach is
no longer valid neither for the equilibrium model nor for the computation of the oscillation properties.

\subsection{Formalisms}
Oscillation properties are computed for a stellar model which is considered in static equilibrium. 
The rotating model  is no longer  perturbative but is assumed to keep the axisymmetry and  
must then be described as a 2D configuration. Hence, the equations 
are separable in $\phi$ but no longer in  $(r,\theta)$  variables
in a spherical coordinate system.   
In order to study the structure of a rotating star, several  works have developped various techniques 
 with the goal of building 2D rotating equilibrium  configurations as mentionned in Sect.\ref{rapideq}.

\subsubsection{Eigenvalue problem}

Once the equilibrium configuration is built, 
the goal is to calculate the adiabatic oscillations   of a given model  defined by the quantities
  $\rho(r,\theta), p((r,\theta)$ etc.... When the star is rotating fast, as the latitudinal and radial 
  dependences (in $\theta$, $r$) are no longer separable, one here again one deals with a 2D computation. 
Linearization of the equations,  $\phi$ variable separation and the  hypothesis of 
a steady state configuration  allow to write  the displacement eigenfunction as: 
$ \xi \propto e^{i(\omega t - m \phi)}$
where $m$ is an  integer. Solving the associated  eigenvalue problem has led to series of different studies starting
with Clement (1981), see Reese (2006) for a detailed bibliography.  One of the techniques is to expand the solution as a  
series of spherical harmonics for the angular dependence:
\begin{eqnarray}
\vec \xi_m =  e^{i\omega t}~ r~\sum^{\infty}_{\ell\geq |m|} & &
~\left(\right.S_\ell(r) Y_{\ell}^m(\theta,\phi)  \vec e_r + H_\ell(r) \vec \nabla_h Y_{\ell}^m(\theta,\phi)
\nonumber \\
& &  +T_\ell(r) ~ \vec e_r \times \vec \nabla_h Y_{\ell}^m(\theta,\phi) \left.\right)
\label{xiht} 
 \end{eqnarray}
and $f '= \sum^{\infty}_{\ell\geq |m|} f'_\ell(r) Y_{\ell}^m(\theta,\phi)  e^{i\omega t}$. 
One obtains a  infinite set of
coupled  differential equations  for the depth dependence of the eigenfunctions. The  
properties of axisymmetry and  
symmetry with respect to the equator  $(\theta \rightarrow \pi-\theta)$  cause a decoupling of the problem
 into 2 independent eigenvalue systems (Unno et al. 1989; Lee \& Saio 1986). For any integer $j \geq 0$:

\ni Even modes  (i.e sym/equator)  are  $\ell= |m|+2j+2 , \ell'=\ell+1$ 
i.e. $ (m=0, \ell=0,2,4..) ;  (m\pm 1, \ell=1,3,5..); ..$

\ni Odd modes (i.e. antisym/equator) are $ \ell=|m|+2j-1, \ell'=\ell-1 $
i.e. $(m=0, \ell=1,3,5,..) ; (m\pm 1, \ell=2,4,5..); ..$

Lee and Saio (1986) used this  technique to study the g-modes of a  10 $M_\odot$ stellar model, the frequencies 
were computed by keeping only the first two harmonics in the series Eq.\ref{xiht}. Note that 
the equilibrium model was obtained by means of perturbation  as developped by 
Kippenhan, Meyer-Hofmeister, Thomas (1970).

More recently, Espinosa (2004)  considered  the effect of the centrifugal 
force only, neglecting the Coriolis force, assumed
 the Cowling approximation  and neglected the Br\"unt- V\"aiss\"al\"a  frequency 
 in the adiabatic oscillation equation $(N^2 << \omega^2)$. 
These  above assumptions are valid for high frequency p-modes.
He also assumed a uniform rotation. The numerical resolution was 
 based on  a finite difference method.
Espinosa(2004) studied first a 
model with a uniform  density   then turned  to 
a realistic model  built by Claret   with $\epsilon^2$ between 0  and 0.3.
An alternative approach has been  developed by 
  Reese \etal~ (2006)  who studied  a polytrope in uniform  rotation, that-is the structure is built according
to 
\begin{eqnarray}
p_0 &=& K \rho_0^\gamma ~~;~~\Delta \psi_0 =  4 \pi G \rho_0  \nonumber \\
0 &=& -\nabla p_0 - \rho_0 \nabla \left(\psi_0-{1\over 2} \Omega^2 s^2 \right)
\end{eqnarray}
The computation of adiabatic oscillation frequencies of  p-modes 
is based  on a 2D approach which uses  spectral methods in both dimensions
 $r,\theta$ with expansions in spherical  harmonics  for the angular part
 and  in Chebitchev polynomials for the radial dependence. For any function, $f$,
$$f(r,\theta,\phi) =  \sum_{\ell=0}^{\infty}   \sum_{m=|\ell|}^{\infty} f_m^\ell(r)
Y_\ell^m(\theta,\phi)
$$
Each of the radial functions  $f_m^\ell$  is written as:
$ f_m^\ell(r) =   \Sigma_{j=0}^{\infty}  a_j^{\ell,m} ~T_j(2r-1)$.
For the velocity field:
$$\vec v(r,\theta,\phi) = \Sigma_{\ell=0}^{\infty} \left(  \Sigma_{m=|\ell|}^{\infty} u_m^\ell(r) \vec R_\ell^m 
+ v_m^\ell(r) \vec S_\ell^m + w_m^\ell(r) \vec T_\ell^m \right)$$
where  $\vec R_\ell^m, \vec S_\ell^m, \vec T_\ell^m $   constitute a basis which  becomes the usual  spherical  basis 
 $Y_\ell^m \vec e_r, \vec \nabla Y_\ell^m,  \vec e_r \times \vec \nabla Y_\ell^m$ (as in Eq.\ref{xiht}) when $\Omega \rightarrow 0$.
According to the authors, 120 points with a spectral  method 
 correspond to 5000 points  with  a finite difference method  
 at least when the structure is smooth as is that of a
 polytrope. Furthermore, the authors chose a coordinate system which is better adapted to the oblate geometry 
$(\zeta, \theta, \phi)$  as proposed earlier in another context by  Bonazzola et al. (1998). 
The relation between the star radius in spherical (r) and oblate ($\zeta$)  coordinate systems is given by: 
$$ r(\zeta,\phi) = \zeta \left(1-\epsilon +{5\zeta^2-3\zeta^4\over 2} (R_s(\theta)-1+\epsilon) \right)$$
where $R_s(\theta)$ is the surface radius, $\epsilon$ is the oblatness parameter.
 The surface boundary condition is taken as $\psi' =0$ 
at large distance of the surface of the star. The numerical computation solves a full matrix  system.

\subsection{Some results and conclusions}

Espinosa  (2004) and  Ligni\`eres et al. (2006) studied the effect of the  centrifugal force.
The first study chose an oblateness parameter in the range $\epsilon  = 0 -0.3$
 while the second one  investigated the range  $\epsilon  = 0 -0.15; \Omega/\Omega_k = 0$ up to 
 0.59 that-is a rotational velocity up to 150 km/s for a  $\delta $ Scuti 
where $\Omega_k = ({G M/ R_{eq}^3})^{1/2}$ with $M$ the stellar mass and  $R_{eq}$ the equatorial
 radius.  The effects of the  centrifugal  force  increase 
  with increasing $\omega$  (Fig.2 of Ligni\`eres et al 2006).
   Frequencies  of even and odd modes behave differently: frequencies of even
 modes tend to increase  whereas those of  odd modes decrease.  
 The effect of the centrifugal force on p-modes is to contract the eigenfrequency spectrum.
 These  effects are  more important  for high  $n$ and low  $\ell$ (see Fig.3 of Ligni\`ere et al 2006). 
 Reese \etal~ (2006)  computed frequencies of a polytrope with a polytropic index $n=3$, using 70-80 $\ell$ degree 
spherical harmonics and 60-80 points for the radial grid in the spectral decompositions.
 Both the centrifugal and  Coriolis  forces are included. 
The Coriolis effects are found to  decrease with  increasing  $\omega$. 
For high frequency modes, the  effects du to the centrifugal  force  dominate. 
Visualisation of oscillations of a distorted  rotating star can be seen 
on  the  web site of D. Reese  or Fig.4 of Espinosa (2004).

\medskip
\ni{\it  Mode classification}
In general, the oscillation modes of  a fast rotating star look like modes for a spherically symmetric star 
and a mode   classification  remains possible:  the  dominant  $\ell$ degree  most often  corresponds to  
 the  degree of the mode  for a zero rotation.
 This is  shown   by following the mode starting from a zero rotation and progressively increasing the rotation rate.
 However following the mode with increasing rotation rate is  made more 
complicated by the presence of mixed
modes.
For the highest rotational velocities,  the dominant degree $\ell$ 
 is no longer the degree  of the mode with zero rotation. The dominant $\ell$ can be as large as 
  $\ell_0+6$  where  $\ell_0$ is  the degree of the  mode  for the nonrotating    model.
In that case, one can wonder whether the mode would  still be visible in disk integrated light from ground?
Would that  contribute to explain the complicated patterns seen in  power spectra of fast
 rotating $\delta$ Scuti stars where no frequency are detected in large domains in between frequency ranges where  
many frequencies are detected for the same star?

\medskip
\ni {\it Mode pairing }
Espinosa  (2004) found that the numerical frequencies computed assuming  
a uniform rotation and  with the aforementioned approximation  obey the relation
$\omega_{n,\ell,m} =  \bar \omega_{n,\ell,m} + m  \Omega   -  f_{n,\ell}(m^2,\Omega) - (-)^{\ell+m} g_{n,\ell}(m^2,\Omega)\nonumber
$
where  $f$ and $g$ are positive definite and monotonically increasing functions with $ m^2$. 
The combined effect of the  3rd and  last terms produces compressions of the frequencies  within the multiplets  which 
 generate pairing of modes with different types of symmetry.
 Indeed for increasing rotation, frequencies  tend to assemble by pair of modes with 
different values of   $m$  (Fig.1 of Espinosa 2004, Ligni\`eres  \etal~ 2006, Reese \etal~ 2006). 
This effect starts to appear   for larger $\ell$ when the rotation is increased.
The origin of such a pairing is attributed to the nonspherical distorsion
 of the structure.
This effect might have  already been observed in 
the Fourier spectra of $\delta$  Scuti star light curves.

\medskip
\ni {\it New regularities in frequency spectra}
Fig.7.3 in  PhD thesis of Reese  (2006) shows that when the radial order $n$ increases,  the large separation 
 still tends  toward a constant value but the  small separation no longer decreases to 0 . 
Fig.7.2 also shows that the small separation between $\ell=0,n$ and $\ell=2,n-1$  modes for instance 
increases for  increasing  rotation as
the frequency of the $\ell=2$ mode decreases whereas that of the radial one remains almost constant until 
some rotation is reached beyond
which the small separation is now between the $\ell=2,n-1$ and the radial mode below ie $\ell=0,n-1$.  This behavior is
reminiscent of the avoided crossing behavior of the same modes when the (non-rotating) stellar model evolves i.e. with decreasing
effective temperature or central hydrogen content.
Reese et al. (2006a,b) modelled the behavior of the numerical frequencies as :
$\omega_{n,\ell,m}= \Delta_n n +\Delta_\ell \ell-m \Omega+\Delta_m |m|+\alpha_\pm$
where the $\Delta$ coefficients  are constant depending on $\Omega$ but not on $n,\ell,m$, $\alpha_\pm$ depends on $\Omega$
and on the $(\ell+m)$ parity of the mode.

\medskip
\ni {\it Concentration of the amplitude toward the equator:}
The centrifugal distorsion is found  to cause  a high concentration of 
the mode amplitude  toward the equator   (Fig.8 of Ligni\`ere et al (2006)).

\medskip\ni {\it From a  rotating polytrope to a more realistic stellar  model}
Most of  the above effects obtained for a polytrope appear to be  also 
verified  when one  turns to the oscillation frequencies of a more realistic stellar model
  (Espinosa, 2004, PhD thesis). However, the effect of 
  mode couplings   seems to be stronger in the realistic case. This perhaps might be expected as the structure of a
   polytrope is smoother than that of a more realistic stellar  model.

\subsection{Validity of the results from perturbation techniques}\label{valid}

From the previous chapter and what has been described in this chapter, one concludes that  
some  results of  perturbative and non pertubative methods are  qualitatively similar. On the other hand, some new types of behavior 
seem to appear when the perturbative methods are no longer valid. For concrete studies, one then
 needs to quantify the domain of validity of perturbation methods.
In order to do so, one must compare the 
results coming from the perturbation methods under the form Eq.\ref{solcub}
with the numerical frequencies computed with a nonperturbative approach.
Reese et al (2006) determined the
dependence of the frequency $\omega$  in function of the rotation rate $\Omega$ 
by means of  a polynomial fit of the form
$$\omega_{fit}(\Omega) = A_0+A_1 ~ \Omega+ A_2 ~ \Omega^2+A_3 ~ \Omega^3+O(\Omega^4)$$ 
The $A_j$ coefficients are determined by fitting the numerical frequencies computed with the non perturbative approach. 
The authors then compared the numerical frequencies 
to their conterpart at a given level of approximation. The comparison is carried out with  
the differences $\delta_j=\omega-\omega_{app,j}$   where 
$\omega_{app,j} = \sum_{1}^j A_j \Omega^j$ .
 When only the centrifugal force  in included, 
the error between second order perturbative and nonperturbative frequencies 
 is   $\sim 11 \mu$Hz   for $\Omega/\Omega_K = 0.24$ for a typical delta Scuti star with a rotational velocity of 100
km/s (Fig.12, Ligni\`eres et al 2006). The relative error reaches $3\%$ at $\Omega/\Omega_K = 0.3$.
 Reese \etal (2006) considered  more appropriate  quantities for the comparison:
\begin{eqnarray}
{\omega_{n,\ell,m}+\omega_{n,\ell,-m}\over 2} 
&= \omega^0_{n,\ell,m}+\omega^2_{n,\ell,m} \Omega^2 + \omega^4_{n,\ell,m} \Omega^4 \\
{\omega_{n,\ell,m}-\omega_{n,\ell,-m}\over 2} 
&= \omega^1_{n,\ell,m} \Omega + \omega^3_{n,\ell,m} \Omega^3 + \omega^5_{n,\ell,m} \Omega^5 
\end{eqnarray}
Frequency differences  between  perturbative and nonperturbative computations are the same for two  modes 
with opposite  $m$  and same $\ell,n$. 
Reese \etal (2006) found that 
the error which is introduced between perturbative frequencies and non perturbative ones is 
 larger than the  accuracy of frequency measurement
 (0.4 $\mu$Hz)   for 150 days  of observation  for a solar like oscillating star.   
(0.08 $\mu$Hz)  (Fig.4 of Reese et al 2006). 
In the frequency range considered and with
 COROT's accuracy,  it is found that    complete calculations  are required 
 beyond $v sin i=50 km s^{-1}$ 
  for a $R = 2.3 R_\odot, M=1.9 M_\odot$ polytropic star. 
  Furthermore, it is shown that the main differences
 between complete and perturbative calculations come essentially from the centrifugal distortion.
A comparison between the results of nonperturbative calculations and 
those from perturbative computations as described in the  previous section 
is also necessary and is currently being carried out.

\section{Observed rotational splittings and forward inferences}

\subsection{Solar-like pulsators}
Several   low to intermediate mass main sequence stars are  known 
to oscillate with solar like oscillations. However, they are
 rotating too slow and in all but one instance and apart for the Sun, their splittings have not  been detected yet.

\subsubsection{$\alpha$ Cen A :} 50 days of  space photometric observations  of  this star with Wire led to the detection of   18 frequencies in the range
1700-2650 $\mu$Hz which were  measured with an accuracy of 1-2 $\mu$Hz  (Fletcher et al. 2006).
Detected modes have $\ell=0$ to 2 and radial order $n=14-18$. These observations allowed to determine an 
average value for the rotational splitting of $0.54\pm 0.22 \mu$Hz.

 \subsubsection {Procyon:} Solar like oscillations for Procyon have been detected in velocity measurements (Claudi et al. 2005, Leccia \etal~ 2007, Heeker \etal ~ 2007). Some
controversy exists as to whether these oscillations can  be detected in photometry even from space. 
As a prototype for a 1.5  $M_\odot$ main sequence star,  Bonanno et
al. (2007) have modelled the   latitudinally differential rotation in the outer convective layers 
for  this star.  Their models give a flatness of order 0.1-0.45  which is comparable with the solar case
0.25. The authors computed   a latitudinal averaged  $\Omega(r)$. 
The splittings   differ by a few nanoHz 
 depending on the latitudinal shear 
 obtained in the models generating the differential rotation.  Such a  small value could be detected with
 Corot after averaging over several multiplets.

\subsubsection{HD 49933:} The space experiment  Corot, which was successfully launched in december 2006, 
 seems to be able to keep its promises and makes  possible to  detect mean splitting values as well as individual splittings at least
in some favorable cases.  As the first solar-like target for this mission, 
HD49933 was observed from ground with the spectrograph HARPS (Mosser et al, 2005). 
These observations confirmed the expectation that this
star is a solar like oscillator. The star shows a high degree of stellar activity and the rotation 
appears to be  quite rapid  with a period of the order of 4-8 days. 
The star   was the object of a hare and hound exercise in order
to see whether splittings can indeed be detected and which accuracy can be expected. 
This exercice indicated that
 for an assumed  rotation of 1.4-2.9$\mu$ Hz,
 many splittings could be detected. Only a few correct ones  however were measured 
 within $0.5 \mu$Hz from the correct value and  with error bars less that
$0.5 \mu$Hz  (Appourchaux et al. 2006).

\subsection{$\delta $ Scuti stars\index{$\delta $ Scuti}}

These stars are nonradial multiperiodic  variables with excited modes in the vicinity of the fundamental.
The main problem  for these stars is  mode identification. As it is not yet possible 
to assign $n,l,m$ values with full confidence  to
the observed frequencies, it is difficult to  get information on the star. As far as rotational splitting is
concerned, a few detections have been claimed:

 \subsubsection {GX Peg\index{GX Peg}}  belongs to a
spectroscopic binary system.
Knowing the surface rotation from the binarity, the identification of a splitting 
led to the conclusion that the core is rotating faster than the outer  layers   (Goupil \etal~ 1993).  The  rotationally split
multiplet was found asymmetric which could be attributed to distorsion of the star by the centrifugal
force.

\subsubsection{FG Vir\index{FG Vir}} is one of the   best studied delta Scuti star.
 Its variability has been extensively observed and  analysed and the results   have been the
subject of many observations and theoretical studies over the past 20 years
 (Breger et al. 2005; Pamyatnykh, Daszynska-Daskiewicz \etal~ 2005; Zima \etal~ 2006). Among about 70 detected frequencies, 
 a  dozen  $\ell$ and $m$ values can be assigned. The star appears to be moderately rotating with a
equatorial velocity of $ 66\pm 16 km/s$. Most of the identified modes are axisymmetric but
Zima \etal ~ (2006) found that  one $\ell=1$ triplet is observed
with a splitting value of 6.13 $\mu$Hz. 
   Failure to find a stellar model which frequencies 
match the observed frequencies of FG Vir subsists even 
 when  taking into account second order rotational effects.   
 
\subsubsection{44 Tauri\index{44 Tauri}}  has been observed for a long time and is known to pulsate with at least 13 frequencies. Mode
identification indicates a few possible splitting components among them which  indicate that the star is a
slow rotator with 26 km/s as an equatorial velocity (Poretti \etal ~ 1992, Antoci \etal ~ 2006 and references
therein).

\subsubsection{HD104237:}  Detecting the  rotational splittings of  oscillating PMS stars 
  ought to be relatively easy as they are fast rotators and their 
  power spectra are expected to be much simpler
  than those of their postZAMS counterparts i.e. delta Scuti stars (Suran et al. 2001).
    Determining their rotation profiles and internal structures 
    can give us clues  about  temporal  evolution of rotation profiles and transport of internal 
    angular momentum.   In this spirit,  the binary PMS system  HD104237 
    is being studied as one of its two 
    components  has been identified as a $\delta$ Scuti star 
    (Alecian \etal ~ 2005, 2007;   Goupil \etal ~ 2005 ; Dupret \etal ~ 2006).

\subsection{$\beta$ Cephei stars\index{$\beta$ Cephei}:}
More massive than  $\delta$ Scuti stars,  B-type  main sequence stars 
oscillate with periods  between 1.5 h and  8 h corresponding to low radial order p and g -modes. 
Their  projected  rotational  velocities can reach  up to 300 km/s. 
Rotational splittings  for $\ell=1$ and $\ell=2$ modes   seem to have been 
detected for a few stars   (see for instance Fig.1 of Pigulski, 2007).

\subsubsection{HD129929, a $\sim 8-9 M_\odot$ $\beta$ Cephei star: }

This star is known to oscillate with at least  6 identified frequencies.
From one   $\ell=1$  multiplet,  the rotational  splitting  yields a rotational velocity of  3.61 km/s 
 when assuming a uniform rotation. On the other hand, 
 2 successive components of a $\ell=2$  multiplet indicate a rotational velocity of 4.20 km/s.
This leads to the likely conclusion that the rotation   varies with depth  with a 
ratio of the angular velocity at the center to that  the surface    
 $\Omega_c/\Omega_s = 3.6$ (Dupret et al. 2004).

\subsubsection{$\nu$ Eridani\index{$\nu$ Eridani}} is another $\beta$ Cepheid which   
shows a rich frequency spectrum   but with only a few of them identified (see Fig. 1 of
Ausseloos \etal 2004). Pamyatnykh \etal (2004)  used the identified 
 $\ell=1$, $g1$  triplet  and two  components 
 of the $\ell=1$, $p1$   to determine 
a non uniform rotation for this star with   $\Omega_c/\Omega_s \sim 3 $, close to what is found for HD  
129929. Using an assumed depth dependent profile, Suarez et al (2007) computed the splittings 
and their associated 
departures from equidistant in order to reproduce the observed 3
$\ell=1$ multiplets   obtained by Jerzykiewicz et al (2005).  The observed asymmetries
are in favor of a non uniform rotation profile although they are too small to allow a definitive 
conclusion.

\subsubsection{$\theta$ Ophuichi\index{$\theta$ Ophuichi}}
 is a $\sim 8-9 M_\odot$  $\beta$ Cepheid  
with  7 identified frequencies (Handler \etal, 2005): a $\ell=1,p1$ triplet  
and 4 components of a  $\ell=2,g1$ quintuplet 
(Fig.6 of Handler \etal,   2005). The splittings indicate a rotation velocity of 
$ 29\pm 7 $ km/s, large enough to
cause detected departure from equal splittings
(Briquet et al. 2007).

\section{Inversion for rotation }\label{inversion}

We consider  the simplified problem of determining 
the depth dependent rotation
law of a star from the measurements of its rotational splittings. 
For more general  2D inversions which yields $\Omega(\theta, r)$ 
as well as for more details on inversion methods and results 
on the solar internal rotation,   we refer the reader to
  general such as 
reviews  Gough (1985); Gough \& Thompson, (1991), Thompson \etal (2003).\\
According to Eq.\ref{split4}, a  1D rotation profile 
is related to  the splittings  by the  integral relation
\begin{equation}
\delta \omega_j + \delta (\delta \omega_j) = \int_0^R  ~ K_j(r) ~\Omega (r) dr 
\label{invrel}
\end{equation}
where  for shortness $j$  represents the set of values $(n,\ell)$  and $\delta (\delta \omega_j)$ is the incertainty
pf the measured splitting values.  
In general, only a finite number, $N$, of 
 discrete splittings values  associated with $N$ modes,  are available.
The  measurements  are  polluted with some observational errors
$\delta S_j$ which are assumed  uncorrelated with a variance $\sigma_j^2$. 
In order to derive the rotation profile from the observed splittings, one must
invert the relation  Eq.\ref {invrel}. As the data are related to  the rotation profile by a linear functional,
 one is confronted to a {\it linear inverse problem}\index{inverse problem}.

\subsection{Splittings of high frequency p-modes  and  the Abel equation}

For modes with high radial orders  or degrees, the eigenfunctions vary more rapidely than the
equilibrium quantities; hence the scale of spatial variations of the eigenfunctions 
being smaller than that of the basic state, a local analysis is valid. One then assumes 
that the solution can be cast under the form of a plane wave: 
$\vec {v'} \propto e^{i(\omega t - \vec k \cdot \vec r)}$
with  $\vec  k$ the wavenumber  and  $|\vec k|^2 = k_r^2 + k_h^2$, with $k_r,k_h$ the radial and
horizontal components. One can also neglect the perturbation of the gravitational field (Cowling
approximation) as well as the derivatives of the
equilibrium quantities. 
For  high frequency p-modes which nature is mainly acoustic, 
the eigenfrequency is much larger than the Br\"unt-V\"aiss\"al\"a frequency 
($\omega^2 >> N^2 $)  which
therefore can be neglected in the problem considered here.
With these approximations, one recovers the expected dispersion relation for an acoustic 
wave $\omega^2= k^2  ~ c_s^2 $
 where $k^2=k^2_r+k^2_h$ and $k_h= {S^2_{\ell}/c_s^2} = {\Lambda /r^2}$ with $\Lambda = \ell(\ell+1)$
and $c_s(r)$ is the adiabatic sound speed profile,
$ S_\ell^2 =   {c^2_s   \Lambda / r^2 }  $ is the squared Lamb frequency. One then has:
$$ k^2_r(r) \approx {1\over c_s^2  } (\omega^2 - S_\ell^2)=   {\omega^2\over c_s^2  } -{\Lambda \over r^2}
 $$

Let recall first that for a 1 dimensional  acoustic wave  with a wavenumber $k$  with a node at one end
 ($r=0$) and a free  boundary at the other end ($r=R$),  the resonant 
 condition  for the existence of  a normal mode  in a homogeneous medium is
$k R =n \pi - \pi/2$ for some  integer $n$. In the stellar 3D  (inhomogeneous) medium, this condition becomes
\begin{equation}
 \int^{R}_{r_t} ~k_r(r) ~dr = (n-{1\over 2}) \pi 
 \label{reso}
 \end{equation}
where $r_t$ is the radius of the mode inner turning point.


Taking for  the mode frequency, the frequency  corrected for the first order effect of rotation, 
$\omega \sim   \omega_0+(\delta \omega+m\Omega)$ where $\omega_0$ 
is the zeroth order frequency (ie the frequency of the mode  in abscence of rotation), one obtains 
$$k_r(r)= {\cal Q}
+ \left({ \omega^{(0)}\over c_s^2  {\cal Q}} \right) (\delta \omega+m\Omega) 
$$
with $${\cal Q}= \left({\omega^{(0)2}\over c_s^2  } -{\Lambda^2 \over r^2} \right)^{1/2} = {\omega^{(0)}\over c_s} \left(1 
-{a^2\over W^2} \right)^{1/2}$$ 
and $a=c_s(r)/r$ et $W=\omega_0/\Lambda$. Recalling that at zeroth order, one has 
$$\int_{r_t}^R {\cal Q}(r) dr = (n-{1\over 2}) \pi  ~~~~~~~~~,$$
the resonant  condition Eq.\ref{reso} becomes: 
\begin{equation}
\delta \omega
  ={1\over S} ~ \int^{R}_{r_t} ~m ~\Omega(r) 
  \left(W^2 -a^2 \right)^{-1/2}   ~{dr\over c_s} 
   \label{Abel} \end{equation}
with 
$$S= \int^{R}_{r_t} \left(W^2 -a^2\right)^{-1/2}   ~{dr\over c_s}  $$
Note that for very high radial order modes, $a/W<<1$ and 
\begin{equation}\delta \omega = m~ 
 \int_{r_t}^R  \Omega(r) {dr\over c_s}    ~ \left(\int_{r_t}^R   {dr\over c_s}\right)^{-1} 
\end{equation}
which shows that the perturbation of the frequency by the rotation 
  can be expressed  as an average of the rotation weighted by the acoustic propagation time 
in the radial direction inside the cavity.\\ 
Eq.\ref{Abel}  is of  Abel  type. Indeed the Abel (integral) equation\index{Abel equation} is defined as: 
\begin{equation}
\int_0^x {f(y)\over (x-y)^\alpha } dy = g(x) ~~~~~0<\alpha<1  ~~~g(0)=0
\label{Abel2}
\end{equation}
and has the formal solution (Gough, 1985):
\begin{equation}
f(y) = {1\over \pi} \sin(\pi \alpha)
 {d\over dx} \int_0^y {g(x)\over (y-x)^{1-\alpha}} dx
 \label{Abel3} 
 \end{equation}
Eq.\ref{Abel} then admits  an analytical inverse solution. The angular velocity is  given by:
\begin{equation}\Omega(r) = - {2a\over \pi} 
{d\over d\ln r}\int_{a_s}^a (a^2-W^2)^{-1/2}
{\cal H}(W) dW  
\end{equation}
where $a_s=a(R)$   and $ {\cal H}(w)$ represents the data set derived from the observed
splittings. 

The Abel  equation, Eq.\ref{Abel2},  is known to be an ill-posed inverse problem\index{ill-posed inverse
problem}
 in the sense of Hadamard (1923).
The existence of an analytical solution to the inverse problem does not suppress the 
ill-posed nature of the numerical inversion, particularly when the observational data are noisy.



\subsubsection{Ill-posed nature of an inverse problem}

  A {\it well-posed problem in the sense of Hadamard} (1923) is defined as follows:
let consider  the equation ${\cal H}(f) = g $
where  ${\cal H}$ is an operator $\in {\cal F} \longrightarrow {\cal O}$
where  ${\cal F}$ et ${\cal O}$ are closed vectorial spaces and more specifically here the integral equation:
\begin{equation} 
\int_0^R K(x,y) f(y)  dy = g(x)     
\label{invpb}
\end{equation}
The three  Hadamard conditions are  :

- existence of a  solution:  $\forall g \in {\cal O}$,
 $\exists f \in {\cal F}$ such that $ {\cal H}(f)= g$

- unicity  of the  solution: $\forall g \in {\cal O},
 \exists$ at most one $f \in {\cal F}$ such that $ {\cal H}(f)= g $

- stability of the  solution with respect to the data:
$\forall f_n$, any sequence $\in {\cal F}$  such that
$$\lim_{n\rightarrow +\infty} {\cal H}(f_n) = {\cal H}(f)  ~~~~~~~{\rm then}~~~~~ 
\lim_{n\rightarrow +\infty} f_n = f $$
 This last condition can be understood  in a more practical sense as
 $$ |{\delta g \over g}| <<1 \rightarrow   |{\delta f \over f}| <<1 $$  

Existence and unicity are the  classical  conditions
for solving an equation and  
the stability condition, on the other hand, is an additional condition
which  requires  that, for a well-posed problem, any small perturbation of the data
can only lead to a small   error on the   solution. This  condition
is  barely  verified in practice for inverse problems. In particular, 
inversions  of integral equations often are  ill-posed  problems: 
noise, even small, in the data can 
 cause  the derived  solution  to be very far from the true one.

\medskip
\ni {\it Nonuniqueness and sensitivity to noise}

\ni One must recall that the inverse problem  Eq.\ref{invpb} 
  admits  the set of solutions $f(y)+h(y)$  
provided that \\
$\int_0^R K(x,y) h(y)  dy = 0$ 

Furthermore, the Riemann-Lebesgue theorem  states that:
$$ ~\int_a^b K(x,y) \cos(my) dy =  0 ~;~\int_a^b K(x,y) \sin(my) dy = 0$$
when $m \rightarrow \infty$,  valid   for any integrable $K(x,y)$. Consequences:

\ni {\it i)}  on one hand, high frequency components of the solution $f(y)$ 
 are not accessible  as they are smoothed by the  kernel $K(x,y)$ 
 to an arbitrary small  amplitude level in the data. 
As an illustration, let 
 some  perturbation of the source  function $f$ in Eq.\ref{Abel2}, be  of the form:
$\delta f(y) = A \sin(2 \sqrt{\lambda y})$ (Craig \& Brown, 1986).
It causes  a  perturbation of the data of the  form:
\begin{equation}
\delta g (x) = A \sqrt{\pi} \Gamma(1-\alpha) 
\lambda^{\beta} x^{(3/2-\alpha)/2} J_{3/2-\alpha}(2 \sqrt{\lambda y}) 
\label{delg}
\end{equation}
where $\beta = (\alpha-1/2)/2$ and $J_{3/2-\alpha}$ is a  Bessel function of 1st kind  of order $3/2-\alpha$. 
For high frequencies (i.e. $\lambda \longrightarrow \infty$), one can use
$\vert J_\nu (z) \vert \sim \vert{2/\pi z}\vert^{1/2}$
and Eq.\ref{delg} becomes
$\vert \delta g(x)\vert \sim \Gamma(1-\alpha) \lambda^{(\alpha-1)/2}  
x^{(1-\alpha)/2}$
so that  for any amplitude $A$ and $\forall \alpha <1$,
 $\delta g(x) \longrightarrow 0 $  when  $\lambda \longrightarrow \infty$. 
 Any discontinuity can be represented by an infinite Fourier series 
but all the high frequency components are smoothed out and  contribute to the 
data at a small amplitude level; they  
can not be 
distinguished from the high frequency noise in the data. 
The smoothing properties of the operators and/or  kernels in the process  $f \rightarrow g$ remove
 any real  discontinuity of the solution  $f$  and which is therefore difficult, 
 if not impossible, to  bring out from the data.


\medskip 
\ni {\it ii)} on the other hand, high frequency  small  perturbations in the data  (noise) 
 appear as large amplitude oscillating components  in the calculated 
   source. These large amplitude contributions can  dominate the inverse  solution 
  even if they are totally spurious as due to small perturbations in the data due to noise. 
  An exemple given by Craig \& Brown (1986) illustrates this issue (cf Fig.1).
  One considers  the particular kernel $ K(x,y)=1$ in Eq.\ref{invpb} (differentiation).
Let 2  functions $g_1(y)$ et $g_2(y)$  differing  only   by 
 the addition of a sinusoidal  component with a given frequency $\Delta g$. We choose for instance 
 $g_1(x)  =1 - e^{ - \alpha x }  ~~;~~g_2(x)  =  g_1(x)+ \beta \sin \omega x $.
 The  difference appears in the 
  source functions $f_1$ and $f_2 $ as  
   $\Delta f(y)  = f_2(x)-f_1(x)  = \omega \beta \cos \omega y $.
Fig.1 shows the data and the source functions  for $\alpha = 0.8 ; \beta=0.04 ; \omega=20$. 
Although the difference $g_1-g_2$ is very small, the corresponding difference in the source functions 
is an oscillatory behavior with a large amplitude.
  

\begin{figure}[h]
\centerline{\epsfig{file=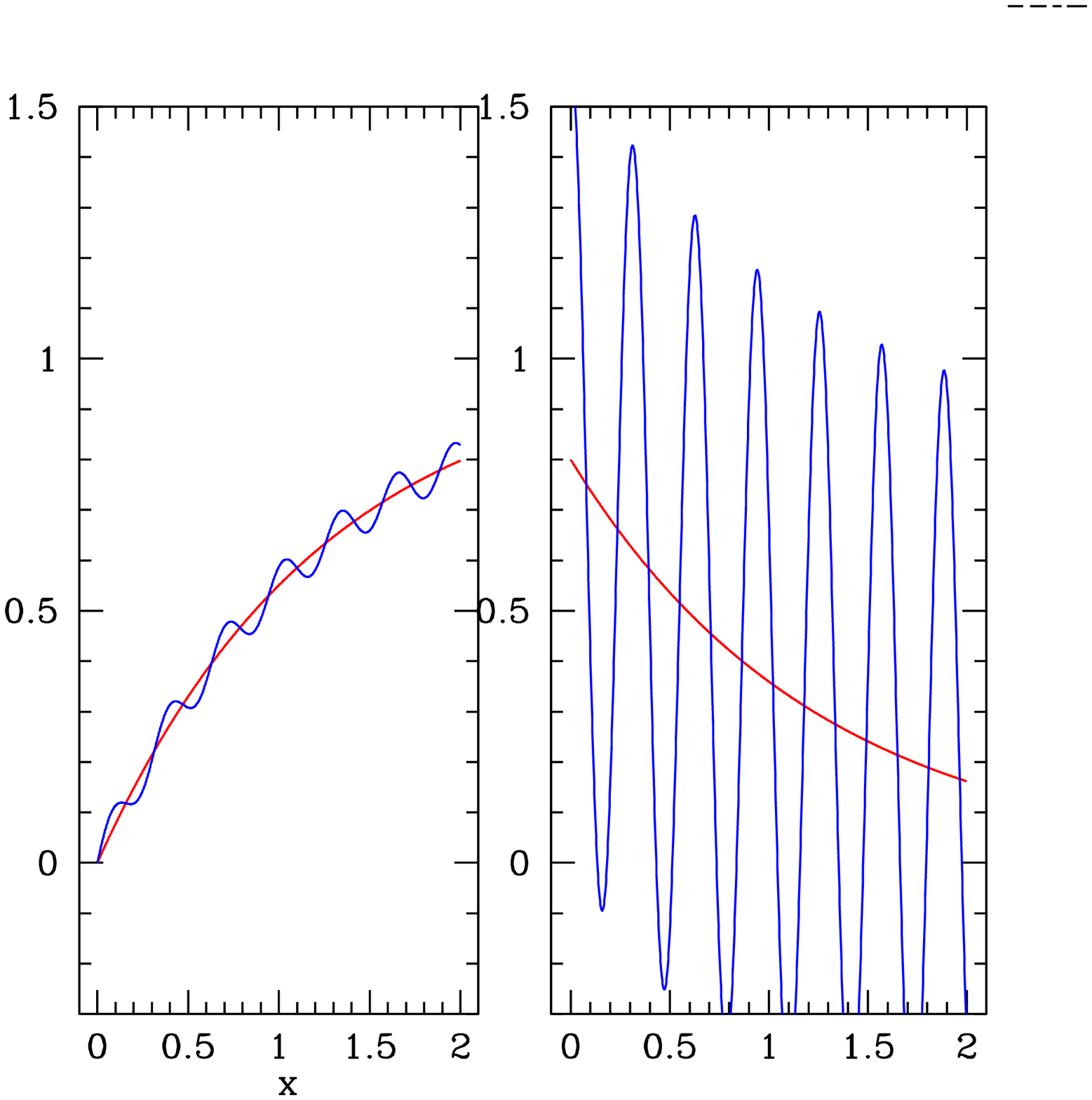,width=6cm}}
\caption{The left pannel displays  the   noise free data function $g1$ and  the  slightly noisy data 
function $g2$ where the perturbed part  is represented by a small (5 \%) amplitude oscillation.
 The right pannel shows the true  inverse solution, $f1$, and   the  inverse solution of the 
 noisy function $ g2$,  $f2$,   which is dominated by a spurious  large  (100 \%) amplitude oscillation {\sl (adapted from Craig \& Brown (1986))}.}
\end{figure}

Hence  since observations cannot in general  be noise free, a simple ' naive'  
inversion is not able to give the correct inverse solution.

\subsubsection{Discretisation and a  measure of the   ill-posed nature of the  problem}

Although the first inversions for the solar rotational splittings using Eq. \ref{Abel}   were successfull enough, 
the asymptotic integral relation remains an approximation.
In practice, one actually  considers the general relation Eq.\ref{split} and  inversions are carried out 
first by  discretizing the integral. Let write the  resulting discretized  relation  as :  
 $\Delta = A X $ 
where $\Delta $ represents  the  data  vector and  $X$  is the rotation vector  to be determined. 
 The stability can be measured with the conditionning of the matrix  $A$. 
 If  $A$ is  regular, its  conditionnement  is   defined as
$cond(A) = ||A|| ~||A^{-1}||$ where  $||A||$ is a matrix norm for close $A$.
 Note that $cond(A) \geq 1$ is always verified.  A well conditionned matrix   
is  such  $cond(A)$  remains close to 1.
Using the norm $||A||= max_{j}(\mu_j)$ where $\mu_j$  are the 
 singular values of $A$ (i.e. the square root of the eigenfrequencies  of the squared symmetric matrix $AA^t$), one shows that
 $cond(A)={\mu_{max}/ \mu_{min}}$.
%
A small perturbation $\delta \Delta $  of the data  generates  
a  perturbation $\delta X$ of the solution
then 
 $\delta \Delta = A(\delta X)$ and
it is easy to deduce from the matrix norm properties  that 
$${||\delta X||\over ||X||} \leq cond(A) {||\delta \Delta ||\over ||\Delta||}$$
Accordingly, if  $A$ is  well conditionned ($cond(A) \sim 1$),
the stability condition is satisfied
whereas for $cond A >> 1$ which is most often the case, 
    $||\delta X||$ can become large compared with  $||X||$
    and the problem is ill-posed.






Once  the ill-posed\index{ill-posed} nature of the  inverse problem\index{inverse problem} is recognized, 
one is then force to carry out a regularization\index{regularization} of the calculated solution. 
As seen above, helio- and astero- seismic  inversions are ill-posed problems which therefore require 
regularisation inverse techniques.

\subsection{Inversion with regularisation}

Several classes of inversion techniques with regularisation\index{regularisation} exist.
  They are  based  on the addition of   
a priori information on the solution, often on the shape of the solution.
One wants a solution which remains   insensitive  
to errors in the data. Most methods involve  one or two  parameters 
which control the  sensitivity  of the solution  to errors in the data. 
We briefly present two of the most currently used ones in the stellar context.

\subsubsection{Inverse methods with  a priori douceur or Tikhonov's methods\index{Tikhonov method } :}
The   inversion {\it strategy}   consists in   determining 
a solution where  high frequency variations 
-which cannot be identified as  real or due to noise  magnification-
have been removed.  One then seeks for a {\bf global} solution which reproduces at best the data  without  reproducing 
the noise   (seen as  the high frequency components). The procedure is to   find 
 a least square solution    submitted to a regularisation  constrain :  
 $min(||A X- \Delta||^2 + \gamma   ~\sigma^2 ~B(X) ) $
where $\sigma$ is the variance of the noise and $\gamma$ a trade-off parameter 
and $B$ a discretized regularising
function. This is equivalent to solve the normal form equation:
$(A^t A+\gamma B^t B ) X = A^t \Delta)$ where $A^t$ is the transposed matrix. 
The regularising function $B(X) $ is often taken as 
the second derivative of the solution since one wishes to eliminate the high frequency part of the solution. 
The trade-off parameter  $\gamma$ must be adjusted to realize the best compromise between
 a  noise insensitive but distorted solution (large $\gamma$) and a
solution which is  not regularised enough (small $\gamma$).  
 
\subsubsection{Optimally Localized  averages: OLA\index{OLA method} and its variants}

One looks for  a {\bf local}  solution that-is- an averaged value of the solution over  
a small interval about a given radius.  The SOLA  method is a variant of the OLA technique (Backus \& Gilbert 1970) 
developed by  Pijpers and Thompson (1992, 1994, 1996).
It  aims at building 
localised kernels  ${\cal K}$ about a predefined  value, $r=r_0$  as    linear  combinations of 
the rotational kernels  $K_j(r)$: ${\cal K}(r,r_0)  =   \Sigma_j  c_j(r_0) ~    K_j(r)$.
The coefficients $c_j$  must be  determined by a minimisation process:
       $$min_{cj} (|\int({\cal K}-T)^2 ~dr|^2  + \gamma E^2)$$  
where $T(r,r_0)$ is a predefined   target, $E$  is the noise covariant  matrix; 
$\gamma$  a  trade-off parameter  in order to obtain a satisfying compromise 
 between magnification of the error and  the width of the target ie the interval over which the solution is averaged. 
 The SOLA method with a predefined target
 presents  the advantage over the original  OLA  method that  the inverse matrix is computed once and for all. 
Once the $c_j(r_0)$ are determined, one computes the average rotation value as 
$<\Omega(r)>_{r0}=\sum_j  c_j(r_0) ~ \delta \omega_j$. Indeed one has 
\begin{eqnarray}
\sum_j  c_j(r_0) \delta \omega_j&=& \sum_j c_j(r_0)  (\int  K_j(r) \Omega(r) dr) \nonumber \\ 
&=& \int (\sum_j c_j(r_0) K_j(r)) \Omega(r) dr \nonumber \\ 
& =& \int {\cal K}(r,r_0) \Omega(r)  dr 
\sim \int  T(r,r_0) \Omega(r)  dr \nonumber  
 \end{eqnarray}
If the target $T$ is a Dirac function i.e. $T =\delta(r-r_0)$  , then 
$\sum_j  c_j \delta \omega_j=\Omega(r_0)$.  However one seldomely has enough information in the data  
to succeed in building Dirac function as  localised kernels.
 Usually the target function  takes the form of  gaussian with  a given width. 
 Again a compromise must be obtained between the 
 magnification of the error and the width of the target ie the interval over 
 which the solution is averaged.

\subsection{Application to stellar seismology}

\subsubsection{The solar rotation profile}\label{suninvers}
With the high quality  helioseismic data available for the past 2 decades or so, 
highly accurate frequency splittings  were obtained and allowed to derive  the 2D rotation profile
in the 3/4 outer part of the Sun.   See Fig.5 of Schou  \etal~ (1998)
 for instance for the result of an inversion based on SoHo data.
Several 2D  inversion methods have been  applied  with various  adjustements and adaptations 
to the available helioseismic data sets over the years
(Schou \etal~ (1994ab, 1998), Antia \etal~(1998)).
This has provided  the
internal rotation profile for the Sun with quite a number of surprises at  the time: 
latitudinal dependence of rotation in the convective outer region 
and uniform rotation in the radiative region. Only the rotation of the inner part is still  not yet
accessible. 

The inversion results  have generated  many studies in order  to explain these seismic results. 
Several studies have  led to identify  the shear of the  Reynolds
stress as responsible for differential rotation in latitude in the convective outer layers
of the Sun; others have studied the importance and the role of the tachocline in the solar 
magnetogydrodynamical processes,  see for instance 
R\"udiger \etal~ 2003,   2005; Miesch \etal~ 2007, Zahn \etal ~2007. 

The discovery of the uniform  rotation of 
the solar  radiative zone drove the developement of 
 models   able to transport angular momentum from the inner to the outer part. 
It was found 
 type I rotationally induced  mixing models are not efficient enough and 
 predict  an  incorrect increase of the rotation rate with decreasing radius 
(Talon \etal~ 1997; Matias \& Zahn, 1998)). 
Internal waves,  on the other hand, are very efficient in transporting the angular momentum from center
upward to the surface and inforce a uniform rotation in the radiative zone.   
(Charbonnel \& Talon , 2005; Talon,  2006 and references therein). 
 Fig.1 of  Charbonnel Talon (2005) 
 shows the  evolution of the rotation profile for a $1.2 M_\odot$ 
 stellar model  from 0.2 Gyrs up to 4.8 Gyrs   when rotational induced mixing of 
 type I  is  included and the successful  approach of uniform rotation when 
 mixing of type II is included.



One current  issue  is the role of the magnetic field, 
another is to   determine to what extent these results 
and explanations are  valid for other stars  than the Sun, particularly young stars which rotate faster
than the Sun. In that context, information from 3D simulations will be of
great help (Brun \etal~ 2006, Ballot \etal~ 2007).

\subsubsection{Solar like  stars}

The observed high frequency solar p- modes do not give access to the core rotation. 
 For stars slightly more massive  and slightly more evolved 
 than the Sun such as $\eta$ Boo for instance,
 a small number of mixed modes does indeed  exist in the high frequency  domain 
of   modes excited by  the turbulent  convection of the outer layers. 
A theoretical study was carried out with  SOLA inversions  for the rotation profile 
   on simulated data  for a 
1.5 $M_\odot$ stellar  model with a 
surface velocity of $\sim 30 $ km/s and a ratio between the rotation rate at the center to that at the
surface $\Omega_c/\Omega_s
\sim 3$  (Lochard \etal ~ 2005).  The incertainties were estimated in function of the expected Corot performances.
Fig.8 of Lochard \etal ~ (2005)  shows that
  the variation for the  rotation profile   with radius in the vicinity of the convective core 
  - varying from $\Omega/\Omega_s = 1.5$ at $r=0.2$,  
  (with a incertainty on the recovered value  $\pm 0.2$)  to 
  $\Omega/\Omega_s = 2.4$ at $r\leq 0.1$  (with a incertainty on the recovered value  $\pm 0.6$)-
  is   accessible. 
 \begin{figure}[t]
\centerline {\epsfig{file=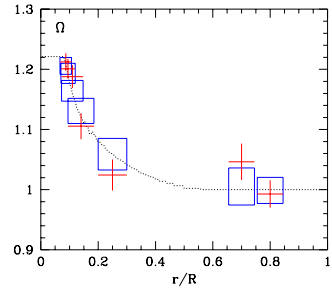,width=5cm}}
\vskip -1.truecm
\caption{Inversion for the rotation profile for a model of 
$\epsilon$ Cep, a $\delta$ Scuti star.  Data are the splittings computed for 
an input stellar model (1.8 $M_\odot$   main sequence star with   
$T_{eff} = 7588 K$ 
  and a surface rotational velocity 
$v=120 $km/s).  The input prescribed  rotation profile
 $\Omega$ normalized to its
surface value is represented by dotted lines. 
Squares show the results of a SOLA  inversion using the input model; 
 crosses represent a SOLA  inversion using a
different model than the input one with a 1.90 $M_\odot$ mass and 
$T_{eff}= 7906 K$  
with no rotation but with the same large separation $ 50 \mu$Hz. {\sl from Goupil et al. 2000, unpublished})  }
\end{figure}

\subsubsection{Inversion for rotation for $\delta$ Scuti like oscillations  }
 
  Other stars oscillate with  modes with lower frequencies in the vicinity of the  fundamental, driven by an opacity mechanism. These stars are  slightly more massive than the Sun and 
 develope a convective  core on the main sequence  which  receeds  with time. This generates  the existence of mixed modes in the excited
 frequency range.  Only one or two such modes are enough to give  access 
 to $\Omega_c$ the core rotation rate (Goupil et al. 1996).  
 One illustrative exemple is  $\epsilon$ Cep.  Assuming that  one  has obtained 
 a seismic model  for  this star that-is  a model
  as close as possible of  reality  with the same  mean large separation,  a SOLA  inversion was  performed with a data set of $\ell=1,6$ 
linearly unstable modes. The variation of the rotation profile with depth is well recovered.

This theoretical study assumed a noise level according to Corot performances. 
The inversion  was  carried out with linear splittings Eq.\ref{split4} although for a real case,  
 distorsion effects on the splittings ought to be  included for such a moderately rapid rotator ($v ~\sin i = 91 $km/s, Royer \etal~ 2002).
Several frequencies have   recently been detected for this star but its complex frequency pattern has not yet been fully elucidated (Bruntt \etal~ 2007).


\section*{Acknowledgment} I gratefully thank Rhita Maria Ouazzani for a careful reading which helped to improve 
grandly  the manuscript.

\clearpage
\addcontentsline{toc}{section}{Index}
\index{vibrational mode}
\index{pulsation} 
\index{rotational splitting}
\flushbottom
\printindex


\begin{thebibliography}{8.}
\addcontentsline{toc}{section}{References}


\ni {Alecian}, E. and {Catala}, C. and {van't Veer-Menneret}, C. and 
	{Goupil}, M.-J. and {Balona}, L.,
 2005, AA 442, 993  \\


 \ni {{Alecian}, E. and {Goupil}, M.-J. and {Lebreton}, Y. and {Dupret}, M.-A. and 
	{Catala}, C.},
 2007, AA 465, 241 \\


\ni {{Antia}, H.~M. and {Basu}, S. and {Chitre}, S.~M.}, 1998, 
    MNRAS 298, 543 \\

\ni Antoci, V., Breger, M., Bishof, K., Garrido, R.,  2006, PASP 349, 181,
eds. C. Sterken, C. Aerts \\


 \ni  {{Appourchaux}, T. and {Berthomieu}, G. and {Michel}, E. \etal~}, 2006, 
{ESA Special Publication}, vol. 1306, p. 377 \\


  {{Ausseloos}, M. and {Scuflaire}, R. and {Thoul}, A. and {Aerts}, C.},
2004, MNRAS, 355, 352 \\


Backus, G. \& Gilbert, F., 1970, Phil. Trans. R. Soc. London, A 266, 123\\

Baglin, A., Auvergne, M., Barge P. Deleuil, M., Catal, C. Michel, E., Weiss, W., and the CoRoT
 team,  2006,  {ESA Special Publication}, vol. 1306, 33 \\
 
 {{Ballot}, J. and {Brun}, A.~S. and {Turck-Chi{\`e}ze}, S.},
  2007, ApJ 669, 1190 \\

{{Bonazzola}, S. and {Gourgoulhon}, E. and {Marck}, J.-A.},
  prd,  {arXiv:astro-ph/9803086}, 58,  \\

{{Bonanno}, A. and {K{\"u}ker}, M. and {Patern{\`o}}, L.}, 2007, AA 462, 1031 \\


 {{Breger}, M. and {Lenz}, P. and {Antoci}, V. \etal~},
 2005, A\&A, {\bf 435}, 955 \\



{{Briquet}, M. and {Morel}, T. and {Thoul}, A. and {Scuflaire}, R. and 
	{Miglio}, A. and {Montalb{\'a}n}, J. and {Dupret}, M.-A. and 
	{Aerts}, C.},
2007, MNRAS  381, 1482 \\ 

{{Brun}, A.~S. and {Miesch}, M.~S. and {Toomre}, J.}, 2006,
   {ArXiv Astrophysics e-prints},
    {astro-ph/0610073}  \\
    
    
Bruntt, H.,  Suarez, J.C. , Bedding, T.R.,  \etal, 2007, AA 461, 619 \\

   {{Burke}, K.~D. and {Thompson}, M.~J.},  2006, 
in {Proceedings of SOHO 18/GONG 2006/HELAS I, Beyond the spherical Sun},
{ESA Special Publication}, 624, \\
   
{{Buzasi}, D.~L. and {Bruntt}, H. and {Bedding}, T.~R. 	\etal ~}, 2005, AA 619, 1072 \\

        
    {{Chaboyer}, B. and {Zahn}, J.-P.}, 1992, AA 253, 173 \\



{{Chandrasekhar}, S. and {Lebovitz}, N.~R.}, 1962a, ApJ 136, 1069 \\
    
{{Chandrasekhar}, S. and {Lebovitz}, N.~R.}, 1962b, ApJ 136, 1082 \\
    
{{Chandrasekhar}, S. and {Lebovitz}, N.~R.},1962c,  ApJ 136, 1105 \\


{{Chandrasekhar}, S.}, 1964, ApJ 140, 599 \\
 

Chandrasekhar, S. and Lebovitz, N.R.,  1968, ApJ 152, 152  \\



 {{Charbonnel}, C. and {Talon}, S.}, 2005,  {Science}, 309, 2189 \\


 {{Charbonneau}, P.}, 1992, AA 259, 134 \\



{{Chlebowski}, T.},1978,
 {Acta Astronomica}, 28, 441 \\

 
 {{Christensen-Dalsgaard}, J. and {Mullan}, D.~J.}, 1994, MNRAS 270, 921 \\
 

{{Christensen-Dalsgaard}, J. and {Thompson}, M.~J.}, 1999, AA 350, 852 \\

 Christensen-Dalsgaard, J., 2003 (CD03), Lecture Note on Stellar Oscillation,
 5th edition,  http://www.phys.au.dk/~jcd/oscilnotes/) \\



 {{Claudi}, R.~U. and {Bonanno}, A. and {Leccia}, S. \etal ~},
2005, AA 429, L17 \\


{{Clement}, M.~J.}, 1965a, ApJ 141, 210 \\

{{Clement}, M.~J.}, 1965b, ApJ 142, 243 \\


{{Clement}, M.~J.}, 1981, ApJ 249, 746 \\ 


Clement, M.J., 1984,  ApJ 276, 724 \\


{{Clement}, M.~J.}, 1989, ApJ 339, 1022\\




  {{Cowling}, T.~G. and {Newing}, R.~A.}, 1949, ApJ 109,  149 \\

{{Cox}, J.~P.}, 1980
   , "{Theory of stellar pulsation}", 
{Research supported by the National Science Foundation Princeton, 
NJ, Princeton University Press \\



Craig, I.J.D, Brown, J.C.,  1986, Inverse problems in Astronomy, Ed Adam Hilger Ltd \\


Daszynska-Daszkiewicz  Dziembowski, W.A., Pamyatnykh, A.A., Goupil, M.J.,  2002, AA 392, 151\\


Daszynska-Daszkiewicz, J.,  Dziembowski, W.A., Pamyatnykh, 2003, AA 407, 999  \\
 
 
 {{Deupree}, R.~G.}, 1990, ApJ 357, 175 \\


 




{{Deupree}, R.~G.}, 2001, ApJ 552, 268 \\

 {{Domiciano de Souza}, A. and {Kervella}, P. and {Jankov}, S. \etal~},
2005, AA 442,  567 \\


 {{Dupret}, M.-A. and {Thoul}, A. and {Scuflaire}, R. \etal~ },
 2004, AA 415, 251  \\

 {{Dupret}, M.-A. and {B{\"o}hm}, T. and {Goupil}, M.-J. \etal ~}, 2006
 Communications in Asteroseismology 147, 72 \\

{{Dyson}, J. and {Schutz}, B.~F.}, 1979, 
  in  {Royal Society of London Proceedings Series A},  368, 389 \\
   

 {{Dziembowski}, W.~A. and {Goupil}, M.-J.}, 1998, 
in  {The First MONS Workshop: Science with a Small Space Telescope, 
held in Aarhus, Denmark, June 29 - 30, 1998,
 Eds.: H. Kjeldsen, T.R. Bedding, Aarhus Universitet, p. 69.,
eds. {{Kjeldsen}, H. and {Bedding}, T.~R.},
p. 69 \\


{{Dziembowski}, W.~A. and {Goode}, P.~R.},
1992 (DG92), ApJ  394, 670 \\

 {{Eddington}, A.~S.}, 1929, MNRAS 90, 54\\

Eggenberger, P., Carrier, F., 2006, AA 449, 293 \\


 {Espinosa}, F. , 2004, PhD thesis

 {{Espinosa}, F. and {P{\'e}rez Hern{\'a}ndez}, F. and {Roca Cort{\'e}s}, T.}, 2004
in {SOHO 14 Helio- and Asteroseismology: Towards a Golden Future}, {ESA Special Publication},
   vol. 559, ed. {{Danesy}, D.}, p. 424 \\


{{Espinosa Lara}, F. and {Rieutord}, M.}, AA 470, 1013 \\

 {{Fletcher}, S.~T. and {Chaplin}, W.~J. and {Elsworth}, Y. and 
	{Schou}, J. and {Buzasi}, D.},
2006, MNRAS 371, 935-944 \\
 
{{Gough}, D.}, 1985, 
Sol. Phys 100, 65 \\

 {{Gough}, D.~O. and {Thompson}, M.~J.},
 1990 (GT90), MNRAS 242, 25 \\


Gough, D.O., Thompson, M.J., 1991, 
`The inversion probleme, 
in Solar interior and atmosphere
Eds A.N. Cox; W.C. Livingston, M.S. Matthews, 
University of Arizona Press
p. 519 \\


Gough, D.O., 1993,  `Linear adiabatic stellar pulsation', in Astrophysical fluid fluid dynamics,
Les Houches Session XLVII 1987, Eds J.P. Zahn, J. Zinn-Justin,
Elsevier Science Publishers B.V., p. 399 \\

{{Goupil}, M.~J. and {Michel}, E. and {Lebreton}, Y. and {Baglin}, A.
       }, 1993, AA 268, 546 \\

{{Goupil}, M.~J. and {Michel}, E. and {Cassisi}, S. \etal~}, 1995, 
 {IAU Colloq. 155: Astrophysical Applications of Stellar Pulsation},
 {Astronomical Society of the Pacific Conference Series},
 83, eds  {{Feast}, M.~W.},  453 \\


{{Goupil}, M.-J. and {Dziembowski}, W.~A. and {Goode}, P.~R. and 
	{Michel}, E.}, 1996, AA 305, 487 \\



 {{Goupil}, M.-J. and {Dziembowski}, W.~A. and {Pamyatnykh}, A.~A. and 
	{Talon}, S.}, 2000
in  {Delta Scuti and Related Stars},
 {Astronomical Society of the Pacific Conference Series},
  210,
   eds.  {{Breger}, M. and {Montgomery}, M.},
    p. 267 \\


{{Goupil}, M.~J. and {Talon}, S.}, 2002,
in {IAU Colloq. 185: Radial and Nonradial Pulsations as Probes of Stellar Physics},
 {Astronomical Society of the Pacific Conference Series},
 259,
eds  {{Aerts}, C. and {Bedding}, T.~R. and {Christensen-Dalsgaard}, J.
	}, 306 \\

{{Goupil}, M.~J. and {Samadi}, R. and {Lochard}, J. \etal~}, 2004, 
in  {Stellar Structure and Habitable Planet Finding},
 {ESA Special Publication},
  538,
 eds  {{Favata}, F. and {Aigrain}, S. and {Wilson}, A.},
133 \\
 {{Goupil}, M.-J. and {Dupret}, M.~A. and {Samadi}, R. and \etal~}, 
	2005, JApA vol. 26, 249   \\

 {{Goupil}, M.~J. and {Moya}, A. and {Suarez}, J.~C. and \etal ~},
2006, {ESA Special Publication}, vol. 1306, p. 51 \\

 {{Goupil}, M.~J. and {Dupret}, M.~A.},2007, 
 {EAS Publications Series},  26, 93 \\

Hadamard J., 1923, 'Lectures on Cauchy's  
Problem in Linear Partial Differential Equations',
 (New Haven, CT: Yale University Press) \\

 {{Handler}, G. and {Shobbrook}, R.~R. and {Mokgwetsi}, T.},
 2005,  MNRAS  362, 612 \\


{{Hansen}, C.~J. and {Cox}, J.~P. and {van Horn}, H.~M.}, ApJ 217, 151 \\
    
 {{Hansen}, C.~J. and {Cox}, J.~P. and {Carroll}, B.~W.}, 1978, 
Apj,  226, 210 \\
   
{{Hekker}, S. and {Arentoft}, T. and {Kjeldsen}, H. \etal~},
{ArXiv e-prints},  {0710.3772} \\

{{Jackson}, S. and {MacGregor}, K.~B. and {Skumanich}, A.}, 2005, ApJS 156, 245 \\


{{Jackson}, S. and {MacGregor}, K.~B. and {Skumanich}, A.},  2004, ApJ 606, 1196 \\

 {{Jerzykiewicz}, M. and {Handler}, G. and {Shobbrook}, R.~R. \etal~ },
2005, MNRAS  360, 619 \\

 {{Karami}, K. and {Christensen-Dalsgaard}, J. and {Pijpers}, F.~P. \etal~ },
 2005,  {ArXiv Astrophysics e-prints}, {astro-ph/0502194}, \\



 {{Kippenhahn}, R. and {Meyer-Hofmeister}, E. and {Thomas}, H.~},  1970, AA 5, 155 \\


{{Kippenhahn}, R.~W.~A.}, 1994, 
Stellar Structure and Evolution,
 Springer-Verlag, Berlin Heidelberg New York.~Also Astronomy and Astrophysics Library} \\
  


 {{Kjeldsen}, H. and {Arentoft}, T. and {Bedding}, T. \etal~ },
in {Structure and Dynamics of the Interior of the Sun and Sun-like Stars},
 1998, {ESA Special Publication},  418, Ed. {{Korzennik}, S.}, p.{385}\\
   


 
 {{Lebovitz}, N.~R.}, 1970a, ApJ 160, 701 \\
    
{{Lebovitz}, N.~R.}, 1970b, Ap\&SS 9, 398 \\


{Leccia}, S. and {Kjeldsen}, H. and {Bonanno}, A. \etal~ }, 2007, AA 464, 1059 \\




{Ledoux}, P., 1945, ApJ 102, 143 \\

{Ledoux}, P., 1951, ApJ 114, 373\\

{{Ledoux}, P. and {Walraven}, T.}, 1958,  {Handbuch der Physik}, 51, 353 \\

 
 {{Lee}, U. and {Saio}, H.}, 1986, MNRAS 221, 365 \\


 {{Ligni{\`e}res}, F. and {Rieutord}, M. and {Reese}, D.},
 2006, AA 455, 607 \\


{{Lochard}, J. and {Samadi}, R. and {Goupil}, M.~J.},
 2005, AA 438, 939 \\

Lovekin, C.C. \& R.G. Deupree, 2006, Mem S.A., It, vol.77,  137 \\

{{Lynden-Bell}, D. and {Ostriker}, J.~P.},
 1967, MNRAS   136,  293 \\

 {{Maeder}, A. and {Zahn}, J.-P.}, 1998, AA 334, 1000  \\

{{Maeder}, A. and {Meynet}, G.}, 2000, ARAA 38, 143
 

{Maeder}, A. and {Meynet}, G. and {Hirschi}, R. and {Ekstr{\"o}m}, S., 2006, 
Chemical Abundances and Mixing in Stars in the Milky Way and its Satellites, 
ESO ASTROPHYSICS SYMPOSIA.~ISBN 978-3-540-34135-2. Springer-Verlag,  p. 308 \\

 {{Mathis}, S. and {Zahn}, J.-P.}, 2004, AA 425, {229} \\


 {{Mathis}, S. and {Palacios}, A. and {Zahn}, J.-P.}, 2004, AA 425, 243 \\
  

{{Mathis}, S. and {Eggenberger}, P. and {Decressin}, T. \etal ~}, 2007, 
in {EAS Publications Series},  Vol. 26, 65 \\

{{Mathis}, S. and {Zahn}, J.~-.}, 2007, 
   {ArXiv e-prints}, {0706.2446}, \\
    
   {{Mathis}, S. and {Palacios}, A. and {Zahn}, J.-P.}, 2007, AA 462, 1063 \\

  {Mathis}, S. and {Decressin}, T. and {Palacios}, A. \etal~, 2006, 
in {Proceedings of SOHO 18/GONG 2006/HELAS I, Beyond the spherical Sun},
 {ESA Special Publication}, 624 \\
 

{{Maeder}, A. and {Meynet}, G.}, 2003, AA 411, 543 \\


{{Maeder}, A.}, 2003, AA 399, 263 \\
      

{{Meynet}, G. and {Maeder}, A.}, 2000, AA 361, 101 \\
      
      
{{Mestel}, L.}, 2003, 
in 'Stellar astrophysical fluid dynamics', ~eds M. J.~Thompson, J. Christensen-Dalsgaard, 
Cambridge University Press, ISBN 0-521-81809-5, 2003, p.~75  \\


{{Michel}, E. and {Chevreton}, M. and {Goupil}, M.~J. \etal~ }, 1995, 
 in 4th SOHO Workshop Helioseismology, vol. 2, 533 \\






 {{Miesch}, M.~S. and {Brun}, A.~S. and {DeRosa}, M.~L. and {Toomre}, J}, 2007, 
 {American Astronomical Society Meeting Abstracts}, 210, 17 \\


{{Monnier}, J.~D. and {Zhao}, M. and {Pedretti}, E. \etal~ },
 2007,  Science 317, 342 \\

{Mosser}, B. and {Bouchy}, F. and {Catala}, C. \etal~ ,
  2005, AA 431, L13 \\

Palacios, A.,  Talon, S., Charbonnel, C., Forestini, M., 2003 AA 399, 603\\


{{Pamyatnykh}, A.~A. and {Handler}, G. and {Dziembowski}, W.~A.},
 2004, MNRAS   350, 1022 \\

{Pamyatnykh}, A.~A., 2003,  PASP 284, 97 \\


Daszynska-Daszkiewicz, J., Dziembowski, W.A., Pamyatnykh, A.A., \etal~  
2005, 438, 653 \\
   
   {Pigulski}, A. and {Kopacki}, G. and {Kolaczkowski}, Z.,
  2001, {Acta Astronomica}, 51, 159 \\
   
    
{{Pijpers}, F.~P. and {Thompson}, M.~J.},
 1992, MNRAS 262, L33

 
{{Pijpers}, F.~P. and {Thompson}, M.~J.}, 1994, AA 281, 231 \\
   
{{Pijpers}, F.~P. and {Thompson}, M.~J.},
 1996, MNRAS    279,    498

 
 {Pijpers}, F.~P., 1997, AA 326, 1235 \\

 {{Pijpers}, F.~P.},  1998, MNRAS 297, L76

 Pigulski, A.,  2007, Coast 150,  159  \\ 
 
Poretti, E., Mantegazza, L., Riboni, E., 1992, AA 256, 113  \\

  {{Reese}, D.}, 2006 , PhD thesis,
  {AA(Universit{\'e} Toulouse III - Paul Sabatier)}\\

 {{Reese}, D. and {Ligni{\`e}res}, F. and {Rieutord}, M.},   2006, AA 455, 621\\

{{Reese}, D.}, 2007, 
in  {EAS Publications Series}, 26, 111 \\
  
  {Rieutord}, M., 2006a, AA 451, 1025 \\
  
  {Rieutord}, M., 2006b,   {ArXiv Astrophysics e-prints},  {astro-ph/0608431} \\


{Rieutord}, M., 2007,   {ArXiv Astrophysics e-prints},    {astro-ph/0702384} \\


{Roxburgh}, I.~W., 2004, AA 428, 171 \\


{{Roxburgh}, I.~W.}, 2006, AA 454, 883 \\

 {{Royer}, F. and {Grenier}, S. and {Baylac}, M.-O. and {G{\'o}mez}, A.~E. and 
	{Zorec}, J.},  2002, AA 393, 897 \\

 {{R{\"u}diger}, G. and {Kitchatinov}, L.~L. and {Arlt}, R.}, 2005, 
 AA 444, L53 \\

 {{R{\"u}diger}, G. and {K{\"u}ker}, M. and {Chan}, K.~L.},
2003, AA   399 743 \\


Saio H.,  1981, ApJ 244, 299 \\

 {{Saio}, H.}, 2002
in {IAU Colloq. 185: Radial and Nonradial Pulsationsn as Probes of Stellar Physics},
{Astronomical Society of the Pacific Conference Series},
 259, eds. {Aerts}, C. and {Bedding}, T.~R. and {Christensen-Dalsgaard}, J.,    p. 177 \\



Schou, J., Christensen-Dalsgaard, J., Thompson, M.J.,  1994a ApJ 433, 389  \\

 {{Schou}, J. and {Brown}, T.~M.}, 1994b, Apj  434, 378 \\

{{Schou}, J. and {Antia}, H.~M. and {Basu}, S. \etal ~},
 1998, ApJ  505, 390 \\


{{Schutz}, B.~F.}, 1980a,  MNRAS 190, 7 \\
    
{{Schutz}, B.~F.}, 1980b, MNRAS 190, 21 \\


{{Simon}, R.}, 1969, AA 2,  390 \\

{{Soufi}, F. and {Goupil}, M.~J. and {Dziembowski}, W.~A.}, 1998, AA
 334, 911 \\



 {{Su{\'a}rez}, J.~C. and {Bruntt}, H. and {Buzasi}, D.}, 2005, AA 438, 633 \\

   
{{Su{\'a}rez}, J.~C. and {Garrido}, R. and {Goupil}, M.~J.}, 2006a, AA 447, 649

{{Su{\'a}rez}, J.~C. and {Goupil}, M.~J. and {Morel}, P.}, 2006b, AA 449, 673 \\


{{Su{\'a}rez}, J.~C. and {Garrido}, R. and {Moya}, A.}, 2007, AA 474, 971


{{Su{\'a}rez}, J.~C.}, 2007,  {EAS Publications Series}, 26, 121 \\


 {{Suran}, M. and {Goupil}, M. and {Baglin}, A. \etal ~ }, 2001, AA   372, 233 \\


 {{Sweet}, P.~A.}, 1950, MNRAS 110, 548 \\

{{Talon}, S. and {Zahn}, J.-P. and {Maeder}, A. and {Meynet}, G.},
 1997, AA 322, 209 \\



 {{Talon}, S.}, 2006, 
 {Proceedings of SOHO 18/GONG 2006/HELAS I, Beyond the spherical Sun},
 {ESA Special Publication},
624,37 \\

 {{Talon}, S.}, 2007, 
in proceedings of the Aussois school "Stellar Nucleosynthesis: 50 years after B2FH" {ArXiv e-prints},   vol. 708, \\

 
 {{Tassoul}, J.-L.}, 1978,  '{Theory of rotating stars}',
 Princeton Series in Astrophysics, Princeton: University Press \\


   
{{Thompson}, M.~J. and {Toomre}, J. and {Anderson}, E. \etal ~},
 {Science}, 1996,  272, 1300 \\


Thompson, M. J., J.Christensen-Dalsgaard, M.S. Miesh, J.Toomre,  2003,  
{\it The internal rotation of the Sun} 
Ann. Rev. Astron. Astrophys. 41, 599 \\


{Unno}, W. and {Osaki}, Y. and {Ando}, H. \etal ~, 1989, Nonradial 
	oscillations of stars, Tokyo: University of Tokyo Press, 1989, 2nd ed. \\

 {{Vogt}, H.}, 1929, {Astronomische Nachrichten}, 234, {93} \\



{{von Zeipel}, H.}, 1924a,  MNRAS 84, 665 \\

{{von Zeipel}, H.}, 1924b,  MNRAS 84, 684 \\







 {{Zahn}, J.-P.}, 1992, AA 265, 115 \\

{{Zahn}, J.-P.}, 2003, 
in 'Stellar astrophysical fluid dynamics',
eds M.J.~Thompson,  J. Christensen-Dalsgaard, 
Cambridge University Press, ISBN 0-521-81809-5, 2003, p.~205 \\


 {{Zahn}, J.-P. and {Brun}, A.~S. and {Mathis}, S.},
 2007, AA   474, 145 \\


Zima, W., Wright, D., Bentley, P.L.  \etal~ 2006, AA 455, 235  \\





  

   
\end{thebibliography}
\end{document}